\documentclass[final,dissertation]{msudoc}
\usepackage{color,graphicx}
\usepackage{amssymb,amsmath,eurosym,psfrag}
\usepackage{mathrsfs}
\usepackage{enumerate}
\usepackage{natbib}
\usepackage{caption}
\usepackage{subcaption}
\usepackage{url}

\newcommand\araa{\rm{ARA\&A}}%
\newcommand\apj{\rm{ApJ}}%
\newcommand\apjl{\rm{ApJ}}%
\newcommand\aap{\rm{A\&A}}%
\newcommand\solphys{\rm{Sol.~Phys.}}%
\newcommand\ssr{\rm{Space~Sci.~Rev.}}%
\newcommand\jgr{\rm{J.~Geophys.~Res.}}%
\newcommand{\figref}[1]{Figure \ref{#1}}
\newcommand{\Frac}[2]{\frac{\raisebox{-0.5ex}{\ensuremath{#1}}}{#2}}
\newcommand{\partials}[3]{\frac{\partial^{#3} #1}{\partial #2^{#3}}}

\newcommand{\vect}[1]{\mathbf{#1}}

\newcommand{\Matrix}[1]{\mathbb{#1}}
\newcommand{\tothe}[1]{\ensuremath{^{\hbox{\scriptsize #1}}}}
\newcommand{\order}[1]{\ensuremath{\mathcal{O}(#1)}}
\newcommand\arcsec{\mbox{$^{\prime\prime}$}}%

\newcommand{\csum}{\ensuremath{\textsf{c}\hspace{-0.75em}\sum}}
\newcommand{\fnc}[1]{\textsf{#1}} 
\newcommand{\unit}[1]{\ensuremath{\, \mathrm{#1}}}

\def\bigvar#1#2{{\hbox{$\left#1\vbox to #2{}\right. \n@space$}}}
\def\n@space{\nulldelimiterspace=0pt \m@th}
\def\m@th{\mathsurround=0pt }

\title{Energetic consequences of flux emergence}
\author{Lucas Adrian Tarr}
\authorshort{Lucas A.~Tarr}
\degree{Doctor of Philosophy}
\department{Physics}
\submissiondate{August 2013}
\copyrightyear{2013}

\chair{Dr.~Dana W. Longcope}
\head{Dr.~Richard J. Smith}
\dean{Dr.~Ronald W. Larsen}

\begin{document}

\frontmatter
\maketitlepage
\makecopyrightpage
\makeapprovalpage
\makepermissionpage

\begin{dedication}
  \begin{verse}
    In memory of my friend, Orrin Gorman McClellan, 1985--2010
  \end{verse}
\end{dedication}

\begin{preface}
  While the research presented in this dissertation is ultimately an attempt to understand basic physical processes in the cosmos, in particular the storage and release of energy by magnetic fields in plasmas, it does have important societal impacts.  It is for this reason that the work was funded, primarily through NASA's Living With a Star (LWS) program and associated missions and grants.  Work presented in the second chapter was supported directly by NASA LWS.  Work in the third and fourth chapter were supported by NASA under contract SP02H3901R from Lockheed--Martin to MSU.  The author also acknowledges the teams responsible for the Helioseismic and Magnetic Imager (HMI) and Atmospheric Imaging Assembly (AIA) instruments onboard the Solar Dynamics Observatory (SDO) and the X--Ray Telescope (XRT) instrument onboard the joint Japanese/US satellite Hinode.  This work would not be possible without the data and support provided by these instruments.
\end{preface}

\begin{acknowledgements}
  Undertaking the type and extent of work presented in a disseration is not a journey that can be accomplished by oneself.  Newton famously quipped that if he had seen farther than others, it is because he stood on the shoulders of giants.  The shoulders of giants may have been high enough for long--sighted Newton, but I needed to be bodily hurled by giants to reach the point I have.  It is hard to list everyone, so below I have listed just a few of those giants who, very luckily for me, decided to throw me at some point.  
  
  The MSU Solar Physics faculty in general, and my advisor Dana Longcope in particular.  Also David McKenzie, who pushed my physical thinking and picked up the tab for my education when necessary.  The members of our weekly research meeting: Silvina Guidoni, Anny Malanushenko, Sean Brannon, and Roger Scott.  Several other graduate students close to me, particularly Adam Kobelski and Paul T.~Baker.  Our department staff is extremely important.  Departments do not function without excellent staff, and our department seems to function exceedingly well.  My family and friends from Whidbey, Portland, and their respective diasporas: they may not have contributed directly to the physics presented in this disseration, but their love and support continues to propel me foward.  Finally, someone very important to me, who I had the pleasure of meeting near the end of all this, Sarah Jaeggli.
\end{acknowledgements}

\maketableofcontents
\makelistoftables
\makelistoffigures

\begin{abstract}
  When magnetic field in the solar convection zone buoyantly rises to pierce the visible solar surface (photosphere), the atmosphere (corona) above this surface must respond in some way.  One response of the coronal field to photospheric forcing is the creation of stress in the magnetic field, generating large currents and storing magnetic free energy.  Using a topological model of the coronal magnetic field we will quantify this free energy.  We find the free energy just prior to major flares in active regions to be between 30\% and 50\% of the potential field energy.  In a second way, the coronal field may topologically restructure to form new magnetic connections with newly emerged fields.  We use our topological model to quantify the rapid restructuring in the case of solar flare and coronal mass ejections, finding that between 1\% and 10\% of total active region flux is exchanged.  Finally, we use observational data to quantify the slow, quiescent reconnection with preexisting field, and find that for small active regions between 20\% and 40\% of the total emerged flux may have reconnected at any given time.
\end{abstract}

\mainmatter

\chapter{Introduction}\label{chap:intro}
\noindent ``If it wasn't for the magnetic field, the Sun would be as boring of a star as most astronomers seem to think it is.''\\
\hspace*{\fill} --- R.W.Leighton

The above quote, attributed to R.W. Leighton by Eugene Parker (both large--looming figures in the early stages of what became the discipline of Solar Physics) highlights two important facts in a straightforward and humorous manner.  First, that the Sun is not at all boring, but rather an aggressively pulsing ball of plasma; and second, that the magnetic field is the beating heart of this pulsing plasma.  To understand the Sun, and the many other stars like it, one must understand both the plasma composing the star and the magnetic field to which the plasma's dynamics inseparably couple.  That coupling spans many different processes in different time and length scales, from reconnection processes lasting a few minutes within current sheets of a meter or less across, to the magnetic dynamo which recycles the Sun's large scale dipole at a gigameter scale every 22 years, to the ``edge'' of the solar system where the solar wind couples to the interstellar medium around 20 terameters or more from the Sun, and may vary on thousand and million year timescales as the solar system passes through different regions of interstellar space.

We will not even briefly contemplate the vast majority of those processes and scales, for such is the work of generations, not dissertations.  But even these disparate phenomena are intimately related: the very local rapid reconnection may change the global topology of the low solar atmosphere and launch a Coronal Mass Ejection (CME) in the process, creating a supersonic disturbance that propagates through the solar wind, generating electrical currents within the Earth's crust and tearing off chunks of Martian atmosphere as it passes by.  To study one aspect of this is essential for informing our understanding of the others.  

And so, we will limit our present query to a single, small corner of this expansive topic, succinctly summarized in this dissertation's title, ``the energetic consequences of flux emergence.''  We will discuss only briefly \emph{why} concentrated chunks of magnetic flux emerge through the visible solar surface, the photosphere, in the first place; or what dynamical processes occur immediately after a rapid topological change and relaxation; or even detailed analysis of the field's geometry during emergence.  Instead, we will focus on the amount of magnetic energy that is injected into the corona during emergence, how much of that energy is available for conversion into kinetic and thermal energy of the plasma, and how those quantities vary with time.

In the several sections which comprise this Introduction, we will discuss the observational and theoretical concepts which place the body of this dissertation in context.  First among these is that, observationally, the Sun has a magnetic field, concentrated patches of which form \emph{active regions}.  Second, we will discuss flux emergence in particular, introducing the dynamical equations along the way.  Finally, we will turn to determination of the coronal magnetic field itself, and how that field can store the energy required to power solar flares.  We end the final portion of the Introduction with an overview of the remaining 4 chapters of this dissertation.

\section{\label{sec:histobs}Observations of the Sun's Magnetic Field}

In 1908, George Ellery Hale published an amazing article in \textsl{The Astrophysical Journal}, entitled ``On the Probable Existence of a Magnetic Field in Sunspots'' \citep{Hale:1908}.  The work drew on a surprising number of contemporary experiments to conclude, even if only tentatively, that sunspots contained magnetic field.  These experimental findings included: that a charged, spinning disk produces a magnetic field; that gases, when ionized, contain charged particles; that many neutral elements at high temperature emit numerous negative ``corpuscles'', and so must also have positively charged particles; that the Sun contains such hot gases; that the Sun  also has rapidly moving ``vortices,'' and so likely generates a magnetic field in places; that Zeeman had demonstrated that radiating gas placed in a magnetic field produces emission doublets, with noted polarization states; and finally that a new spectrograph on the Mount Wilson telescope allowed for precise measurements of the solar spectrum at various locations on the solar disk, and various polarization states.  Combining all these, Hale succeeded at comparing observations of line splitting and polarization in sunspots to laboratory observations of emitting gases in magnetic fields, finally concluding that the sunspots likely contained magnetic field of about a kilogauss in magnitude.

Hale spent the next few decades attempting to measure magnetic phenomena on the Sun, and coordinating global efforts to do so: he secured funding to send spectrographs to telescopes around the world to take solar spectra.  Because of limited spatial and spectral resolution, problems with personal bias in measuring Zeeman splitting, and weakness of the fields, Hale and contemporaries were unable to conclusively measure the magnetic field strength outside of sun spots, even though such field was inferred by the existence of coronal streamers observed during solar eclipses \citep{Babcock:1963, Stenflo:1970}.

\begin{figure}[ht]
  \begin{center}
    \begin{subfigure}[b]{0.48\textwidth}
      \centering
      \includegraphics[width=\textwidth]{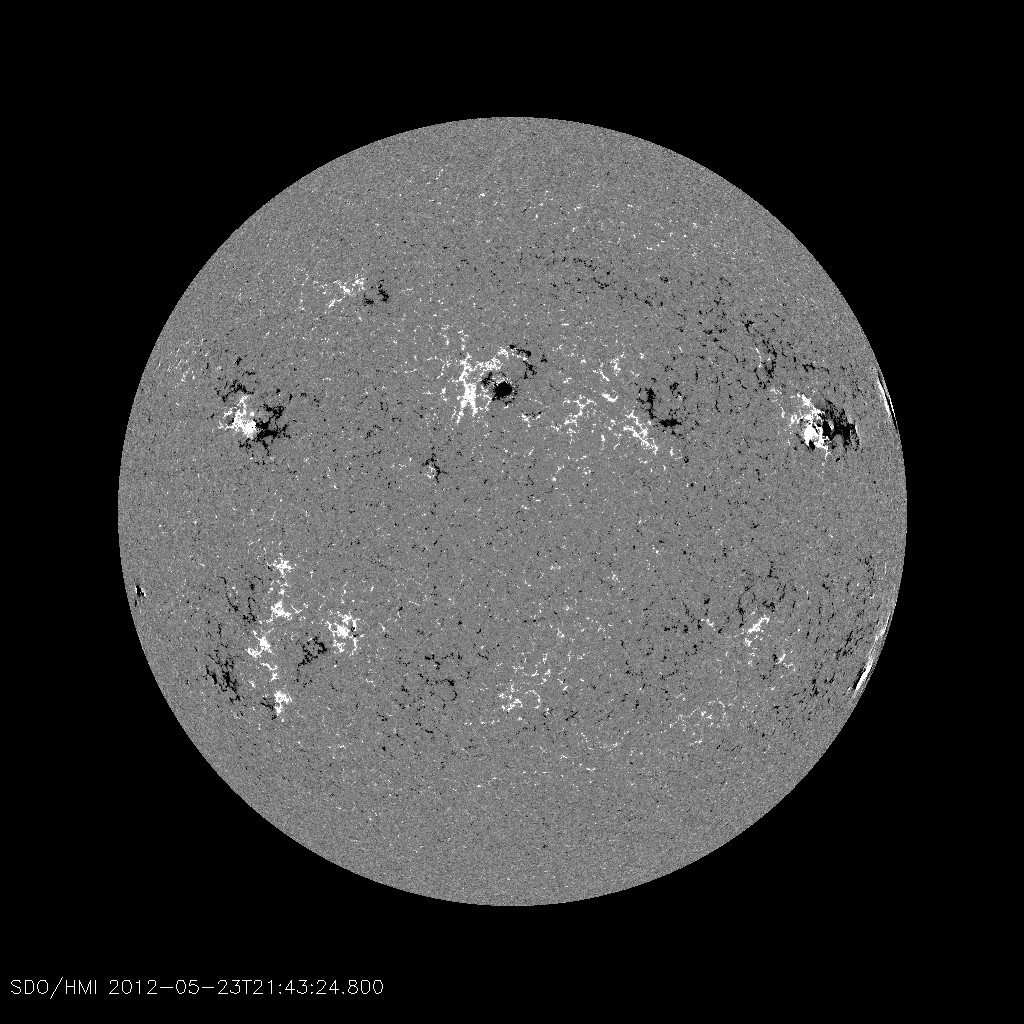}
    \end{subfigure}
    \begin{subfigure}[b]{0.48\textwidth}
      \centering
      \includegraphics[width=\textwidth]{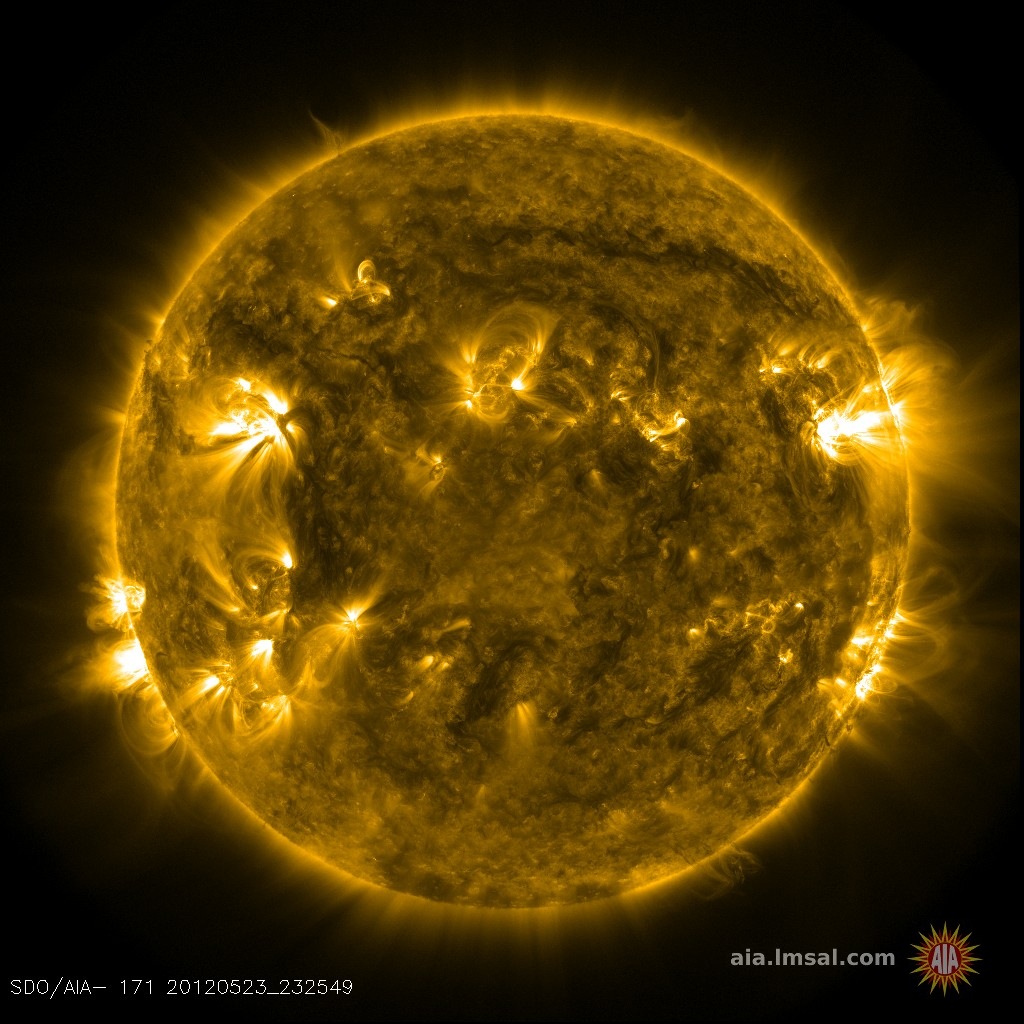}
    \end{subfigure}
  \end{center}
  \caption[Full--disk Magnetogram and 171\AA{} EUV]{\label{fig:fd} Left: Full--disk magnetogram from SDO/HMI, showing the strength of the line--of--sight magnetic field in each pixel.  Dark pixels point away from the observer, light pixels towards.  Right: Full--disk EUV image (171\AA{}) from SDO/AIA, at the same time.  Note that the concentrated patches of magnetic flux lie at the footpoints of the bright EUV loops, with loops connecting positive to negative flux regions (light to dark in the magnetogram).}
\end{figure}

Unfortunately, Hale died in 1938 and therefore missed the development of the magnetograph in the 1950s.  This device allows for relatively rapid generation of maps of the (usually line--of--sight component) magnetic field strength at the solar surface, creating an image known as a magnetogram \citep{Babcock:1963}.  On these early instruments, a full disk magnetogram took about 1 hour to create, with a resolution of about $23\arcsec$ (as seen from Earth, the Sun's radius when measured in visible light is about $970\arcsec$).  In the last half century we've increased both spatial and temporal resolution by about two orders of magnitude: the left panel of \figref{fig:fd} depicts a modern magnetogram from the Helioseismic and Magnetic Imager (HMI) onboard the Solar Dynamics Observatory (SDO), created in 45 seconds with a resolution of about $1\arcsec$.  

Following the initial magnetographs, a new method using photographic plates gave better spatial resolution but made less sensitive field strength measurements.  Even still, using the new magnetographs, Babcock and Babcock (1955) established the existence of the general dipolar field at this time, as well as the reversal of that field during the most recent solar maximum.  This latter discovery came as something of a revelation of the truly dynamic nature of the solar magnetic field \citep{Babcock:1963}.

The newly discovered ``solar dynamo''---the periodic reversal of the Sun's dipolar field---was quickly described by the Babcock--Leighton model, where polar field reversal is due to the decay of active regions over the course of a solar cycle.  This dynamo model fell out of favor for a couple of decades, but has recently been revived due to recent long term magnetic field observations of solar cycles 21 and 22, namely that polar field reversal is indeed intimately related to the decay of active regions \citep{Charbonneau:2010}.  One aspect of this model is that the magnetic field of the Sun becomes wound--up around the rotation axis due to differential rotation.  The now stretched out field forms \emph{flux tubes}, which buoyantly rise after the onset of Parker's magnetic buoyancy instability \citep{Parker:1955}.   This finally leads to the emergence of flux tubes as they pierce the solar surface, each leg of which forms a sunspot.  The combined effects of sunspot tilt---the leading polarity being closer to the equator than the trailing---gradual diffusion, and flux cancellation with the field of decaying spots from the other side of the equator leads to a situation where the toroidal field of the spots has become preferentially poloidal and concentrated at the poles, and the cycle may then repeat.

Flares were noted to occur within or adjacent to complex (multi--polarity) and rapidly changing sunspot regions, which are also often properties of young regions: flux tubes that had very recently broken through the solar surface.  As of 1963, both observers and theorists favored the idea that flares were associated with the magnetic field of sunspots, as opposed to the surrounding network field \citep{Babcock:1963}, and that they were primarily driven by magnetic field energy.  Precisely how that energy was transferred from the magnetic field to thermal, radiative, and kinetic energy was, and remains, a continuing topic of debate.

10 years later, and nearly 70 years since the first magnetic observations of Hale, it became observationally clear that emerging magnetic flux represented a large injection of energy into the coronal plasma \citep{Rust:1975}, and that this was a direct source of energy for flares.  Rust and Bridges used a new array of photodiodes to rapidly create time series of solar data, including both magnetograms and emission spectra in continuum, H$_\alpha$, Ca II, He I, H I, and Fe I.  In studying that initial data, they found a strong correspondence (8/12 cases) between emerging flux and flares.  In 3 of the remaining 4 cases, observational constraints made unambiguous association between flares and emerging flux impossible.  Based on this evidence, they concluded with the rather strong statement that ``\emph{flares start within $5\arcsec$ of some point where new magnetic fields are emerging through the photosphere}'' (emphasis theirs).

The observations of \citet{Rust:1975} finally led to a rough model of flux emergence by \citet{Heyvaerts:1977} that could naturally explain the plethora of observational features found in flaring regions: electromagnetic radiation from radio to X--ray wavelengths, particle acceleration, mass expulsion, and so on.  Their model still serves as the foundation for theoretical and computational models of flux emergence and its consequences today---even after 4 decades, the paper receives 10 or 15 citations each year.

Since the 1970s, both data collection and modeling capabilities have rapidly advanced.  Many of the numerical modeling ideas originally introduced in the 1970s have only recently stepped into the limelight with modern computing resources (cf. reviews by \citet{Fan:2009, Archontis:2008, Wiegelmann:2012}).  Despite this, some very fundamental questions remain concerning the properties of emerging flux regions: \emph{Will it produce flares and/or CMEs?  How much energy will they release?  Will they be single events or a string of homologous events? will it sympathetically disturb other regions present on the Sun?  Exactly how much energy does it have in the first place?}  In this dissertation of course, we will focus on the last of these.  Actually answering that question should help us out with the others, as well.

\section{\label{sec:emerge}Solar Magnetic Fields and the Dynamical Equations}

Almost all of the \emph{stuff} that we can see in the universe lives in a plasma regime such that its dynamics are governed by the magnetohydrodynamic (MHD) equations: the stuff is a gas of unbound ions and electrons, whose independence endows the plasma with a large electrical conductivity.  In Earth--centric, physical space terms, this regime basically starts in the magnetosphere and extends outward, encompassing the dynamics of the solar wind, the Sun, the interstellar medium, galactic jets, and so on.  In these environments, the physical variables are the plasma velocity field $\vect{v}$ and magnetic field $\vect{B}$: one solves the MHD equations in terms of these variables, and then, if one needs the current density and electric field, $\vect{J}$ and $\vect{E}$, they may be calculated using Maxwell's equations after the fact.\footnote{Parker published ``Conversations on Electric and Magnetic Fields in the Cosmos'' in 2007 essentially as a manifesto to drive home this point.}  The important point here is that the magnetic field $\vect{B}$ is a real physical variable, not simply a convenient mathematical construct, whose properties determine the dynamics of the system.

In contrast to the``reality'' of $\vect{B}$ away from Earth as described above, the situation is reversed from the ionosphere down.  In this regime, we normally think of applying a voltage to an electrical system which drives a current and in turn generates a magnetic field via Biot--Savart/Amp\`ere's Law: the magnetic field is an effect due to electrical causes, ions and electrons are bound into atoms, and the magnetic field's dynamics are decoupled from those of the gas.  That the MHD conditions do not normally exist on Earth except in specifically prepared laboratory environments (like a Tokamak) makes many characteristics of the subject unfamiliar and counterintuitive.  From the perspective of almost everywhere else in the known universe, however, our little ball of space where the magnetic field usually plays a subservient role is the odd man out.

With this in mind, the conditions for basic MHD are that the system's (i) gradient length scale is larger than the Debye length; (ii) timescale is larger than the inverse of the plasma frequency; and (iii) velocities are nonrelativistic, neglecting terms of \order{v^2/c^2}.\footnote{Many texts deal with MHD; three that I have found very useful are \citet{Parker:2007, Schrijver:2009, Choudhuri:1998}}  (i) and (ii) enforce quasi--neutrality of the plasma, and ensure that all particle species (electrons and the various ions) have the same temperature, so the system is described as a single fluid.  Condition (iii) may be broken to accommodate the more exact relativistic models, and sometimes certain aspects of (i) and (ii) may be relaxed to describe multi--fluid MHD models, but great care must be taken in such relaxations.  The system must always be describable as a fluid.  Otherwise, one must use kinetic theory to calculate the system's dynamics, and the resulting problems, especially on large scales, become even more intractable than they already are. 

Thus introduced, the equations that generally govern the cosmos are mass conservation (continuity), momentum conservation (force balance), energy conservation, and magnetic induction: 
\begin{align}
  \label{eq:cont} \partials{\rho}{t}{} & = - \nabla\cdot(\rho\vect{v})\\
  \label{eq:mom} \rho\partials{\vect{v}}{t}{} & = -\rho(\vect{v}\cdot\nabla)\vect{v} -\nabla p + \Frac{1}{4\pi}(\nabla\times\vect{B})\times\vect{B} + \rho\vect{g}+\mu\nabla^2\vect{v}\\
  \label{eq:energy} \rho\partials{e}{t}{} & = -\rho(\vect{v}\cdot\nabla)e - p\nabla\cdot\vect{v} + \nabla\cdot(\kappa\nabla T) + \dot{Q}\\
  \label{eq:induct} \partials{\vect{B}}{t}{} & = \nabla\times(\vect{v}\times \vect{B}) - \nabla\times(\eta\nabla\times\vect{B})
\end{align}
where $\rho$ is plasma mass density, $\vect{g}$ gravitational acceleration, $\mu$ dynamic viscosity, $e$ specific energy, $\kappa$ thermal conductivity, $T$ temperature, $\dot{Q}$ a generic bulk heating rate, and $\eta = c^2/4\pi\sigma$ the magnetic diffusivity with $c$ the speed of light and $\sigma$ the electrical conductivity.  In the momentum equation, we generally neglect the gravitational and viscous terms, but include them above for completeness.  For completeness, these equations must be closed through an equation of state, expressing $e$ and $T$ each in terms of $\rho$ and $p$.

The induction equation \eqref{eq:induct} follows from Faraday's Law after expressing the electric field in the frame of reference of the plasma moving with velocity $\vect{v}$ and neglecting terms of order $v^2/c^2$.  If $\vect{E}^\prime$ is the electric field in the rest frame of a fluid parcel and $\vect{E}$ is in the lab frame then 
\begin{align}
  \label{eq:transformE}  \vect{E}^\prime & = \vect{E}+\frac{\vect{v}}{c}\times\vect{B}
  \intertext{while for the magnetic field}
  \label{eq:transformB}   \quad \vect{B}^\prime & = \vect{B}-\frac{\vect{v}}{c}\times\vect{E}\qquad .
\end{align}
For a highly conducting plasma, $\vect{E}^\prime\approx 0$ \citep{Parker:2007}, so that $E\sim \frac{v}{c}B$.  This makes $\vect{B}^\prime = \vect{B}$ to \order{v/c}.

Therefore, in studying flux emergence, we are really studying the dynamics of the velocity and magnetic fields $\vect{v}$ and $\vect{B}$ governed by the induction equation.  Taking the ratio of the two terms on the RHS of Equation \eqref{eq:induct} for typical length, velocity, and field strength L, V, and B, we define the magnetic Reynolds number:
\begin{equation}
  \mathcal{R}_M = \frac{V B/L}{\eta B/L^2} = \frac{VL}{\eta}\qquad .
\end{equation}
Let us apply this to the Sun.  For a plasma of ionized Hydrogen at $T=10^4\unit{K}$ the diffusivity is about $\eta = 10^7 \unit{cm}^2\unit{s}^{-1}$ \citet{Choudhuri:1998}.  A solar convection cell at the photosphere is about 1Mm across and has velocity of $\sim 1\unit{km}\unit{s}^{-1}$ \citet{Rast:1993}, so we find that the magnetic Reynolds number is $\mathcal{R}_M\approx10^6$: the ideal term completely dominates the dissipative term.  

For this reason, we will work almost exclusively with ideal MHD where the conductivity $\sigma$ is so large that we set $\eta\rightarrow 0$.  In that case, Alfv\'en's Theorem holds\footnote{As described by \citet{Choudhuri:1998}, the theorem goes back to Helmholtz in 1858, and was proved by Lord Kelvin in 1867.  The more general proof, given in \citet{Choudhuri:1998}\S4 and \citet{Schrijver:2009}\S3, appeals directly to the Fundamental Theorem of Calculus and is therefore widely applicable.  It states that if a vector field $\vect{Q}$ satisfies the equation $\partial_t\vect{Q}=\nabla\times(\vect{v}\times\vect{Q})$ then $\frac{d}{dt}\int_S\vect{Q}\cdot d\vect{S} = 0$.  Alfv\'en noted that the ideal induction equation provided such an instance.} and the lines of magnetic force move in tandem with the flow field $\vect{v}$ and are essentially embedded within that field; it is therefore also known as the ``frozen--in--flux'' condition.  We consider this to hold except in small regions of locally enhanced, anomalous resistivity.  In such locations, magnetic field lines ``slip'' relative to velocity field, by which we mean reconnection occurs, and the topology of the magnetic field correspondingly changes.  There is mounting evidence that this so--called ``patchy reconnection'' process takes place continually on the Sun \citep{Savage:2012a, Savage:2012b}.

Equations \eqref{eq:cont} through \eqref{eq:induct}, together with an equation of state, form a set of coupled nonlinear partial differential equations, and are certainly difficult, if not impossible, to solve in any general fashion.  This is especially true in the case of flux emergence where the relevant dynamics span many orders of magnitude in length and timescales.  \citet{Archontis:2008} reviews the numerical MHD simulations of flux emergence over the years, with a focus on emergence through the photosphere and into the corona, while \citet{Fan:2009} provides a review of the field with many more observational details, and more focus on emergence through the convection zone.  Full 3D MHD simulations of emergence from convection zone to corona are not yet possible for a variety of reasons.  Typically, such simulations can treat about 3 density scale heights.  The scale height changes so rapidly near the photosphere that even bridging the upper $1\%$ of the convection zone to coronal region is not possible, although very recent simulations are approaching this realm \citep{Moreno:2013}.

Very briefly, the overall magnetic field of the Sun is generated through a dynamo process, as first discovered in the 1950s.  The dynamo oscillates the field between an overall poloidal field and toroidal field.  Starting from the dipole configuration, the differential rotation of the convection zone (varying rotation rate in both latitude and radius) winds the magnetic field toroidally around the Sun by virtue of the frozen--in--flux condition: the plasma $\beta \gg1$\footnote{$\beta$ is the ratio of magnetic to gas pressure, and defined more precisely in the following section.} and the magnetic diffusion rate is slow compared to the convective timescale, so the magnetic field is stretched and driven around by the convective plasma motions.  At some point, sections of this twisted field become buoyantly unstable to perturbations and rise through the convection zone.  

The onset of instability may occur through any number of processes, as discussed in the reviews \citet{Fan:2009}, and \citet{Archontis:2008}.  Whatever the method, the end is the same: a loop shaped like an $\Omega$ (so--called ``Omega--loops'') which rises buoyantly.  When such a loop pierces the photospheric surface we see the observational signature of flux emergence: the rapid appearance of a mixture of opposite polarity flux patches that move apart to coalesce into two larger concentrations, one of each polarity.

The interaction of the emerging flux with preexisting coronal field is driven primarily by the dynamics of the current sheet that forms between the two magnetic domains.  As mentioned above, the original model of this interaction was developed in \citet{Heyvaerts:1977}, informed strongly by the observations of \citet{Rust:1975}.  They put forth a number of simple calculations of relevant timescales, reconnection rates, and heating rates for different phases of emergence.  Many of the properties they introduced were later analyzed in exhaustive detail in theoretical and numerical treatments scores of authors, a work which continues to this day.  

Much of this work has focused on the dynamics of flux emergence, energy release after emergence, and energy conversion within current sheets.  For example, \citet{Tur:1978} provide a very detailed, if ultimately qualitative, treatment of energy balance within the current sheet as it rises into the atmosphere, and specifically as it reaches a threshold temperature and jumps from one stable branch solution to another.  In a later 2D model, \citet{Forbes:1984} define 4 phases of reconnection between new and old flux as a flux tube emerges, with each phase marked by different reconnection rates.  Recent 3D models see similar successive phases of reconnection marked by different reconnection rates and the onset of different types of instabilities \citep{Moreno:2013}.

Comparatively little effort had focused on the amount of energy storage.  Even the very recent simulation of \citet{Moreno:2013}, which spends some amount of space discussing free energy, does not do so in any quantitative fashion.  Many that do focus on quantifying this energy do so by considering the Maxwell stresses generated by photospheric motions, or magnetic shear \citep{Longcope:2004, Baumann:2013}.  What we will focus on in the following section is stress induced by the emergence itself.  The effects of shear are of course non--negligible, and they are implicitly part of the calculations of Chapters 3 and 4.  What we will not consider, however, is any storage mechanism other than magnetic stresses.

\section{\label{sec:cor_fields}Coronal Magnetic Fields}
We now focus on the fundamentals of determining the free magnetic energy of the coronal magnetic field.  In the corona, the magnetic field is likely \emph{force free}.  In this situation, the magnetic field energy density dominates plasma pressure.  With no other comparable forces around, the Lorentz force must balance itself, hence the term force free.  In static equilibrium, $\vect{v}=\partials{}{t}{}=0$ in \eqref{eq:mom}.  Neglecting gravity and viscosity and taking the ratio of the two remaining terms in that equation, for a given characteristic length scale $L$, we have $\frac{p/L}{B^2/4\pi L} = \frac{4\pi p}{B^2} = 2\beta$.  This term is known as the plasma $\beta$ and is defined in terms of the ratio of gas to magnetic pressure, the latter derived by breaking the Lorentz force into the gradient of a scalar portion and a field--aligned gradient: $\frac{1}{4\pi}(\nabla\times\vect{B})\times\vect{B} = -\frac{1}{8\pi}\nabla B^2 + \frac{1}{4\pi}(\vect{B}\cdot\nabla)\vect{B}$.  The first term, $\nabla B^2/8\pi$, acts in all directions and is known as the magnetic pressure.  In many regions of the solar corona the magnetic pressure completely dominates the gas pressure \citep{Gary:2001}.  Even where it does not dominate (which is actually in many places, and is really the point of \citet{Gary:2001}), it is such a useful approximation to make that many people do it anyway.  In any case, it typically is true that in the low corona in the vicinity of active regions that $\beta \ll 1$, in which case the static equilibrium self--balancing Lorentz force momentum equation \eqref{eq:mom} becomes 
\begin{equation}
  (\nabla\times\vect{B} )\times\vect{B} = 0\, .
\end{equation}
This, in turn, implies that 
\begin{equation}
  \label{eq:fff}
  \nabla\times \vect{B} = \alpha(\vect{r})\vect{B}
\end{equation}
where $\alpha(\vect{r})$ is some scalar field.  If $\alpha = 0$ then we may write the magnetic field in terms of a scalar potential $\vect{B} = -\nabla\Phi_M$.  The divergence of $\vect{B}$ is always zero, so the scalar potential solves Laplace's equation: $\nabla^2\Phi_M = 0$.  

The scalar potential also arises from another short calculation.  The energy density (magnetic pressure) of the magnetic field is $W = \frac{1}{8\pi}\vect{B}\cdot\vect{B}$, and the total magnetic energy $U_M$ is the integral of $W$ over all space.  With $\nabla\cdot\vect{B}=0$ always and everywhere, we may write the magnetic field in terms of a vector potential: $\vect{B} = \nabla\times\vect{A}$.  Extremizing $U_M$ through the variation $\vect{A}\rightarrow\vect{A}+\vect{a}$ and requiring that the variation vanish at suitable boundaries forces $\nabla\times\vect{B} = 0$.  This implies that, once again, the magnetic field may be written in terms of the scalar potential, and we see that the potential field is also the field of minimum energy.  Because extremas of the variational calculus, as well as solutions to Laplace's equations, are unique, the potential field is also unique.

Returning to \eqref{eq:fff}, solutions for which $\alpha$ is uniform but nonzero are known as linear force free fields (LFFF).  LFFFs have more energy than the potential field.  Solutions for which $\alpha$ is a spatially varying scalar field are nonlinear force free fields (NLFFF), and also have more energy than the potential field.  If the magnetic field in the solar corona is to provide the energy to drive solar flares (an idea already well established in the 1960s \citep{Babcock:1963}), then initially the field must be in some stressed state and relax towards a more potential configuration while giving up some of its free energy in the process.  Determining the degree of nonpotentiality is how one determines the free energy of the field and therefore how much energy is available for coronal heating, solar flares, particle acceleration, and so on.  For the remainder of this section we will discuss several methods for determining this free energy.

Calculation of nonpotential fields began 40 years ago with the analysis of \citet{Tanaka:1973}.  Being the first excursion away from potential field models, they used the next simplest field extrapolation, the linear force free field, for which $\alpha$ is uniform through the volume.  The particular value of $\alpha$ is related to the amount of twist in the field, with positive and negative values representing right and left handed twist.  The average twist in a field is related to a quantity known as magnetic helicity \citep{Longcope:2004}, and the amount of helicity in turn is strongly related to the free energy (see the monograph \citet{Brown:1999} for an extensive review of the subject).

Because $\alpha$ is a measure of how twisted the magnetic field is, the prescription of a constant $\alpha$ field is quite a strong requirement, in the following sense.  A magnetic field line is defined as a line $\vect{s}$ which satisfies the equation $\frac{\partial\vect{s}}{\partial r} = \frac{\vect{B}(r)}{\vert\vect{B}(r)\vert}$.  You can show that field lines are single--valued in $\alpha$, so that $\alpha$ is a constant along any given line \citep{Malanushenko:2011}.  Recent work has confirmed by multiple avenues that a single active region contains field lines of both positive and negative values of varying magnitude $\alpha$ \citep{Malanushenko:2011, Sun:2012}.  Thus, far from being single--valued everywhere as required by linear force free fields, $\alpha$ is quite generally a spatially varying scalar function, specified by the properties of the magnetic field as measured at the lower boundary \citep{Wiegelmann:2012}.

\citet{Tanaka:1973} showed that one may determine how much free energy is available in the volume around an active region by either a static or time difference method.  If one performs a field extrapolation from the lower boundary, either linear or non--linear (see review by \citet{Wiegelmann:2012}), the free energy may easily be found by integrating $B^2$ over the volume and subtracting off the potential portion, $B_p^2$:
\begin{equation}
  \label{eq:wfree}
  U_f = \frac{1}{8\pi}\int\vect{B}\cdot\vect{B}d^3x - \frac{1}{8\pi}\int\vect{B}_p\cdot\vect{B}_pd^3x\, .
\end{equation}
The extrapolated field in the volume is derived from a single vector magnetogram, and the above calculation provides both the total energy (first term) and free energy (the difference) in the system at a given snapshot in time.  Even if one has a time sequence of vector magnetograms, as has recently become routinely available with the SDO/HMI and Hinode/SOT instruments, the calculated fields at each timestep are essentially unrelated.  Variations in the extrapolated field from one time to the next lead to fluctuations in the free energy of the same order as major (GOES M--class) flares \citep[see][Figure 4.]{Sun:2012}.

\citet{Tanaka:1973} did have a series of vector magnetograms for a flaring active region, six total over eight days.  They performed linear force free field and potential field extrapolations at each timestep to generate the estimate of free energy \eqref{eq:wfree}.  They also developed a very simple model of the relative motions of flux concentrations within the region.  Such a model may be used to determine the rate of energy increase in the following way.  Starting with the induction equation \eqref{eq:induct} and assuming a constant diffusivity $\eta$, the induction equation becomes
\begin{gather}
  \partials{\vect{B}}{t}{} = \nabla\times(\vect{v}\times\vect{B}) + \eta\nabla^2\vect{B}\, .
  \intertext{From this, we can find the time rate of change of the energy density:}
  \partials{}{t}{} W  = \frac{1}{4\pi} \partials{\vect{B}}{t}{}\cdot\vect{B} = \nabla\times(\vect{v}\times\vect{B})\cdot\vect{B}\, ,
\end{gather}
where we have now dropped the resistive term from the induction equation, but it may be included if desired.  After integrating both sides, a few lines of algebra, and application of Stokes' Theorem, we find that the total energy changes as
\begin{gather}
  \label{eq:wgrow}
  \partials{}{t}{}\int \frac{\vert\vect{B}\vert^2}{8\pi} = \frac{1}{4\pi}\int_S [(\vect{v}\times\vect{B})\times\vect{B}]\cdot d\vect{S}
\end{gather}
where the surface $\vect{S}$ is the lower boundary, typically the photospheric $x-y$ plane, and we assume the field falls to zero at the other boundaries.  \citet{Tanaka:1973} used their model of the evolution of the flow field $\vect{v}$ to calculate the energy change between successive timesteps and compared this to the direct calculation from the extrapolated fields.  The method relying on photospheric motions essentially estimates energy injected into the coronal field due to a Poynting flux through the photospheric boundary.  They found the energy rate derived from photospheric motions to be about a factor of three lower than that from differencing the daily values of Equation \eqref{eq:wfree}, and consider the former a more realistic calculation.  As they note, the actual field has a spatially varying $\alpha$, with large values (strong shear; high twist) confined to a small region close to the polarity inversion line (PIL) and weaker shear (closer to potential field) farther away.  This makes much of their extrapolation incorrect, more non--potential than is likely, and therefore results in an unrealistically high free energy.

Of the two methods \citet{Tanaka:1973} proposed for deriving the energy of the coronal field, the method corresponding to Equation \eqref{eq:wfree} forms the foundation of the most often attempted class of energy storage calculations.  For the last 20 years or so, the field has typically been extrapolated using a spatially (and often temporally) varying force--free parameter $\alpha(r,t)$ to generate a nonlinear force free field (NLFFF) \citep[see][and references therein]{DeRosa:2009, Wiegelmann:2012}.

As pointed out in \citet{Mackay:2011}, most studies are inherently static, creating an independent extrapolation for each vector magnetogram in a data series.  A different approach uses a magnetofrictional method together with the induction equation to evolve the lower boundary while continuously relaxing the extrapolated coronal field \citep{Chodura:1981, Yang:1986, Craig:1986, VanBallegooijen:2000, Yeates:2008, Yeates:2010, Mackay:2011}.  The magnetofrictional method relies on a fictitious fluid velocity that is parallel with, and proportional to, the Lorentz force.  This ficticious velocity field relaxes the system to a force free state reguardless of its initial state, monotonically decreasing the energy in the process \citep{Craig:1986}.

In terms of an energy calculation, the introduction of a velocity may seem to follow the \eqref{eq:wgrow} path.  However, it must be remembered that this velocity is ficticious.  When these authors do estimate the magnetic free energy, they necessarily do so from their extrapolated fields via the \eqref{eq:wfree} method.  The fields at different times are related because the relaxed, force free state at one time is used as an initial condition for the next.  Thus, while the field retains a memory of its past state, it is unclear to what extent the loss of energy during relaxation relates to energy loss in the real corona.  

As \citet{Tanaka:1973} briefly noted, and \citet{Metcalf:1995} later considered in detail, the measured field in the photosphere is far from force free, so the boundary condition one uses to perform the extrapolation is inconsistent with the extrapolation itself.  Yet, an advantage of using an extrapolation is that you may trace out representative magnetic field lines in the calculated field and compare their geometry with that of coronal loops observed in EUV data, such as those seen in \figref{fig:euvsmall}.  If you then modify the extrapolated field to more closely match observed loops, presumably the end result is a more realistic extrapolation.  In that case, the modified lower boundary may resemble the field at some height above the photosphere where the field becomes approximately force free.  \citet{Malanushenko:2012} have done just this, using the EUV data as an additional constraint in determining a NLFFF.  It is a static representation of the field, but this could presumably be modified in the future to study both field and free energy evolution.

Direct use of the induction equation via \eqref{eq:wgrow} to calculate coronal free energy is much less common, though such efforts are currently in the works.  \citet{Welsch:2006} provides a detailed derivation of Poynting fluxes at the observed photospheric boundary to determine the change of both total and free magnetic energy.  The method contains its own pitfalls, but avoids those encountered during extrapolation.  \citet{Fisher:2012} discuss the actual implementation of such a calculation with an emphasis on using the recent deluge of vector magnetic field data from SDO/HMI, and it will be very exciting to compare these results when they arrive to those we discuss in the following three chapters.

\section{\label{sec:mcc}The Minimum Current Corona}

\begin{figure}[ht]
  \begin{center}
    \includegraphics[width=0.5\textwidth]{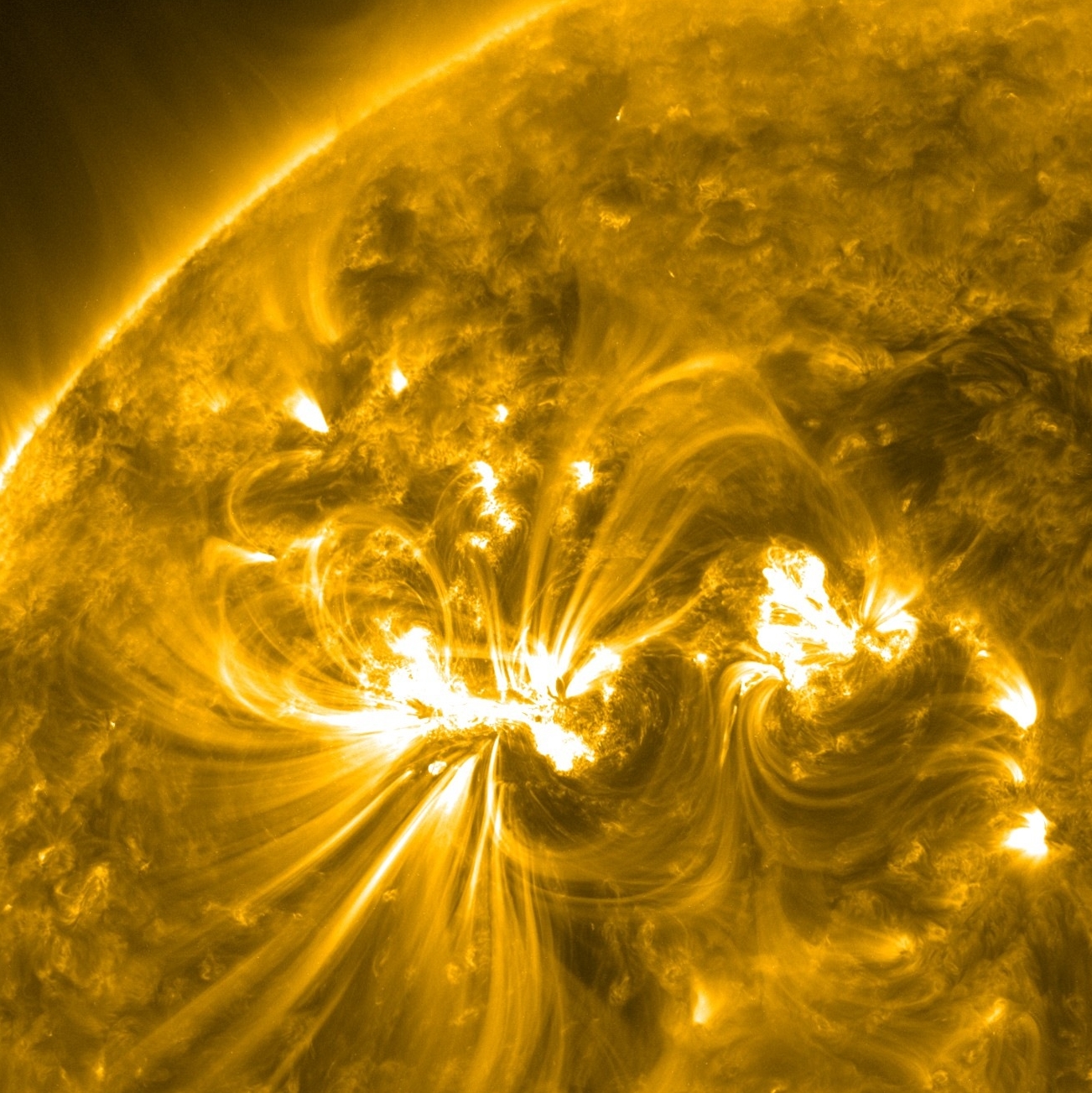}
    \caption[EUV flux domains in 171\AA{}]{\label{fig:euvsmall} Active region complex near the Northeast limb observed on 2012/03/06 in 171\AA{} light by SDO/AIA.  Note the apparent clumping of the coronal loops into distinct flux domains.}
  \end{center}
\end{figure}

In contrast to the two methods just described for determining coronal magnetic free energy, in the following chapters we will employ a very different conceptual framework.  Our work focuses on the topology of the system, with the topology defined by the bounding surfaces that separate different flux domains.  Flux domains are defined through the mapping of magnetic field lines.  Typically, a field line is mapped from one photospheric positive flux concentration to a photospheric negative concentration.  Domains are distinguished by their photospheric footpoints: field lines in one domain connect two concentrations of photospheric flux, while those in another domain connect a different pair of sources.  The subset of fieldlines that do not connect two sources define the system's topology.  Instead, these field lines either originate or terminate  at a nullpoint, a location where $\vect{B}=0$.  The set of such field lines for a given null form \emph{spine} field lines and \emph{fan} surfaces, which divide the coronal volume into its constituent flux domains.

Finding the location of the surfaces bounding each domain and the locations of the surfaces' intersection was work pioneered by \citet{Baum:1980}, though as they say the concept of the importance of distinct flux domains originates with Sweet in the late 1950s\footnote{\citet{Baum:1980} cite an article from 1958 which I have not been able to track down, although Sweet did publish on related topics at the same time: see \citet{Sweet:1958}.}.  The state of the system is then described by its connectivity: the amount of flux connecting each photospheric concentration to every other one \citep{Longcope:1996}.  This is a very natural way to describe the corona, given observations of distinct bundles of coronal loops in EUV data, as shown in \figref{fig:euvsmall}.

\begin{figure}[ht]
  \begin{center}
    \includegraphics[width=\textwidth]{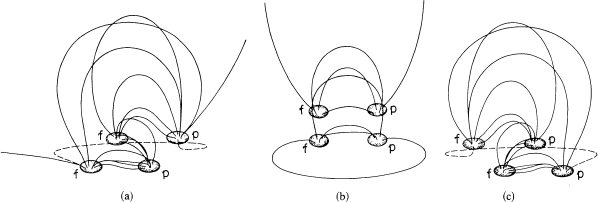}
    \caption[Baum \& Bratenahl (1980) Figure 1]{\label{fig:baumfig1} Representative field lines in a potential field extrapolation for differing geometries.  Adapted from \citet{Baum:1980} Figure 1.}
  \end{center}
\end{figure}

Following the original framework developed by \citet{Baum:1980}, we represent extended photospheric flux concentrations as point sources, a representation known as a Magnetic Charge Topology (MCT).  Here, the state of the field is determined only by the amount of flux connecting each charge to every other charge.  The potential field is unique (as always) and therefore has a unique distribution of flux between all the charges.  As illustrated in Figure 1 of \citet{Baum:1980}, reproduced in \figref{fig:baumfig1}, the exact distribution of flux in the potential field depends on the geometry.  The three cases shown in \figref{fig:baumfig1} represent two bipolar flux distributions in the northern hemisphere, one poleward (top) and one equatorward (bottom).  Each bipole includes one preceding ($p$) and one following ($f$) polarity, and a shown are a few representative field lines originating in each polarity.  In (b) the two bipolar regions are aligned, and in this degenerate case no field lines connect the equatorward to poleward bipoles.  As the two bipoles are shifted relative to one another along the direction of rotation, the amount of flux interconnecting to four polarity changes.  In scenario (a), the equatorward $p$--poleward $f$ domain lies underneath the equatorward $f$--poleward $p$ domain, while the reverse is true for scenario (c).  This demonstrates how the potential field configuration itself changes under relative shear motion of photospheric flux concentrations.

Any departure from the unique potential field distribution of flux represents a nonpotential field that contains the free magnetic energy and the currents associated with it.  This idea forms the heart of the Minimum Current Corona Model, developed in a series of paper \citep{Longcope:1996, Longcope:1998,Longcope:2001,Longcope:2002}.  We illustrate the idea using the simple example of the quadrupolar field of \figref{fig:baumfig1}.  Suppose the set of four bipoles has flux initially distributed as represented in (a).  Under coronal conditions where the electrical conductivity is very high, the coronal field will respond to photospheric motions quasi--statically and ideally, with negligible diffusion of the magnetic field: the footpoints of fieldlines remain fixed as the system evolves.  Therefore, the amount of magnetic flux in each coronal domain remains constant as the photospheric field changes.  This is the Flux Constrained Equilibria (FCE) constraint of the MCC.  If equatorward bipole moves westward relative to the poleward bipole, shearing from configuration (a) to (c), while the amount of flux in each domain is held fixed, then the photospheric field is in configuration (c) while the coronal field is still in configuration (a) and the field is therefore nonpotential.  Currents exist somewhere in the system and the energy is higher than the potential field case.  Note that the constraint that the flux in each coronal domain is constant does not allow for any flux emergence or cancellation.  That constraint must be relaxed to deal with changing total amounts of photospheric flux, as we will consider shortly.


If we have some observational reason to believe the flux is distributed in some way---for instance by observations of EUV loops as in \figref{fig:euvsmall} or by the tracking of emergence of photospheric flux as described in Chapters 2 and 3---and that distribution differs from the potential case, then we may go on to calculate that free energy.  The analysis has been carried out in the past for both idealized and data--driven models, with the various goals of studying both active region and network field evolution, as well as analyzing general properties of magnetic field topology \citep{Longcope:1999, Longcope:2004, Close:2005, Longcope:2005b, Beveridge:2006, Longcope:2007, Longcope:2007d, Kazachenko:2009,DesJardins:2009, Lee:2010, Kazachenko:2010, Longcope:2010,Kazachenko:2012}.  Only two of these studies, \citet{Close:2005} and \citet{Longcope:2005b} allowed for time--varying magnetic flux, and only the first developed any formalism to deal with flux emergence in general.  The others are therefore limited to studying cases of negligible flux emergence or cancellation.  Cast in more typical language, these studies focused only stresses caused by the braiding and twisting of the magnetic field through photospheric shear and rotational motions.  While this type of motion alone can produce enough free energy to power large X--class flares \citep{Kazachenko:2009, Kazachenko:2010}, and the free energies thus determined do agree with those inferred from MHD simulations \citep{Longcope:2004, Beveridge:2006, Longcope:2007d}, dismissing any event with substantial flux emergence severely limits the number and type of events which may be studied.

Basic MCC does not allow the flux in any coronal domain to change as the photospheric field evolves.  Flux emergence obviously breaks the FCE constraint, so the question then arises how to relax that constraint while still retaining the analytic utility of the model to quantify free energy.  Two attempts have been made to use the MCC model to understand coronal energetics in the case of substantial emergence.  As noted in \citet{Longcope:2005b}, coronal loops connecting two active regions have long been held as compelling evidence for reconnection within the corona.  Those authors therefore used EUV observations of coronal loops between a freshly emerged active region and a nearby preexisting active region to quantify reconnection between the two regions.  Using some simplifying assumptions about the amount of flux represented by each observed coronal loop, they were able to estimate the flux contained in a single coronal domain of interest.  In a potential field model, this domain contained a certain finite amount of flux, while the observational history of the active region complex, namely the emergence of one entire active region, suggests that, at least initially, that domain contained identically zero flux in the actual coronal field.  Any loops observed in that coronal domain then demonstrate reconnection in the corona that has transferred flux into that domain.  This modifies a single flux constraint of the MCC in a straightforward way, allowing the authors to determine a reconnection rate and compare that to estimates of heating, radiation, and energy storage and release.  This modification was achieved by hand, and the authors did not develop a method for dealing with more general scenarios of flux emergence.

In contrast \citet{Close:2005} did develop a general method for including flux emergence in the MCC, with a goal of determining the recycling time for coronal fields attached to quiet sun regions and the coronal heating associated with the implied reconnection.  Flux continually emerges through the photosphere, but the total amount of flux in the photosphere is relatively constant.  Thus, eventually enough flux will have emerged to completely replace existing flux.  This is the photospheric flux recycling time, and current estimates place the rate of recycling at $1-2$ hrs, although the cadence of observation can strongly affect the determined rate \citep{Hagenaar:2008}.  

The coronal recycling time is a different but related measure, defined by the time required to completely remap the footpoints of all coronal flux.  This can occur simply through shear motions of photospheric footpoints, but as \citet{Close:2005} determined, is heavily modified by emergence and cancellation.  These authors tracked photospheric flux concentrations in the quiet sun, allowing for merging, splitting, emergence, and cancellation of these sources.  The change in domain fluxes from timestep to timestep determined by an MCT analysis provides the estimate of the coronal recycling time, which they determine to be $\sim 1.4\unit{hr}$.  They assume a potential field at every timestep, so that any stresses induced by photospheric changes dissipate via reconnection between each time.  The MCC model provides the estimate of stresses induced under FCE evolution, and therefore the energy that must be expelled to achieve a potential field at each timestep.  

The model of \citet{Close:2005} makes sense for quiet sun regions where we expect to remain close to potential and stresses to be relaxed at the same rate they are induced, so that the quiet sun remains in a statistical steady state.  It does not make sense for studying active regions, though.  During emergence, an active region slowly builds up stress in the coronal field over the course of several days.  This energy is then converted rapidly during a flare or series of flares, each typically lasting less than 1 hour.  It is precisely the buildup of stresses disallowed in the analysis of \citet{Close:2005} that creates the necessary conditions for flares in the first place.  To model emerging active regions, we must therefore further modify the FCE constraint of the MCC model to allow for both the buildup and subsequent release of this energy.

In the following body of work, I develop a formalism for including flux emergence in MCT models of active region evolution.  The evolution of the model from one timestep to the next is achieved through the MCC model of \citet{Longcope:2001}, and is described in detail in Chapter 2 \citep{Tarr:2012}.  This work relaxes the constraint that the amount of magnetic flux be fixed during an active region's evolution, and we are therefore able to model flux emergence.  We apply the new formalism to the case of NOAA AR11112, where a modest sized active region emerges into the relatively strong field of an older, decaying active region.  The emergence results in a GOES M3.0 flare, whose observed energies may be compared to those resulting from the MCC model.

Chapter 3 \citep{Tarr:2013} applies the formalism to the very important, and now much studied, AR11158, which produced the first GOES X class flare of Solar Cycle 24 and the SDO era, and whose entire emergence was captured by the spacecraft.  In addition to calculating the free magnetic energy of this region, we also propose a method to quantify the amount of flux exchanged during a series of flares much more realistically within the MCT/MCC framework.  This allows us to estimate the energy conversion, magnetic to kinetic and thermal, during each flare using a relaxation method that minimizes free energy through flux exchange.

Some interesting aspects of that relaxation lead to the observationally driven work presented in Chapter 4 \citep{Tarr:2013b}.  Here, we return to the case of AR11112 to study the reconnection between emerging and preexisting field in much more depth.  We observationally determine the amount of emerging flux that reconnects with preexisting flux in an essentially continuous fashion.  This reconnection is the slow response of the surrounding field to the intrusion of new flux, rather than the rapid flaring response studied in the earlier papers.

We end in Chapter 5 with a discussion of how the research developed in this dissertation may be advanced, or even reformulated to apply to rather different areas of solar physics.

\chapter{Calculating energy storage due to topological changes in emerging active region NOAA 11112}\label{ch:ar11112}
\begin{manuscriptauths}
  Manuscript in Chapter 2
  \newline
  \newline
  Author: Lucas A. Tarr
  \newline
  \newline
  Contributions: Conceived and implemented study design.  Wrote first draft.
  \newline
  \newline
  Co--Author: Dana W. Longcope
  \newline
  \newline
  Contributions: Helped to conceive study.  Provided feedback of analysis and comments on drafts of the manuscript.
\end{manuscriptauths}
\pagebreak
  
\begin{manuscriptinfo}
  \noindent Lucas A. Tarr and Dana W. Longcope\\
  The Astrophysical Journal\\
  Status of Manuscript:\\
  \uline{\phantom{5eM}}Prepared for submission to a peer--reviewed journal\\
  \uline{\phantom{5eM}}Officially submitted to a peer--reviewed journal\\
  \uline{\phantom{5eM}}Accepted by a peer--reviewed journal\\
  \uline{\phantom{5eM}}\makebox[0pt]{\hspace{-2em}x}Published in a peer--reviewed journal\\
  \newline
  \newline
  Published April, 2012, ApJ 749, 64
\end{manuscriptinfo}
\begin{abstract}
   The Minimum Current Corona (MCC) model provides a way to estimate stored coronal energy using the number of field lines connecting regions of positive and negative photospheric flux.  This information is quantified by the net flux connecting pairs of opposing regions in a connectivity matrix.  Changes in the coronal magnetic field, due to processes such as magnetic reconnection, manifest themselves as changes in the connectivity matrix.  However, the connectivity matrix will also change when flux sources emerge or submerge through the photosphere, as often happens in active regions.  We have developed an algorithm to estimate the changes in flux due to emergence and submergence of magnetic flux sources.  These estimated changes must be accounted for in order to quantify storage and release of magnetic energy in the corona.  To perform this calculation over extended periods of time, we must additionally have a consistently labeled connectivity matrix over the entire observational time span.  We have therefore developed an automated tracking algorithm to generate a consistent connectivity matrix as the photospheric source regions evolve over time.  We have applied this method to NOAA Active Region 11112, which underwent a GOES M--2.9 class flare around 19:00 on Oct.$16\tothe{th}$, 2010, and calculated a lower bound on the free magnetic energy buildup of $\sim 8.25 \times 10^{30}$ergs over 3 days.
 \end{abstract}
 
 \section{\label{sec:intro2}Introduction}
 
 It is now widely believed that solar flares are powered by magnetic energy which had been stored in the corona through slow stressing applied from the photospheric boundary.  In an idealized model the energy builds up as the coronal magnetic field responds without resistance (every field line line--tied and unbroken).  The flare then occurs as coronal reconnection exchanges those field line footpoints to achieve a lower energy state.  In this process, the footpoints are changed by the reconnection, but the vertical photospheric field in which the field lines are anchored is not.  The potential field from this fixed photospheric field has the minimum magnetic energy possible.  The maximum energy available for release is the amount by which the initial field exceeds this potential field energy, called the {\em free energy}. 
 
 Quantitative simulation of the above scenario has proven to be extremely challenging owing to the vast range of scales involved. Magnetic fields of even modest complexity develop, when stressed, current structures many orders of magnitude thinner than the global length scale \citep{Parker:1972,VanBallegooijen:1985,Longcope:1994}.   A numerical solution therefore requires additional magnetic diffusion to prevent the development of unresolvable current structures.  The corresponding diffusive time is necessarily much shorter than actual diffusive times and generally shorter than the multi--day times governing the stressing phase.  As a consequence, direct simulation, for example by time--dependent MHD solution, includes artificial (diffusive) energy losses competing with the energy build-up.  Few such computations have been capable of demonstrating pre--flare energies comparable to those released by the ensuing flare \citep{Linker:1999}. 
 
 An alternative means of estimating pre--flare energy storage is offered by the Minimum Current Corona model \cite[MCC:][]{Longcope:1996,Longcope:2001}.  This is a quasi-static technique using equilibria, called flux constrained equilibria (FCE), minimizing magnetic energy subject to a set of topological constraints composing a subset of all line-tying constraints.  Rather than constraining every pair of footpoints, the MCC groups footpoints into unipolar photospheric regions and constrains the net coronal flux connecting each pair of regions, called {\em domain fluxes}. Since it uses only a subset of the actual constraints, its fields provide a lower bound on the actual free energy.   As the photospheric regions move relative to one another the potential field above changes as do the fluxes by which it would link region pairs (potential domain fluxes).  Since the actual field is constrained from changing these fluxes, it becomes increasingly different from the potential field and thus gains free energy.  Significantly, this energy is available for release by the violation of topological constraints, a process which can occur on very small spatial scales.
 
 The MCC has been used to estimate pre--flare energy storage in a number of flares \citep{Longcope:1998,Longcope:2007,Kazachenko:2009,Kazachenko:2010, Longcope:2010,Kazachenko:2012}.  A partitioning algorithm was developed to automatically group photospheric flux into distinct regions \citep{Barnes:2005,Longcope:2007b,Longcope:2009}.   Provided it is permissible to neglect submergence or emergence of flux, the regions can be taken to move relative to one another but with constant flux.  This simplified scenario was deemed adequate to model pre--flare evolution in several cases  \citep{Longcope:1998,Longcope:2007}, including the landmark Halloween event \citep{Kazachenko:2010}.  When footpoints within a given region move internally, such as during sunspot rotation, additional constraints must be introduced involving the arrangement of footpoints within regions \citep{Beveridge:2006,Kazachenko:2009}.  
 
 Flux emergence is a well--known precursor for flares and CMEs \citep{Archontis:2008} and is thus likely to play a role in energy storage for many flares.  It has not yet been accounted for in a complete MCC energy estimate owing chiefly to technical hurdles.  Doing so would require the constraints to be somehow modified to account for flux emergence.  Two prototypical cases have been treated, where in each case emergence made an obvious modification to just a single constraint \citep{Longcope:2005, Longcope:2010}.  A step toward a more general application was made by \citet{Close:2004}, who accounted for emergence and submergence in a computation of coronal reconnection times in the quiet Sun.  Changes in the fluxes of photospheric regions were used to generate a list of flux changes due to emerging and submerging domains.  The algorithm used had several drawbacks, including a tendency to assign both emerging and submerging domains to the same photospheric region. 
 
 The present work introduces a new algorithm by which emerging flux regions may be automatically accommodated in the MCC constraints leading to a pre--flare energy estimate.  Flux changes in  photospheric regions are used to generate a list of domain flux changes due to emergence and submergence, as in \citet{Close:2004}.  In this case, however, a single region is linked  to only emerging or submerging domains according to the sense of its own change.  The possibility of artificial photospheric flux changes due to variations in sequential partitioning can be accommodated at the same time using a closely related algorithm.  (This step is independent of the partitioning algorithm.)  We also present a new algorithm for automatically associating domains with separators, a crucial step in the generation of an energy estimate. 
 
 The new methodology is illustrated by applying it to a flare which occurred on 16 Oct.\ 2010 and was observed by instruments on the Solar Dynamics Observatory (SDO) spacecraft. New flux emerges into an existing active region (NOAA AR 11112) for two days prior to the flare (GOES class M3).  We derive photospheric magnetic fluxes from a series of 123 line--of--sight magnetograms from the HMI instrument \citep{Scherrer:2012,Schou:2012,Wachter:2012} at cadences of $\sim30$ minutes, as discussed in the next section.  In Section \ref{sec:photosphere}, we detail our algorithms for partitioning the set of magnetograms into unipolar regions.  The flux variations in these regions are used to define a set of emerging and submerging domains which are automatically generated by an algorithm explained in Section \ref{sec:change}.  The emerging domains largely resemble those we would have expected based on inspection of the time series, but still contain some physically dubious assignments that must be fixed by hand.  In the future, we would like to completely automate each of these algorithms.
 
 In Section \ref{sec:erg}, we use the magnetograms to derive a coronal topology for the post--flare potential magnetic field and use this to place a bound on the free energy of the pre--flare field.  The post--flare potential field includes a coronal null point whose fan surface encloses one of the newly emerged polarities.  This is a novel feature in the MCC and requires the development of one additional method.  We apply our energy estimate to NOAA AR 11112 in Section \ref{sec:ergapp}, which reveals that most of the pre--flare energy is due to currents passing through the coronal null point. 
 
 \section{\label{sec:partic}Particular Case}
 
\begin{figure}[ht]
   \begin{center}
     \includegraphics[width=\textwidth]{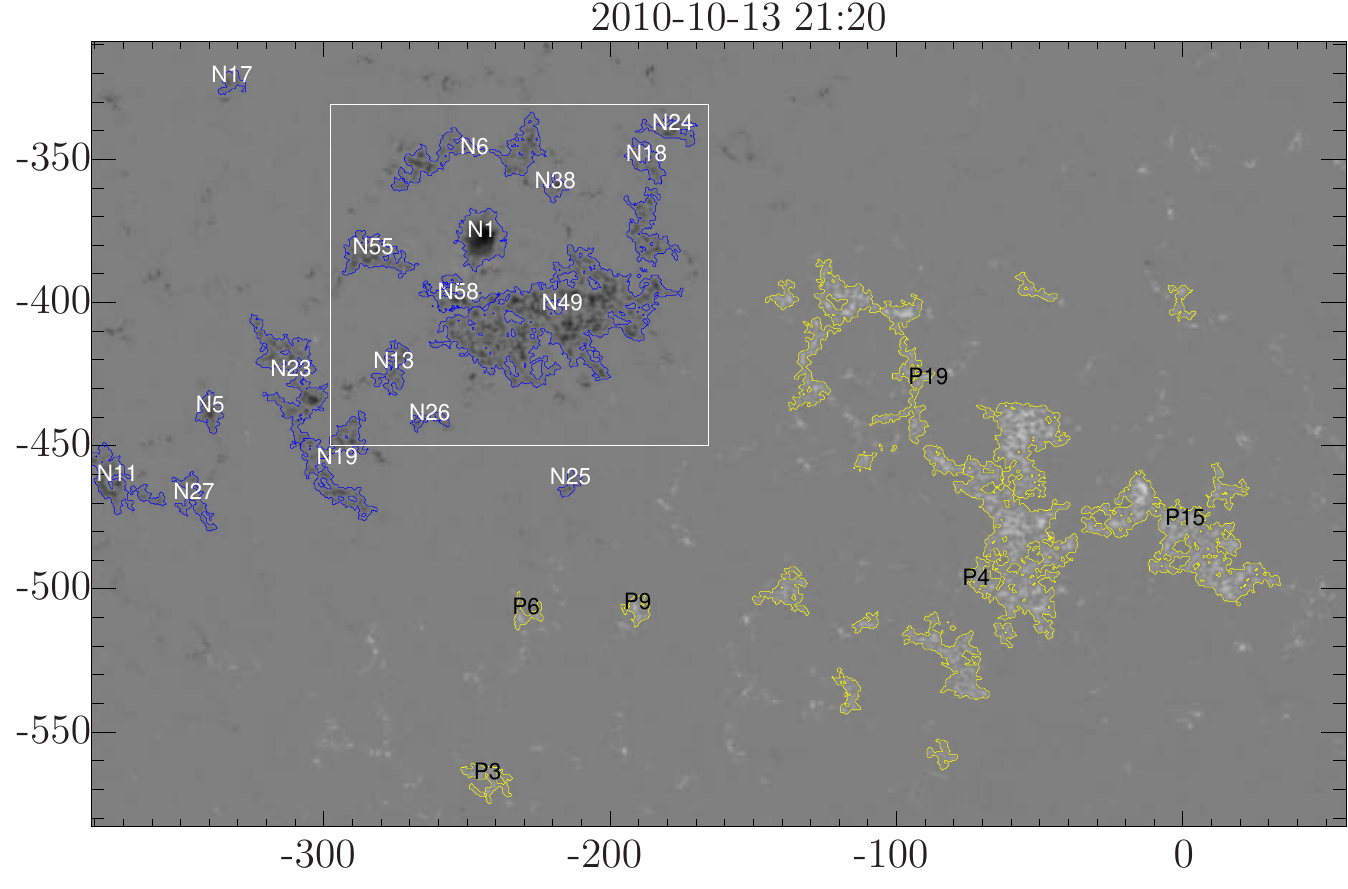}
     \caption[NOAA AR 11112 on 2010-10-13]{\label{fig:oldflux}  HMI LOS magnetogram of NOAA AR 11112 prior to new flux emergence.  The grayscale saturates at $\pm max(\vert B\vert )=\pm 1500\unit{G}$, and the axes are in arcseconds from disk center.  The boxed area encloses the region of flux emergence, which begins $\sim 11$ hours after this magnetogram.}
   \end{center}
 \end{figure}

 Though the methods presented here are of general utility, we apply them, for concreteness, to NOAA Active Region 11112.  We base our calculations on magnetograms taken by the Helioseismic and Magnetic Imager (HMI) on board the SDO spacecraft.  \figref{fig:oldflux} shows an example line--of--sight (LOS) magnetogram from Oct.~13th, prior to flux emergence.  Lighter pixels show positive flux, dark pixels negative flux, and gray pixels zero flux; the grayscale saturates at $\pm 1500$G.  The active region originally contains only previously emerged flux as it crosses the eastern limb in the southern hemisphere on October 9th, 2010.  New flux begins emerging around 08:00 UT on the 14th, and is associated with a GOES M2.9 flare several days later, at 19:07 UT on the Oct.~16.  The region shows little activity in the 6 days before the flare.

 Our dataset consists of 123 line--of--sight magnetograms, each consisting of $869\times 544$ pixels with 1.0 arcsecond resolution and approximately 0.5 arcsecond/pixel\citep{Scherrer:2012}.  They have a cadence of $\approx$ 2 hours for the first 21 timesteps, from 2010-10-13 00:04 UT to 2010-10-14 18:20 UT, and approximately half hour timesteps thereafter, from 2010-10-14 18:20 UT to 2010-10-16 23:37 UT.  Flux emergence begins around timestep 21.  As shown by the animation of \figref{fig:oldflux}, the photospheric changes are well characterized by these two cadences.  Our data window has flux imbalance $< 10\%$ over the three days leading up to the October $16^{\hbox{\scriptsize th}}$ flare.

\begin{figure}[ht]
   \begin{center}
     \includegraphics[width=0.7\textwidth]{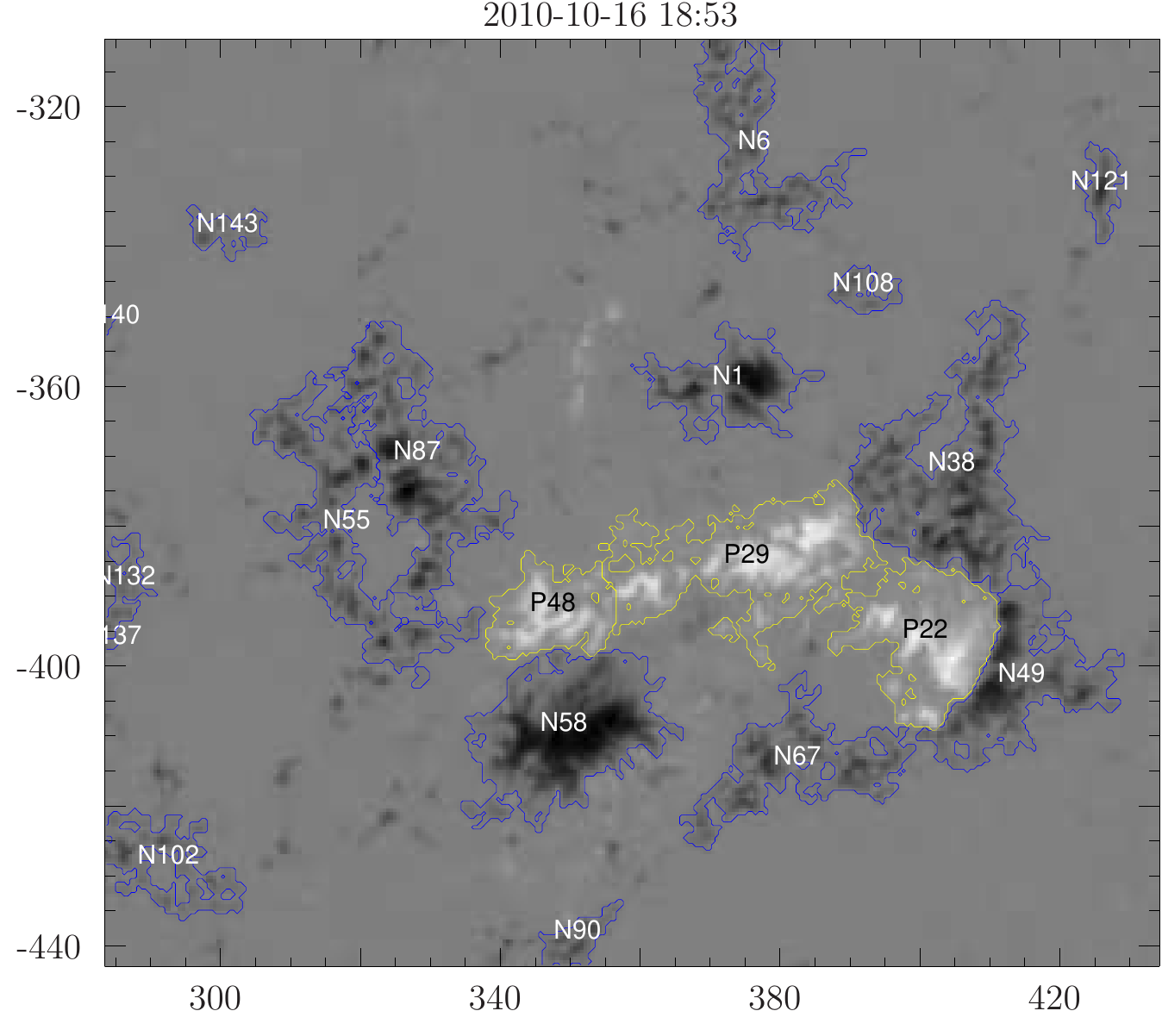}
     \caption[Newly emerged flux within the ring]{\label{fig:newflux}  A zoomed view of LOS magnetogram of the emerging flux region, boxed in \figref{fig:oldflux}, on Oct 16th, 2010, 18:53UT, $\sim$ 10 minutes prior to the M2.9 flare.  The grayscale saturates at $\pm -1176.10$G, and axes are in arcseconds from disk center.}
   \end{center}
 \end{figure}

 The old flux region has a polarity inversion line, running from southeast to northwest, which coincides with a filament as seen in data from the Global High--Resolution H$\alpha$ Network, as well as SDO EUV images at 94, 131, 171, 193, 211, 304, and 335 \AA.  This filament appears unaffected by the 19:07 UT flare.  To the north of the filament, the diffuse, old--flux, negative polarity field includes a curious ring of flux surrounding a strong core, boxed in \figref{fig:oldflux}.  The new flux of both polarities emerges completely within this ring, sweeping the old negative flux out of its way as it carves out a space for itself.  This results in the (zoomed in) field shown in \figref{fig:newflux}, roughly 10 minutes before the flare.

 As the new flux emerges it snow--plows the old flux in front of it, the old flux concentrates and a strong horizontal gradient in the LOS field develops between old and new flux, as seen near S 400'', W 410'' in \figref{fig:newflux}.   The flare is centered on this strong polarity inversion line.  Post--flare loops connect the newly emerged negative flux (N55, N58, N87) within the ring to the diffuse, positive flux to the West (P4, P15, P19).  We therefore believe that any analysis of this flare's energetics must explicitly account for flux emergence.

 Our analysis naturally splits into two parts.  The first parts works directly with the photospheric magnetic field measurements.  We first partition the magnetic field at all times into unique, unipolar regions.  Second, we track these regions and enforce a consistent labeling scheme as they change geometry over time.  Geometric changes include changes in size, shape, orientation, and total flux; merging and splitting of regions; and the complete emergence or submergence of regions.  The final step quantifies flux emergence and submergence between pairs of photospheric sources using a novel algorithm.

 In the second part of our analysis, we use our characterization of the photospheric field as a set of unique unipolar sources to describe the topology of the coronal field.  Each distinct photospheric region is replaced by a point source of the same flux located at the region's flux--weighted centroid.  We then find the flux within each potential field domain, and the null points and separators of the potential field.  Changes in the potential field's domain flux relative to the actual domain fluxes, assumed fixed, indicates the storage of magnetic energy in excess of the potential field.  In an energy--minimized field, the free energy stored by the field is manifest by current ribbons that develop along separator field lines.  For brevity, we will use the phrase ``energy stored in the separator'' as a shorthand for ``free energy stored in the magnetic field which manifests itself as a a current ribbon at the location of a separator.''  The amount of current flowing along these ribbons places a bound on the free magnetic energy of the system.

 \section{\label{sec:photosphere}Characterizing the Photospheric Field}
 
 We begin work on the photospheric data by partitioning the magnetogram using the gradient--based tessellation algorithm described by \citet{Barnes:2005}.  The initial LOS magnetograms are converted to a vertical field by assuming a radial field at each pixel, which amounts to dividing by $\cos(\theta)$, where $\theta$ is the polar angle from disc center.  When determining the flux in each region, we must also account for foreshortening within each pixel, dividing by a second factor of $\cos(\theta)$.  We then convolve all vertical field, $B_z=B_{LOS}/\cos^2(\theta)$, with the Green's function for a potential extrapolation up to a height $h$ from an unbounded plane:
 \begin{equation}
   K_h(x,y) = \Frac{h/2\pi}{(x^2+y^2+h^2)^{3/2}}\qquad .
 \end{equation}
 This effectivly smooths the data.  To reduce the effects of noise, we neglect all the convolved field below a threshold of $\vert B_{th}\vert = 50\unit{G}$.  Using the smoothed field, we assign a unique label to all local maxima and every pixel strictly downhill with respect to $\vert K_h\star B_z\vert$ from each maxima.  Internal boundaries in unipolar regions are eliminated---multiple regions are merged into a single region---when the saddle point $\vert B_z\vert > \hbox{min}(\vert B_{pk}\vert)-B_{sad}$. $B_{pk}$ is the greatest vertical field strength of the saddle's surrounding peaks, and $B_{sad}$ is a threshold value.  At the end of this process, each data pixel has an associated label.  We call the set of all labels at a given time a mask, and the set of masks at all times a mask array.

 Our goal in this section is to make the mask array consistent from timestep to timestep.  In particular, we wish to determine how the newly emerged flux interacts with the old flux, and how that process affects the energetics of the active region.  Our primary concern is therefore to distinguish between new and old flux, which is a marked departure from previous, similar investigations \citep{Kazachenko:2009,Kazachenko:2010,Longcope:2007b}.

\figref{fig:newflux} shows an example where the boundary of each unique mask region and its label (P1, N1, P2, \ldots) have been plotted over the line of sight magnetogram data.  All pixels and regions containing flux below our thresholds are given a label of 0.  This figure covers the ring feature of old flux---concentrated in regions N1, N6, N38, N49, N55, and N67---and the newly emerged flux in regions P22, P29, P48, N58 and N87.  The animation of \figref{fig:oldflux} shows that these definitions are not necessarily strict.  For instance, N58, N67, and N87 initially all break off of N49.  Both N58 and N87 are quickly dominated by newly emerged flux, while N67 remains primarily distinct.  We refer to ``new flux'' or ``old flux'' regions based on the dominant type of flux determined by inspection of the timeseries.

For the present analysis, we have set $h=1.0\unit{Mm}$, $\vert B_{th}\vert =50\unit{G}$, and $B_{sad} = 0.8\times B_{pk}$.  We also established a minimum flux of $2.6\times 10^{19}\unit{Mx}\ (20000$ gauss $\times$ pixel area, after accounting for pixel foreshortening) for a solitary region.  Our field of view contains unsigned flux of order $10^{22}\unit{Mx}$ above the 50 gauss threshold, so, a region must contain at least 0.2\% of the total unsigned flux before inclusion in our algorithms.  For a discussion of how the parameters $h$, $B_{th}$, and $B_{sad}$ affect the final partitioning, see \cite{Longcope:2009}.
 
 After partitioning each magnetogram, we attempt to associate the partitions at one timestep to those in the next.  To begin with, each region is characterized by its net signed flux and centroid location:
 \begin{equation}
   \label{eq:psi}
   \psi = \int_\mathcal{R} B_z(x,y)\, dx\, dy \qquad \bar{\vect{x}} = \psi^{-1}\int_\mathcal{R} \vect{x} B_z(x,y)\, dx\, dy
 \end{equation}
 For short, we call a region's total flux at the flux--weighted centroid the region's associated \emph{pole}.
 
 As a first pass at creating consistently identified regions, we calculate the distance between each pair of centroids at two consecutive timesteps.  If Centroid A at time $i$, $\vect{x}_A^i$, is closest to Centroid B at time $i+1$, $\vect{x}_B^{i+1}$, and B has A as its closest neighbor, and that distance is less than a threshold ($10\unit{Mm}$), then we conclude that Region B is Region A.  

Both this simple method of association, and exploration of the partitioning parameter space, result in mask arrays that are not to the quality required for MCT analysis of an emerging flux region.  We therefore developed two more sophisticated automatic procedures.  \fnc{rmv\_flick} and \fnc{rmv\_vanish}, described below, provide refinements to the minimum--distance identifications.  Both were heuristically developed to address a prevalent type of inconsistent region labeling.  The first algorithm deals with a region that exists for just a single timestep, but is clearly part of another region.  The second deals with regions that change labels from one timestep to the next.  These problems arise, for instance, when the choice of parameters poorly represents a subset of the timeseries, or when two regions merge or split, and their centroid locations between two timesteps are very different.
 
 For clarity, our examples in the following subsections do not directly use masks from our time series, and are only meant to illustrate the action of each algorithm.  Both algorithms operate directly on pixels in the mask structure, so that there are no physical scales involved.  The algorithms only relabel nonzero elements of the mask, so that no external mask boundaries are modified.

 \subsection{\label{sec:rmvf}\fnc{rmv\_flick}}
 
 \begin{figure}[ht]
   \begin{centering}
     \includegraphics[width=0.5\textwidth]{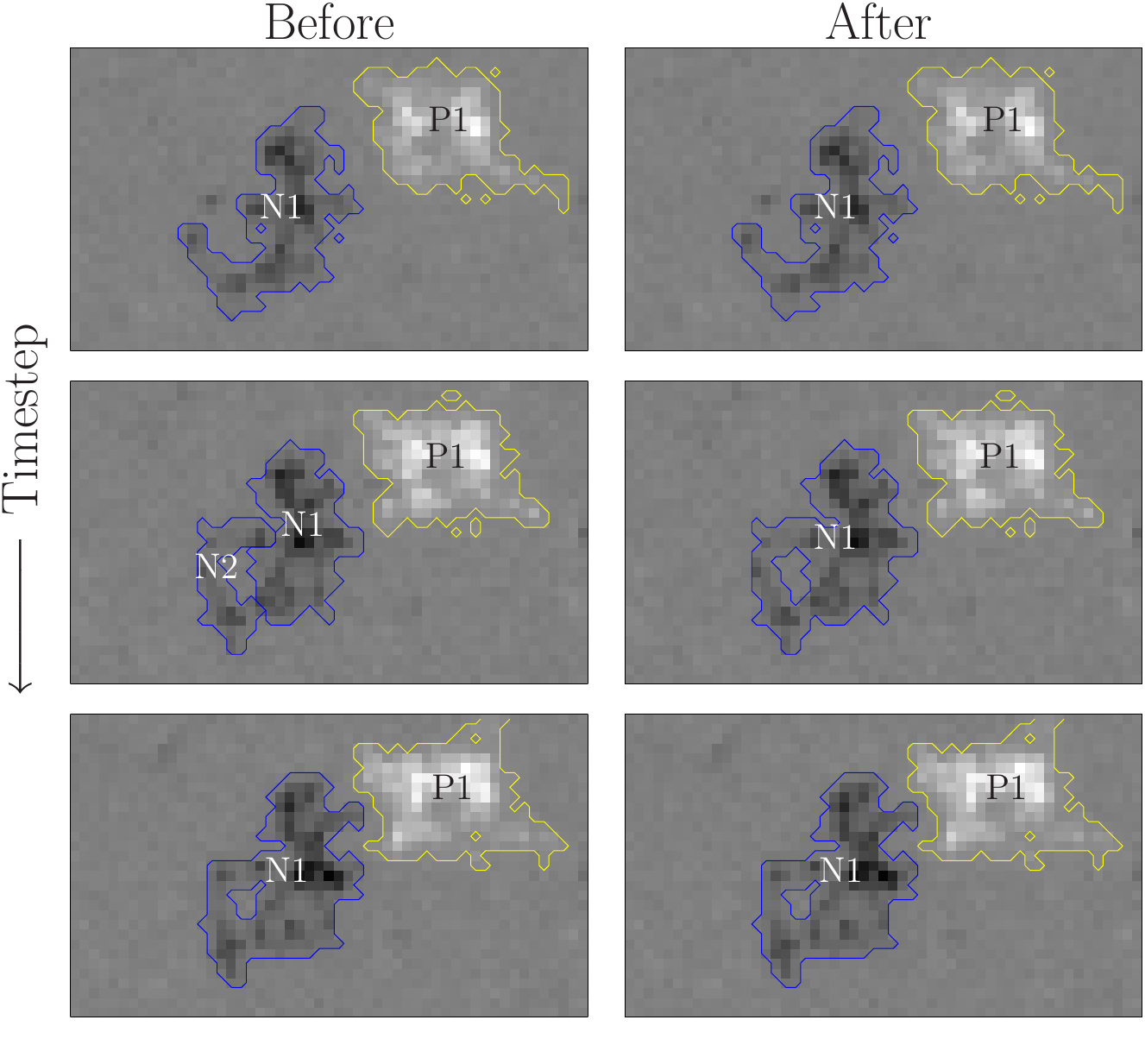}
     \caption[Example of \fnc{rmv\_flick}]{\label{fig:rmvf}The first column demonstrates the problem of ``flickering.''  The middle timestep contains a new created region, $N2$, inside of the area controlled by region $N1$ in the surrounding timesteps.  $N2$ only exists for that single timestep, its area reverting back to $N1$ in the third timestep.  The right column shows the effect of running \fnc{rmv\_flick} on this region: in the middle timestep, $N1$ has been completely restored in place of $N2$.}
   \end{centering}
 \end{figure}
 
 \fnc{rmv\_flick} attempts to smoothe the temporal structure of our mask arrays by better associating regions that last for just a single timestep.  This happens when the saddle points in the tessellation algorithm slosh back and forth over the threshold value.  \figref{fig:rmvf} shows an example of this problem.  Region N2 pops into existence in the middle timestep, occupying some of the territory ascribed to region N1 in the first and third timestep.  We see that this is not a ``real'' new region, but should instead be considered part of N1 for the entire time.  \figref{fig:rmvf} column two shows the result of running \fnc{rmv\_flick} on these data.
 
 The \fnc{rmv\_flick} algorithm works on a sliding, three timestep window (initial, middle, and final timestep), and in two steps.  In the first step, we find all regions $\Theta_i$ that have nonzero flux only in the middle timestep, and find all mask pixels in the initial and final timestep which overlap $\Theta_i$.  We take each overlapping region $\Theta_A$ and relabel those pixels of $\Theta_i$ which are overlapped by $\Theta_A$ in the initial and final timesteps as $\Theta_A$.

 Because our source regions change shape and size (pixel count) over time, we will usually have leftover pixels after performing the above bulk relabeling.  While leftover pixels remain, we relabel the pixel of $\Theta_i$ that currently borders the greatest number of pixels of a single neighbor to that neighbor's label.  We have found that, when multiple regions overlap a flickering region, this method divides those pixels between the overlapping regions in decent proportion to their past and future ``control'' of flickering territory.

 \subsection{\label{sec:rmvv}\fnc{rmv\_vanish}}

 \begin{figure}[ht]
   \begin{centering}
     \includegraphics[width=0.5\textwidth]{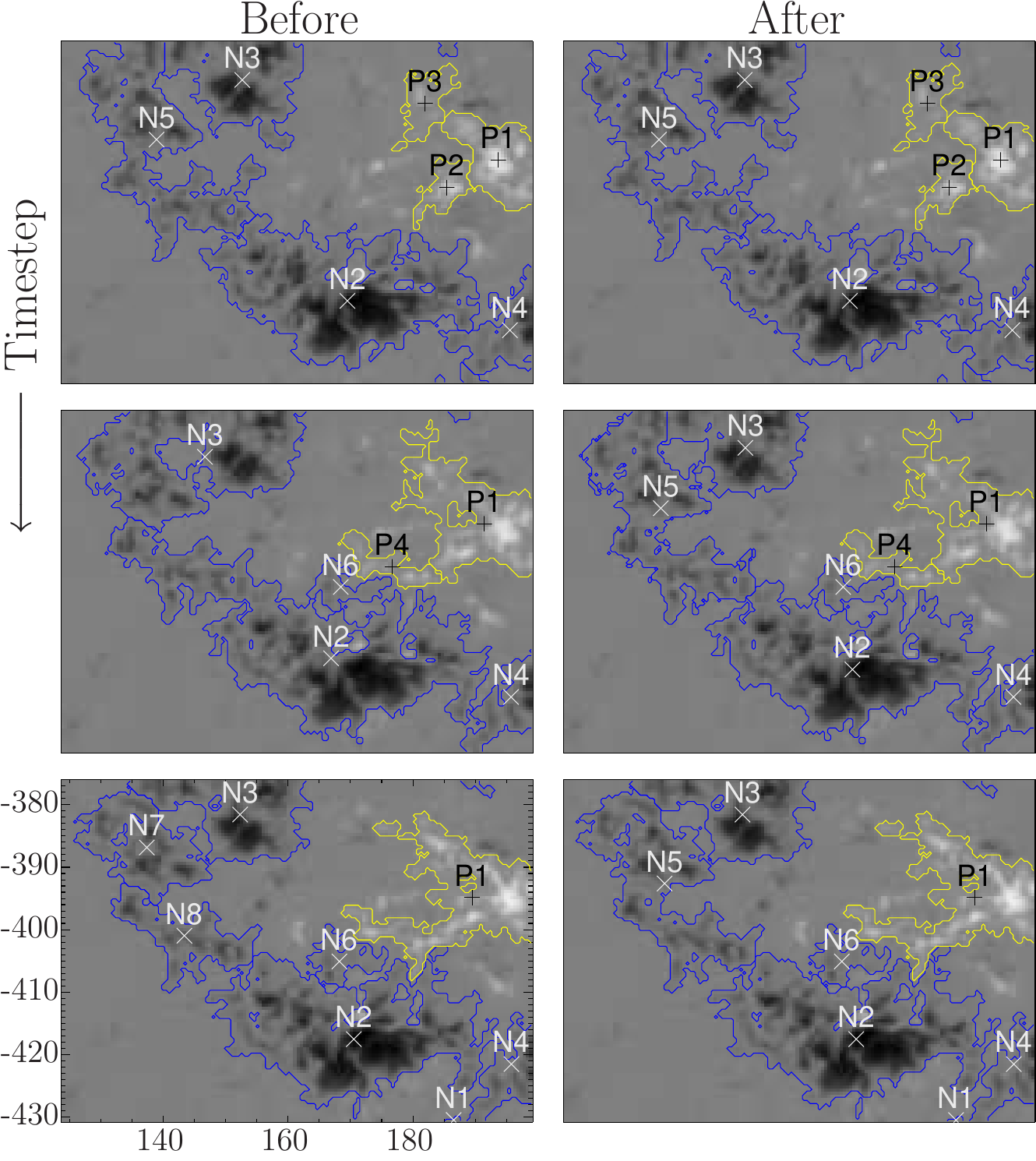}
     \caption[Example of \fnc{rmv\_vanish}]{\label{fig:rmvv}The left column demonstrates the problem of ``vanishing,'' in which a single region---N5 in the top left panel---cycles through series of different labels.  The right column shows the effect of running \fnc{rmv\_vanish} on these data.  The $x$ and $y$ axes are in arcseconds from disk center, though the physical scale is irrelevant to the algorithm's operation.}
   \end{centering}
 \end{figure}
 
 With \fnc{rmv\_flick} we dealt with regions that existed for just a single timestep.  We now deal with the conceptual inverse problem, where a region exists for some time and then suddenly changes names in the next timestep.  \figref{fig:rmvv} demonstrates the problem, as the region labeled N5 in the top left panel (and previous, undepicted times) cycles through a series of names and partitions within a few timesteps.
 
 We again begin by considering the mask array in a sliding, three timestep window.  We find all non--zero flux regions within the window and determine which regions exist in which timesteps.  For every region that disappears in the middle timestep, we project its area from the first timestep into the middle and last timesteps.  We then find all new regions, in both the middle and final times, that overlap with that projected area.  Finally, we relabel any new region whose centroid lays within this projected area.  \figref{fig:rmvv}, right column shows the effect of running \fnc{rmv\_vanish} on data.
 
 \subsection{\label{sec:mskresults}Results of the Consistency Algorithms}
 
 \fnc{rmv\_flick} and \fnc{rmv\_vanish} each use a sliding, three timestep window, so that it takes many repetitions of each for updated information to propagate from the beginning to the end of the mask array.  We must therefore repeat each of the algorithms many times.  Taking these processes into account, we find that the mask arrays best approximate the photospheric field evolution when we switch back and forth between the two algorithms at the beginning of the process, and run just \fnc{rmv\_vanish} many times at the end of the process.  For our NOAA AR 11112 data, we ran \fnc{rmv\_flick} a total of 35 repetitions, and \fnc{rmv\_vanish} a total of 1058 repetitions.
 
 Together, these two algorithms accomplish about nine--tenths of the work in creating a consistent set of tessellated masks.  Part of the remaining tenth may be accounted for by a boundary--shift algorithm, described below, between abutting regions of the same polarity.  Even this does not accurately represent the data, and the final changes to the mask array required for a consistent time series must be done by hand.  These changes again mostly involve the placement of the boundary between abutting regions of the same polarity, caused by a failure to distinguish between old flux and actively emerging flux.  These boundaries are adjusted manually until each region's flux evolves smoothly.  

\begin{figure}[ht]
   \begin{centering}
     \includegraphics[width=0.8\textwidth]{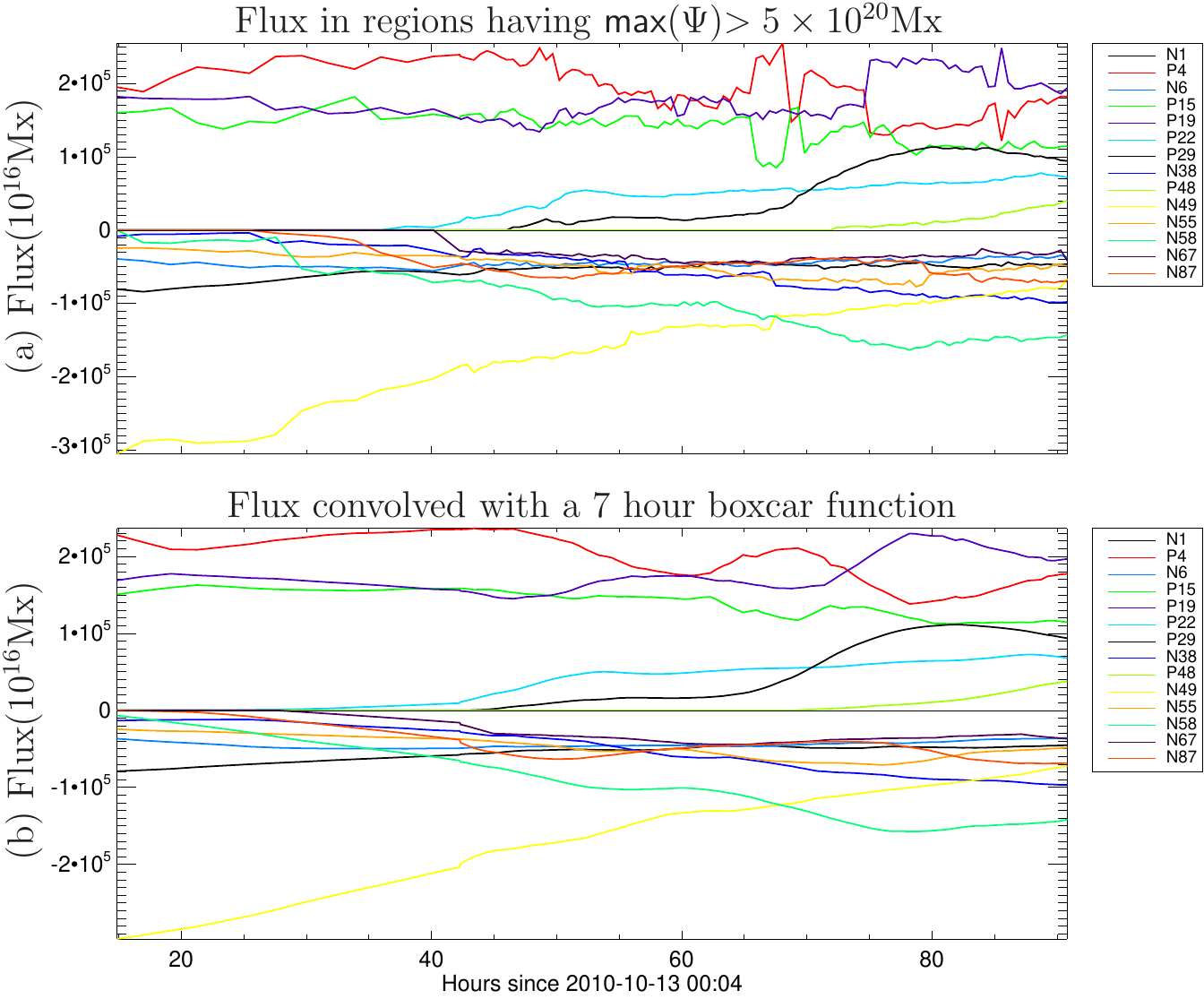}
     \caption[Flux within each region of NOAA AR 11112]{\label{fig:fluxplot} (a) Flux in regions with more than $5\times 10^{20}\unit{Mx}$.  Large, opposing spikes between pairs of curves, for instance between P15 and P19, show boundary shifts between adjacent regions.  (b) Result of smoothing the data by convolution with a 7 hour boxcar function.}
   \end{centering}
 \end{figure}

 \figref{fig:fluxplot} top shows the vertical flux in each (high--flux) region.  This includes editing of the masks by both the automatic algorithms and by hand.  Note that, even after a set of consistent masks have been created, the flux within each region is still rather noisy.  To counteract this, we smoothe our data with a 7 hour (13 timestep) boxcar function.  The result is shown in \figref{fig:fluxplot} bottom.
 
 \section{\label{sec:change}Change--in--Connectivity Algorithm}
 At this point, we transition from a mostly qualitative treatment of the photospheric field's geometry to a quantitative, topological analysis of a model coronal field: the Minimum Current Corona (MCC) Model developed in \citet{Longcope:1996, Longcope:2001}.  We begin by replacing every region in the photospheric field with a magnetic pole defined by the region's total flux and centroid location, as in \eqref{eq:psi}.  With one exception, the rest of this work deals only with the poles and connections between them.  

 The total flux in each pole must be connected to some number of other poles of opposite polarity.  The distribution of flux between each pair of poles may be represented as a graph.  This graph defines the system's connectivity: the undirected, weighted graph, where each vertex of the graph is a pole, and the weight of each edge defines the connectivity between the two vertices.  It happens that the graph of NOAA AR 11112 is \emph{simple} on the eve of the M3 flare, so that each pair of vertices has no more than one associated edge\footnote{This is not generally true for solar magnetic topologies, where two poles may have multiple topologically distinct edges divided by redundant separators \citep{Beveridge:2005,Parnell:2007}.  At every timestep throughout the time series for which we calculated AR11112's connectivity, we found it to be simply connected.  The next section will define these topological terms in more detail.}.  We refer to edges of the graph as domains.  We may additionally define a connectivity graph for the flux \emph{change} between two timesteps: if a pole's flux increases, that increase must be distributed among its domains with opposite polarity poles.  This is the topological entity we now quantify.

 The flux in a single photospheric region, and hence that region's associated pole, may vary in two ways.  The change is either a true evolution of the field by submergence or emergence through the photosphere, or it is a transfer of flux to/from another like--signed region via a shift in the boundary between the regions.  That this latter type still exists in our ``consistent'' data can be seen in regions P4, P15, and P19 in the smoothed data of \figref{fig:fluxplot}.  These three regions compose the large, diffuse positive region to solar west, as seen in \figref{fig:oldflux}.  

\begin{figure}[ht]
   \begin{centering}
     \includegraphics[width=\textwidth]{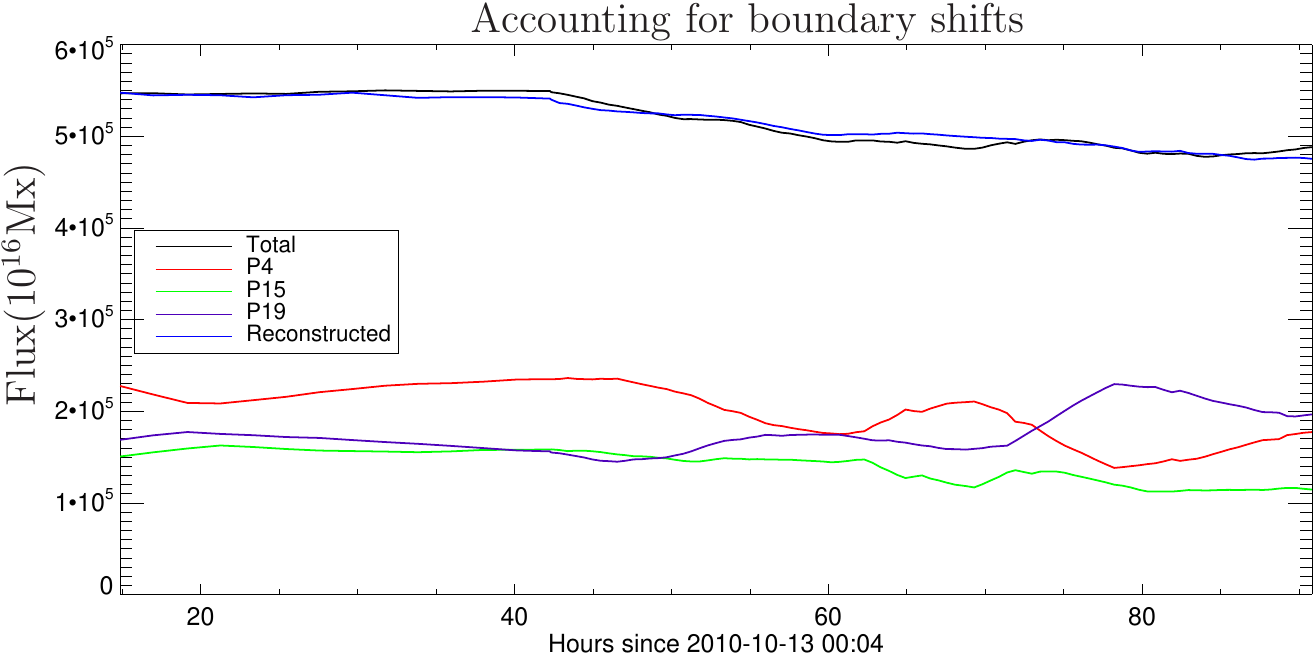}
     \caption[Result of \fnc{bndry\_shift}]{\label{fig:bndry}  The submergence of regions P4, P15, and P19 (red, green, purple), shown in the total flux of the regions (black).  Blue shows the calculated submergence by removing flux change due to boundary shifts from the total flux change.}
   \end{centering}
 \end{figure}

 If we focus on each of these regions individually, the red, green, and purple lines in \figref{fig:bndry}, we see several rapid changes in the flux of each.  However, if we look at the total flux in these three regions---the black line in \figref{fig:bndry}---it steadily decreases: the rapid increases and decreases in \figref{fig:fluxplot}b are almost solely due to shifting boundaries between the three.  Further, the steady decrease shown in the black line is matched by decreases in the old--flux part of the flux ring, regions N1, N6, N38, N49, N55, and N67 in \figref{fig:newflux}: the old positive flux is ``submerging'' with the old negative flux.  The blue line is a demonstration of the effectiveness of the forthcoming algorithms, and we will revisit it shortly.

\subsection{Quantifying Flux Change Due to Boundary Shifts}

 In order to isolate the different varieties of flux change, we have developed an algorithm which estimates the change in each region's flux due only to boundary displacements between adjacent, like--signed regions.  These changes necessarily come in pairs: what one region loses in a boundary shift, another gains.
 
 Our algorithm works as follows.  Consider a set of like--signed poles $\{P\}$, and split $\{P\}$ into submerging and emerging sets, $\{P_\downarrow\}$ and $\{P_\uparrow\}$ respectively.  Each pole has an associated flux change, $\Delta^i\psi\equiv \psi^{i+1}-\psi^i$, with $\psi^i$ the flux at time $i$ given by \eqref{eq:psi}.  We iteratively find the pole with smallest unsigned flux change, $P_s$, and the pole with closest centroid $\mathbf{x}$ and opposite sense flux change, $P_c$.  The flux change between these two poles is canceled, $\Delta\psi_c\rightarrow \Delta\psi_c-\Delta\psi_s$ and $\Delta\psi_s \rightarrow 0$, and the cancellation is recorded in a change--in--connectivity matrix: $\Delta^i\Matrix{M}_{c,s}=-\Delta^i\Matrix{M}_{s,c} = \Delta\psi_s$.  This process is repeated until no more connections can be made.  

 $\Delta^i\Matrix{M}$ is antisymmetric, and the $\{j,k\}$\tothe{th} element is the flux--change of the edge joining $j$ and $k$.  Each row $j$ describes the flux change for a given pole, and each column $k$ in a row records how much of pole $j$'s total flux change is in partnership with $k$.  If all of pole $j$'s flux change is due to boundary shifts, then $j$'s total flux change is given by summation along the $j$\tothe{th} row of $\Delta^i\Matrix{M}$:
 \begin{equation}
   \psi^{i+1}_j = \psi^i_j + \sum_{k}\Delta^i\Matrix{M}_{j,k}.
 \end{equation}
 In this way, we have paired as much shrinking flux with like--signed increasing flux as possible, but have only used the total flux and centroid location of each region.
 
 We now refer back to the photospheric mask for the last time.  For each region in the mask array we find every like--signed region with which it shares a border: call this subset of poles $\{b\}$.  To allow for regions separated by a few pixels to share a boundary\footnote{cf. \figref{fig:newflux} and animation of \figref{fig:oldflux}: at various timesteps, N38 and N49 are separated by several pixels of low field, yet there is a clear transfer of flux across their shared boundary.}, we pad each region by 5 pixels and look for overlap with other regions.  The value of $\Delta^i\Matrix{M}$ for each pair of such regions is then the amount of flux they exchanged through a shift in the mask boundary.  
 
 We have chosen to distribute the flux--change and then restrict to the subset $\{b\}$, rather than the reverse, in order to avoid the ill--defined problem of assigning shifts when one region shares boundaries with multiple other regions, which in turn share boundaries with multiple other regions, and so on.  We see this, for instance, in the triple boundary between regions P4, P15, and P19.  Qualitatively, any discrepancy is small given that we connect regions in a nearest--to--farthest order, and nearby regions tend to share boundaries.

 The sum of all such changes for a given pole gives that pole's total flux change due to boundary shifts.  All other change must be ascribed to submergence or emergence through the photosphere.  We may describe this mathematically as
 \begin{equation}
   \label{eq:update}
   \psi^{i+1}_j = \psi^i_j + \sum_{b}\Delta^i\Matrix{M}_{j,b} + \sum_k\Delta^i\Matrix{S}_{j,k}.
 \end{equation}
 The middle term on the RHS is the pole's boundary change, determined in this section; the final term quantifies the pole's flux change due to submergence and emergence through the photosphere, which we now quantify.

 \subsection{\label{sec:qsef}Quantifying Flux Change Due to Submergence and Emergence}

 Our algorithm for determining the flux change due to submergence and emergence rests on two assumptions.  First, pairs of poles submerge and emerge together, and second, a single region can either consist of submerging or emerging flux, but not both.  These assumptions naturally break our connectivity--change graph into two disconnected subgraphs: one consisting of submerging vertices, the other emerging.  Each subgraph is composed of both positive and negative polarity poles, as flux connects only regions of opposite polarity.  

 In determining the flux change in each domain, we use the same algorithm as for boundary shifts, only with the roles of polarity and sense of change reversed: these domains connect opposite polarity poles with same sense flux change, and represent pairs of poles submerging or emerging together.  The graph generated in this manner reflects our physical intuition about how these systems should interact.  Namely, we expect poles to be more connected to those poles closest to them, rather than those further away.   We also expect that poles with little flux change should not have that small change spread between a large number of partners.
 
 \begin{figure}[ht]
   \centering
   \includegraphics[width=\textwidth]{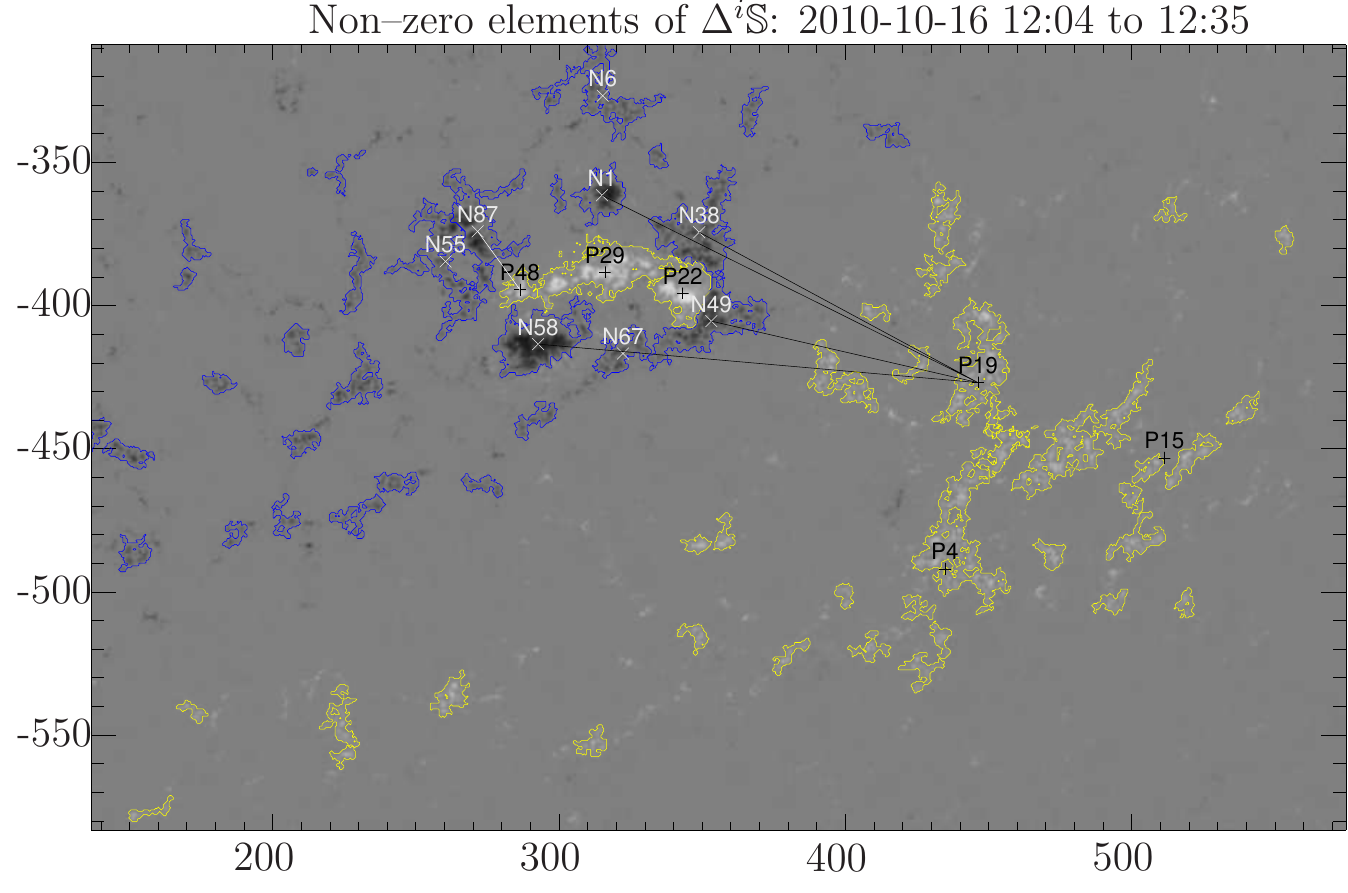}
   \caption[Progression of the connectivity--change algorithm]{\label{fig:smat} Example map of sub/emerging flux, as generated by the algorithm of \S \ref{sec:qsef}.  Black lines show submerging regions, white lines emerging.  Labels mark the centroid locations of the high--flux regions of \figref{fig:fluxplot}, which also contribute to the topological analysis of \S \ref{sec:ergapp}.}
 \end{figure}
 
 We begin by finding the flux difference for each pole between times $i$ and $i+1$, from which we subtract the change due to boundary shifts, found previously.  The remainder is each pole's flux change budget, which must be paired with other poles.  To accomplish this, we again iteratively find the least changing pole and cancel its flux change with its closest appropriately signed and sensed neighbor.  This cancellation is the weight of each edge and is recorded in the source--change matrix, $\Delta^i\Matrix{S}$.  Just like the boundary shift matrix, $\Delta^i\Matrix{S}$ is defined such that, if all of a pole $P_j$'s flux--change between times $i$ and $i+1$ is due to real photospheric changes, then $\psi_j^{i+1} =\psi_j^i + \sum_k\Delta^i\Matrix{S}_{j,k}$; that is, summation over all columns of a given row returns the total flux--change for that row's pole.

 While the combination of our algorithms dealing with boundary shifts and emergence do quite well in most circumstances, occasionally they produce unphysical connections.  Consider the timestep shown in \figref{fig:smat}.  Here, the emerging positive regions (P22, P29, P48) each have some flux change, and our knowledge of the location of emergence says that this change is most likely a combination of boundary shifts and emergence for P48 and P29, as both N87 and N58 are also actively emerging, and primarily boundary shifts for P22.  However, if the boundary shift between P22 and P29 is larger than the emergence of P48, then our boundary algorithm ascribes the boundary shift to P29--P48, leaving the emergence mostly between P22 and N87, which is clearly unphysical.  For this reason, we allow for ``preferred connections'', which accounts for emergence between a subset of poles first, and then proceeds to calculate boundary shifts and submergence and emergence for all poles.  For this region, we first pair emergence between regions P22, N55, N58, and N87 for the first 3 days (77 timesteps), and then between regions P48, N55, N58, and N87 during the remaining time.  As with the generation of consistent mask structures in the previous section, we ultimately wish to automate this entire process.  However, until we develop algorithms that accurately capture the physical evolution of emerging regions, we will rely on a combination of automatic procedures and ad hoc prescriptions to most accurately model these systems.

 \figref{fig:smat} shows a graphical representation of the changing domain fluxes due to submergence and emergence between 12:04 and 12:35 UT on Oct 16th, 7 hours before the flare.  Black lines show submerging domains, white lines emerging.  Table \ref{tab:smat} shows the flux change in each domain.  Regions P4, P15, P22, P29, and N55 have flux change (for these timesteps) due solely to boundary displacements with neighboring regions.

 As discussed above, we begin pairing P48, which increased by $1506.4\times 10^{16}\unit{Mx}$.  The closest (and only) negative pole with the correct sense of flux change is N87, which increased by $849.9\times 10^{16}\unit{Mx}$.  These two poles are paired, the domain flux--change is set to $849.9\times 10^{16}\unit{Mx}$, and P48's flux budget is reduced to $565.5 \times 10^{16}\unit{Mx}$.  There are no more available connections between our preferred poles, so we next account for boundary shifts.  After doing so, the least changing region is N38, which loses $30.6\times 10^{16}\unit{Mx}$.  The closest positive pole with the correct sense of flux change is P19, which lost $1528.9\times 10^{16}\unit{Mx}$.  These two poles are paired, the domain flux--change is set to $30.6\times 10^{16}\unit{Mx}$, and P19's flux budget is reduced to $1498.3\times 10^{16}\unit{Mx}$.  We again select the least changing region (N49), and the process continues until no more pairings can be made.  The final results for this timestep are shown in Table \ref{tab:smat}.

 \begin{table}[ht]
   \centering
   \caption{\label{tab:smat}Non--zero changes in domain fluxes between 12:04 and 12:35 on 2010-10-16}
   \begin{tabular}{lcr}
     \tableline\tableline
     Domain & Sense &  $\Delta^{12:04}\Matrix{S} (10^{16}\unit{Mx}$) \\
     \tableline 
     P19---N1 &  submerge &     245.764   \\
     P19---N38 &  submerge &     30.6279 \\
     P19---N49 &  submerge &     145.569 \\
     P19---N58 &  submerge &     257.030 \\
     P48---N87 &  emerge   &     849.874 \\
     \tableline
   \end{tabular} 
 \end{table}

 The pairings represent a compromise between reasonable inference and mathematical necessity.  Due to their proximity, and common increase, it is reasonable to assume P48--N87 are feet of the same emerging flux tubes.  P19, on the other hand, is relatively isolated but has steadily decreasing flux.  It is necessary to pair it with decreasing negative poles, which turn out to be arranged along the periphery of the old polarity.  Being separated by $\sim100''$, these regions cannot be ``submerging'' in any real sense.  However, we believe this decrease represents steady cancellation with the small scale field surrounding these old--flux regions.  This process should be studied in detail in another investigation, but for the moment we note that it is well represented by a reduction in, for instance, the P19--N49 domain flux.
 
\begin{figure}[ht]
   \centering
   \includegraphics[width=\textwidth]{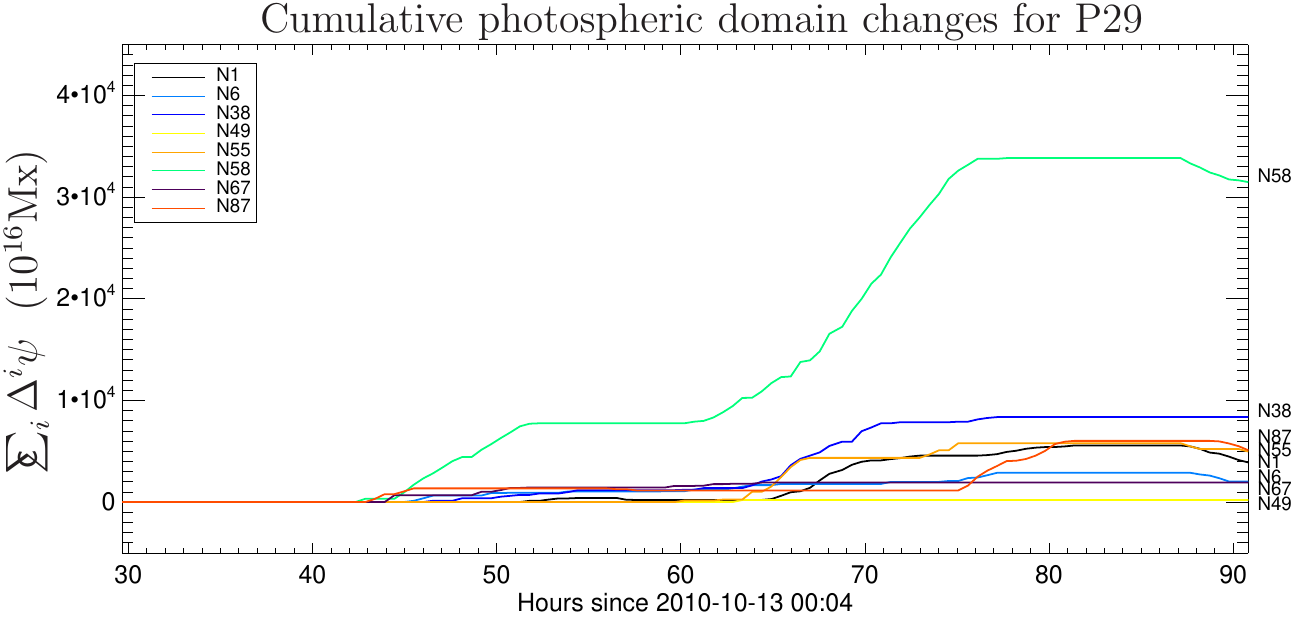}
   \caption[Cumulative changes in P29's domain fluxes]{\label{fig:p29dom}Cumulative changes in each of P29's domain fluxes, as calculated with the algorithm of \S \ref{sec:qsef}.  We find that P29 primarily emerges with N58 and N38.}
 \end{figure}

 We repeat this process for each pair of consecutive timesteps.  \figref{fig:p29dom} provides a representative example, showing the cumulative domain flux changes (designated by the symbol $\csum$) due to the emergence, and slight amount of submergence near the end of our analysis, of P29, one element of the newly emerging flux.  We see here that it emerges primarily with region N58, part of the newly emerging flux.

 The effectiveness of the combined algorithms can be appreciated in \figref{fig:bndry}, showing the three components of the diffuse western region.  The blue line illustrates the effectiveness of our algorithms at capturing the system's behavior due to submergence and emergence.  As the field shears, the boundaries between the three subregions have a tendency to discontinuously jump, resulting in large variations in the flux in each region, accounted for by the boundary algorithm.  However, the combined flux of the three subregions, shown in black, instead displays a fairly steady decrease: this whole region is submerging.  The blue line shows our reconstruction of this submergence using the source--change matrix at each time.  Each pole has an initial flux value, to which we add the elements of $\Delta^i\Matrix{S}$, so that each pole's flux at time $i$ is given by $\psi_j^i = \psi_j^0 + \sum_{l=0}^{i-1}\sum_k \Delta^{l}\Matrix{S}_{j,k}$.  This measure completely discounts the (artificial) boundary shifts between photospherically adjacent regions.  We then sum the flux in each pole thus reconstructed, and this is our estimate for the submergence of diffuse western region.  We note that the blue and black lines follow each other reasonably well, indicating that our algorithms accurately capture the submerging trend of this region.
 
 \section{\label{sec:erg}Calculating the Energy}
 \subsection{Topological Definitions}
 In the second section of our analysis, we employ the Minimum Current Corona (MCC) model \citep{Longcope:2001,Longcope:1996} to use the amount of flux change in each domain to calculate the amount of free magnetic energy stored in our system as it evolves away from a potential configuration.  To that end, we must define a set of topological elements.  These definitions, briefly summarized below, are described more fully in \citet{Longcope:2002,Longcope:2005} and references therein.  The general idea is to model distributed photospheric sources as point sources.  The potential field generated by these sources will have null points, $\vect{x}_\alpha$, where the magnetic vector field vanishes: $\vect{B}(\vect{x}_\alpha) = 0$.  We may perform a linear expansion of the field about these points \citep[\S 2.4]{Parnell:1996,Longcope:2005}
 \begin{equation}
   \vect{B}(\vect{x}_\alpha+\delta\vect{x})\simeq \Matrix{J}^\alpha\cdot\delta\vect{x},
 \end{equation}
 where the Jacobian matrix $\Matrix{J}^\alpha_{ij}\equiv \partial B_i/\partial x_j$ is both symmetric and traceless because $\nabla\times\vect{B}=\nabla\cdot\vect{B}=0$.  As such, $\Matrix{J}^\alpha$ has three orthogonal eigenvectors with three real eigenvalues: the eigenvectors of the two like--signed eigenvalues define a plane (the fan); the eigenvector of the opposite--signed eigenvalue defines a line orthogonal to this plane (the spine).  From this foundation the rest of the topological description follows.  We define a \emph{pole}, the photospheric point source, located at the flux--weighted centroid of a mask.  \emph{Nulls} are the zeros of the potential magnetic field, and \emph{spines}, the fieldlines connecting a pole to a null along the single eigenvector.  A \emph{fan} (synonym \emph{separatrix}) is the set of fieldlines ending in a null's like--signed eigenvector plane.  A \emph{domain} is a simply connected space filled with field lines connecting a given pair of poles.  Finally, a \emph{separator} is the field line connecting two nulls, formed at the intersection of their respective fan surfaces.  In a nonpotential field, the separator may broaden into a two dimensional ribbon.  Coronal current sheets form along the separators of the field.  

 We determine the locations of all nulls in two steps.  First, we find those laying in the photospheric plane via the Newton--Raphson method of \citet{Barnes:2005}.  Second, we check the Euler characteristics given in \citet{Longcope:2002}, which provide both 2D and 3D relations between the number of nulls of each type and the number of sources of each polarity.  When the 3D characteristic is not satisfied, at least one coronal null is missing.  We then supply a ``by--eye'' list of initial $(x,y,z > 0)$ guesses to the same Newton--Raphson root finding method until coronal nulls have been found and the Euler characteristics are satisfied.  Other more automated coronal null detection methods have been developed \citep{Barnes:2007}, but the ad--hoc method described here has worked well in this case.

 After determining the location of all nulls, we located separators via the method of \citet{Barnes:2005}, with one modification.  Those authors only treat photospheric nulls, whose separators lay within separatrices above the photosphere.  For coronal nulls, we search for separators in all $2\pi$ directions within the null's fan surface.
 
 Our model photosphere consists of a plane $(z=0)$ of isolated sources surrounded by regions where the normal component of the magnetic field is zero.  This boundary condition may be satisfied using the method of images, as in \citet{Longcope:2002}\S 6.  With the normal field reflectionally symmetric in $z$, we have that $B_z(x,y,-z) = -B_z(x,y,z)$.  Whenever we introduce current in the corona along a separator, we must introduce its reflection in the mirror corona to maintain this symmetry.  In this way, our separators form closed current loops.

  \begin{figure}[ht]
   \centering
   \includegraphics[width=0.75\textwidth]{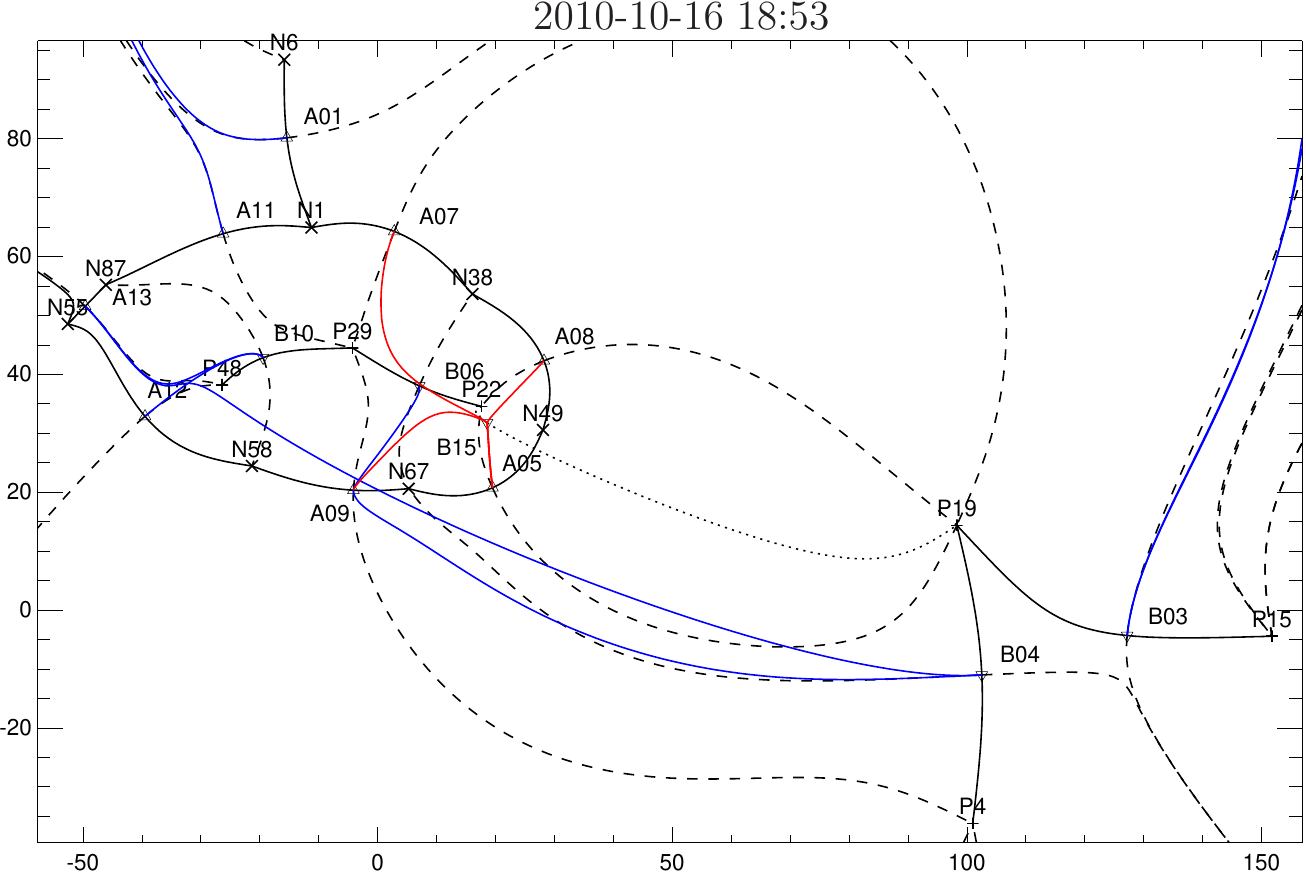}
   \caption[Topology of NOAA AR 11112]{\label{fig:topo}Topology of NOAA AR 11112 on the eve of the flare, depicted in a local tangent plane with coordinates.  The point of tangency is taken as the center of charge in the initial magnetogram, translated through solar rotation to the present time, and the axes are in Mm.  Pluses and crosses are positive and negative poles, respectively; triangles are positive ($\vartriangle$) and negative ($\triangledown$) nulls; solid black lines depict spines, dashed lines the trace of fans within the photosphere, and the dotted line is a coronal spine, attached to the coronal null B15.  Blue and red lines are projections of separators into the photosphere, with red lines those separators connected to the coronal null.}
 \end{figure}
  
  \subsection{Specifics of Domain Fluxes}
  Once we have created the set of matrices describing domain flux changes due to submergence or emergence, we can create similar matrices describing the flux change due to a changing potential field.  We calculate the potential field connectivity at each timestep via the Monte Carlo method of \citet[\S3.1]{Barnes:2005}.  This produces a set of connectivity matrices $\{\Matrix{P}\}$, where $\Matrix{P}^i_{j,k}$ is the flux connecting sources $j$ and $k$ at time $i$.  From these matrices we calculate the change in connectivity due to the changing potential field: 
  \begin{equation}
    \Delta^i\Matrix{P} \equiv \Matrix{P}^{i+1} - \Matrix{P}^i.
  \end{equation}
  Summation along rows of the connectivity matrix returns the total flux of the corresponding pole; eg. $\psi_j^i = \sum_k\Matrix{P}_{j,k}^i$.  We define $\Delta^i\Matrix{P}$ to be antisymmetric, as with the matrices $\Delta^i\Matrix{S}$ and $\Delta^i\Matrix{M}$ before.
  
  The MCC model derives an energy from the discrepancy between the actual domain fluxes $\Matrix{F}^i$ at time $t_i$ and the potential fluxes then, $\Matrix{P}^i$.  In the absence of flux emergence or submergence, the actual fluxes are fixed using the potential field domain fluxes at some initial time: $\Matrix{F}^i = \Matrix{P}^0$, as in \citet{Kazachenko:2010}.  Any flux emergence or submergence through the photosphere, quantified as $\Delta^i\Matrix{S}$ in the previous section, modifies these actual domain fluxes, so that 
  \begin{equation}
    \label{eq:freal}
    \Matrix{F}^i = \Matrix{P}^0 + \sum_{j = 0}^{i-1}\Delta^i\Matrix{S}.
  \end{equation}
  
  Because we assume a field, initially potential, whose fluxes are fixed under future evolution, we determine how far removed the field is from a potential field configuration by answering the question, ``What flux must be added to a potential field domain $D$ at time $0$ to get a potential field domain at time $i$?''  This question is answered by
  \begin{equation}
    \label{eq:fquest}
    \Matrix{P}_D^i = \Matrix{P}_D^0 + \sum_{j=0}^{i-1}\Delta^j \Matrix{S}_D + \sum_{j=0}^{i-1} \Delta^j \Matrix{R}_D.
  \end{equation}
  On the right hand side, the first term is the potential field at the initial time.  The second is the total flux change through the photosphere.  The final term is flux change due to coronal redistribution, which must be achieved by modifying the connectivity matrix.  Together, the first two terms give the actual domain flux $\Matrix{F}^i_D$ at time $i$.  All flux changes from coronal reshuffling are then given by 
  \begin{equation}
    \label{eq:drx}
    \sum_{j=0}^{i-1} \Matrix{R}_D^j = \Matrix{P}_D^i - \Matrix{F}_D^i.
  \end{equation}
  Equation \eqref{eq:drx} holds for all domains, of course, and we represent this as a matrix equation by dropping the domain subscript $D$.  $\Delta^i\Matrix{P}$ is antisymmetric, and we have defined $\Delta^i\Matrix{S}$ to be antisymmetric, so $\Delta^i\Matrix{R}$ is also antisymmetric.  While antisymmetry of these matrices carries no physical information, we will later show that it does endow $\Delta^i\Matrix{R}$ with a nice mathematical property.
  
  Note that the difference in \eqref{eq:drx} between the actual domain fluxes and the potential field fluxes depends on the choice of the initial time $t_0$.  When dealing with actual data, we must pick an initial time to apply the MCC flux constraint when we believe the field is in a potential configuration in order to find a meaningful difference relative to a later potential configuration.  Ideally, we would track a single region from its inception up to an event which redistributes the coronal flux, as in a flare.  While NOAA AR 11112 does not fit this scenario, it likely matters little in this case.  As is apparent from our animation, and also as we will show below, NOAA AR 11112 has a very stable flux configuration for $\sim 40$ hours prior to flux emergence.  This newly emerged flux, fixed in set domains as the photospheric field continues to evolve, rapidly diverges from a potential field configuration at later times.
  
  \subsection{\label{sec:seps}Separators, Separator Currents, and Energy Storage}
  So far this discussion only involves the sources themselves and their interconnections.  Let us now introduce a set of separators $\{\sigma\}$.  Separators are field lines that run from a null point of one type to a null of the opposite type.  In the MCC model, they are the sites of current sheets within the corona \citep{Longcope:2005,Priest:1996}.  Because current may only flow in loops, the separators themselves must close along some path in the mirror corona.  We can then speak of separator fluxes: the flux of domains linked by some separator $\sigma$ with its closure.  We will discuss closures and explain the concept of linking in detail below, but the general idea is that field lines in a linked domain cannot reconnect to another domain except by passing through the separator.  

  Let $\psi_\sigma^i$ be the flux linked by $\sigma$ at time $i$ in the actual field, and
  \begin{equation}
    \label{eq:linkpotl}
    \psi_\sigma^{(v)i} = \sum_D\Matrix{P}_D^i = \sum_{\{(j,k)\}}\Matrix{P}_{(j,k)\in D}^i
  \end{equation}
  be flux in the potential field domains $D=\{(j,k)\}$ linked by $\sigma$.  For a given closure, a separator always links the same set of domains throughout time.  The fluxes in those domains will generally change over time, however.  When a domain's flux goes to zero, the domain no longer exists.

  The flux--constrained--equilibrium (FCE) assumption of the MCC model states that domain fluxes are fixed as the field evolves.  Minimization of the field's energy subject to these constraints shows that the coronal field will be current--free except along the separator, where a current ribbon forms \citep{Longcope:2001}.  This occurs whenever $\psi_\sigma^i\ne\psi_\sigma^{(v)i}$.  We therefore define
  \begin{gather}
    \label{eq:fce}
    \psi_\sigma^{(cr)i} \equiv \Delta\psi_\sigma^i = \psi_\sigma^i - \psi_\sigma^{(v)i} = \sum_D\Matrix{F}^i_D - \sum_D\Matrix{P}^i_D,
    \intertext{which, from equation \eqref{eq:drx}, is}
    \label{eq:totfcr} \Delta\psi^i_\sigma= - \sum_D\sum_{j=0}^{i-1}\Delta^j\Matrix{R}_D.
  \end{gather}
  The double sum over elements of the redistribution matrix $\Matrix{R}$ is the difference between the flux linked by the separator in the actual and potential fields at time $i$.  It is also the self--flux generated by current flowing along the separator, and therefore, the flux that must be added to the real field to relax the FCE constraint of MCC: $\psi_\sigma^{(cr)i} +\sum_{D,j}\Delta^j\Matrix{R}_D = 0$.
  
  Our method for calculating energies therefore rests on two assumptions.  The first is that at some initial time $0 < i$ we assume that the domain fluxes are given by the potential field's domain fluxes, $\psi^0_{(j,k)} = \Matrix{P}^0_{(j,k)}$, so that all the separator fluxes $ \psi_\sigma^{(cr)0} =0$.  The second assumption is that the evolution of the current ribbon's self--flux depends on the evolution of the both the actual and potential field configurations:
  \begin{align}
    \label{eq:dfcr}\Delta^i\psi_\sigma^{(cr)} = \Delta^i\bigl( \Delta\psi_\sigma^i  \bigr) & = \Delta\psi_\sigma^{i+1} - \Delta\psi_\sigma^{i}\\
    & = \Delta^i\psi_\sigma - \Delta^i\psi_\sigma^{(v)}\\
    & = -\sum_D \Delta^i \Matrix{R}_D.
  \end{align}
  Summed over time, this is just the restatement of \eqref{eq:fce} itself, namely that the current ribbon's flux at some time $i$ is the sum of all changes in each linked domain's flux over the life of the separator.
  
  When $\psi^{(cr)i}_\sigma$ is small we can estimate the properties of the current ribbon, as in \citet{Longcope:2004}.  For a separator of length $L$ carrying current $I$, the authors determine the self--flux to be
  \begin{equation}
    \label{eq:crpsi}
    \psi^{(cr)i}_\sigma = \Frac{I L}{4 \pi}\ln\Biggl(\Frac{e I^*}{\vert I\vert}\Biggr), 
  \end{equation}
  where $I^*$ is a measure of magnetic shear in the separator's vicinity.  From this the excess energy of the MCC field relative to the potential field, i.e. the free energy, is \citep{Longcope:2004,Longcope:2001}
  \begin{equation}
    \label{eq:wmcc}
    \Delta W_{MCC} = \Frac{1}{4\pi}\int_{\Psi_{potl}}^{\Psi}I d\Psi = \Frac{L I^2}{32 \pi^2}\ln\Bigl(\Frac{\sqrt{e} I^*}{\vert I\vert}\Bigr).
  \end{equation}
  Note that \eqref{eq:crpsi} and \eqref{eq:wmcc} only take into account energy due to the self inductance of each current loop.  In general, we must include the effect of mutual inductance, though for physically separated loops we expect self inductance to dominate the total energy.
  
  The FCE assumption \eqref{eq:fce} requires that $\psi_\sigma^i = \psi^{(v)i}_\sigma + \psi^{(cr)i}_\sigma$, which allows us to solve \eqref{eq:crpsi} for the current in the current ribbon as a function of $\Delta\psi_\sigma^i$:
  \begin{gather}
    \label{eq:cr}
    I(\Delta\psi_\sigma^i) = I^*\Lambda^{-1}(4\pi\Delta\psi_\sigma^i/LI^*)
  \end{gather}
  where $\Lambda^{-1}(x)$ is the inverse of the function $\Lambda(x)\equiv x\ln (e/\vert x\vert)$.  Equation \eqref{eq:cr} provides a method for determining the current residing in, and hence the energy stored by, each separator: it tells us the current required to change the domain flux enclosed by a separator from that of a potential field to some other value.  In the present case that value is given by the non--zero elements of $\Delta^i\Matrix{R}$.  A reverse, cumulative sum over the time index $i$ for a given flux domain $P_j$--$P_k$ gives the time--history of departure from a potential field configuration.
  
  \subsection{\label{sec:gauss}Using the Gauss Linking Number to Find Linked Domains}
  As is implicit in \eqref{eq:totfcr}, each separator may link more than one domain: the appropriate $\psi^{(cr)i}_\sigma$ for a separator is the sum of all changes in domain fluxes for each domain linked by that separator. Since all field lines in a given domain are topologically equivalent\footnote{This is \emph{not} to say that all flux connecting two poles is equivalent.  A single pair of poles may have more than one distinct domain: see \citet{Beveridge:2005,Parnell:2007}.  This situation may arise, for instance, when you have purely coronal domains.}, we may establish separator linkage using a single representative fieldline and the separator itself.  After each of these open curves is closed, their linkage is found using the Gauss linking number \citep{Berger:1984}, $L_{12}$, for each field line--separator pair: 
  \begin{align}
    L_{12} = L_{21} = \Frac{1}{4\pi}\oint_{\ell_1}\oint_{\ell_2}\Frac{\vect{r}_1 - \vect{r}_2}{\vert \vect{r}_1 - \vect{r}_2\vert^3}\cdot(d\vect{r}_1\times d\vect{r}_2)
  \end{align}
  The linking number not only determines whether a separator links a field line ($L_{12} \ne 0$), but also the sense in which it does so ($L_{12} = \pm n,\ n$ an integer): to calculate the correct $\psi_\sigma^{(cr)i}$, you must sum over all linked domains, multiplied by their respective linking numbers.

  In order to use Gauss' linking formula, we must have two closed curves, whereas our curves (usually\footnote{It is possible for a separator to attach to a coronal null, in which case, in order to create the appropriate loop, you must follow a second separator from the coronal null back to the photosphere, as discussed below.}) begin and end at two separate sources (nulls for separators) in the photosphere, never penetrating beneath.  So, in order to apply the linking formula in this case we must add to each coronal curve some curve in the mirror corona to form a complete loop.  There are, in general, many ways to form the closure for each curve, and not all of them will lead to the same linking number.  One requirement is that the separator be closed \emph{above} the field line's closure: any closure below the field line will always give a linking number of zero.  Therefore, we have chosen to close the separator with a straight line between the footpoints in the photosphere, and the field lines with a rectangular path formed by following the footpoints down and connecting them with a straight line in the $z=-0.5\unit{Mm}$ plane.  Another method would be to close the separator, say, with its projection into the $z=0$ (photospheric) plane.  Either method is valid, so long as the same method is used for all field lines and separators.

  Another requirement for using the linking number to find linked domains is that we follow field lines and separators in a consistent direction.  In general, changing the direction of \emph{one} of the integrals takes $L_{12}\longrightarrow \bar{L}_{12} = -L_{12}$.  For definiteness, we always trace separators from negative (A--type) nulls to positive (B--type) nulls above the photosphere, and thus close separators from B$\to$A in the photosphere.
  
  Because the reconnection matrix $\Delta^i\Matrix{R}$ is antisymmetric, this same care is unnecessary when tracing the field lines, at least for the purpose of the $\psi_\sigma^{(cr)i}$ calculation.  Let us designate by $\Delta^i\bigl(\Delta\psi_{P1,N2}\bigr)$ the difference in flux--change for the real versus potential field between times $i$ and $i+1$ (as in \eqref{eq:dfcr}), when we trace a field line from P1 to N2 in the corona, and close the field line from N2$\to$P1 below the photosphere.  Then, 
  \begin{align}
    \Delta^i\bigl(\Delta\psi_{P1,N2}\bigr) & = \Delta^i\Matrix{R}_{1,2} \times L_{12}
    \intertext{Now trace the field line from N2$\to$P1 in the corona, with the appropriate closure for $z<0$.  We have}
    \Delta^i\bigl(\Delta\psi_{N2,P1}\bigr) & = \Delta^i\Matrix{R}_{2,1} \times \bar{L}_{12}\notag\\
    & = \bigl(-\Delta^i\Matrix{R}_{1,2} \bigr)\times \bigl(-L_{12}\bigr)\notag\\
    & = \Delta^i\bigl(\Delta\psi_{P1,N2}\bigr),
  \end{align}
  We have therefore been justified in simply writing $\Delta^i\psi_\sigma^{(cr)}$, with no designation of tracing a field line from positive to negative or negative to positive in the corona, because $\Delta^i\psi_\sigma^{(cr)}$ is just a summation of multiple such domains.  This holds provided we always trace the separator in the same direction.

  Once determined, the linking number for each separator/field line pair provides the correct method for calculating each separator's total $\psi_\sigma^{(cr)i}$, including all linked domains, and hence the energy stored by each separator.
  
  \subsection{Two Complications}
\begin{figure}[ht]
    \begin{center}
      \includegraphics[width=\textwidth]{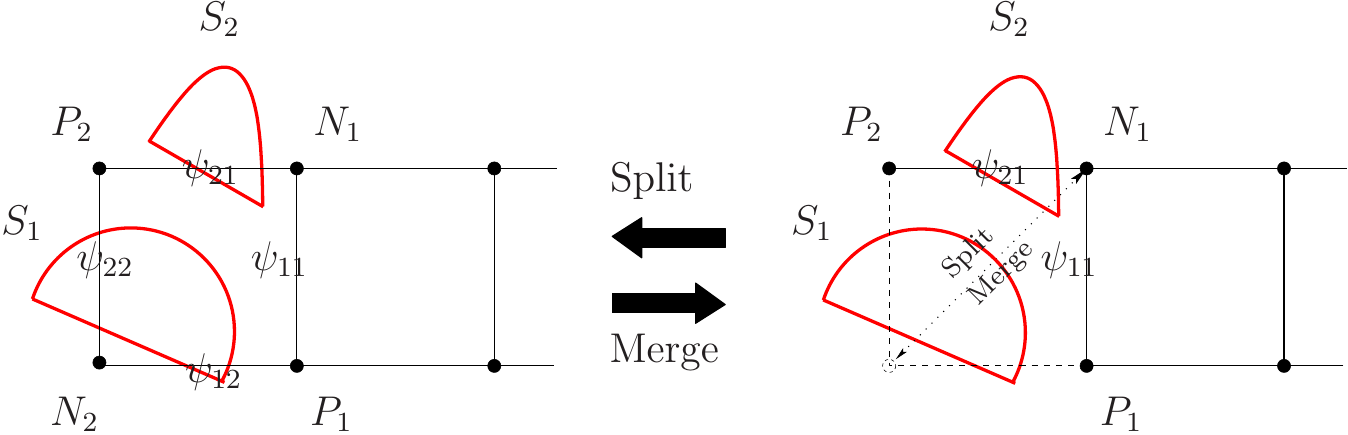}
      \caption[Topology of splitting poles]{\label{fig:constrict} Topological depiction of two poles either splitting or merging: $N_1\longleftrightarrow N_2$.  Dots depict the sources, lines the domains, and red arcs the separators.}
    \end{center}
  \end{figure}

  We must account for instances where two photospheric regions merge together or split apart, as illustrated in \figref{fig:constrict}.  Therein, we focus on 4 poles---$P_1, P_2, N_1, \& N_2$---with four flux domains---$\psi_{1,1}, \psi_{1,2}, \psi_{2,1}, \& \psi_{2,2}$---and two separators, $S_1$ and $S_2$, which link the domains in some fashion.  Suppose two of the poles are each connected to the same null via its  spines.  If the flux of one pole is zero before(after) a split(merger), then the null between them does not exist at that time, and the flux in the domains linking the sources goes to zero.  Now, if either $S_1$ or $S_2$ is attached to that null, then the separator will arise or cease to exist with that null, and the flux linked by it is appropriate.  Then again, if these separators are attached to some other set of nulls and happen to link these flux domains, our calculated linked flux is still appropriate.  Say flux domain $\psi_{2,2}$, for instance, either folds into or breaks out from domain $\psi_{2,1}$.  The actual break or merger constitutes a boundary shift (with $N_2$ having zero flux either before or after the event) and so does not contribute to the flux linked by either separator.  Any additional flux change on top of this is either appropriately picked up by $S_1$ or $S_2$, but not both.

  A second complication arises when we encounter a coronal null, as we see in the topology of NOAA AR 11112 in \figref{fig:topo} near local tangent plane coordinates (20,30)Mm.  If a coronal null has any separators connected to it then it must have at least two: one carrying current up, the other down.  If many separators connect to the coronal null, then some must carry current up, and others down, but all balanced according to Kirchhoff's rules.  In order to calculate the flux within a separator--induced domain, we must close the separator somehow.  Above, we addressed this by prescribing a photospheric closure for separators.  Now, however, we must first stitch pairs of separators together in order to get back to the photosphere.  If we have $n$ separators connected to the coronal null, we require $n-1$ separator pairs, or ``isolating loops'' in the terminology of \citet{Longcope:2001}.  We might assume that the currents flowing along the $n-1$ loops would distribute to minimize the energy due to self and mutual inductances.  This problem remains to be solved, and for the moment we simply designate one separator linked to a coronal null as the ``shared'' separator.  We have found that the energies we calculate do not depend heavily on the choice of common leg, laying within $\approx 20\%$ of each other.

  As shown in \figref{fig:topo}, we have a single coronal null, B15, with spine lines leading to $P22$ and $P19$.  Four separators (in red) connect to the coronal null: from null A09 between $N58$ and $N67$, A05 between $N67$ and $N49$, A08 between $N49$ and $N38$, and A07 between $N38$ and $N1$.  Again, because all current flowing from the photospheric nulls up to the coronal null must be balanced by current flowing back down to the photosphere, we mush combine these four separators into three, all of which share one leg; it does not matter which one.  We have chosen A05--B15 leg to be common to all three separators.  As we will see below, the A05--B15--A08 separator generates the greatest amount of energy in this system.  This is expected, given the high horizontal shear in the photospheric field between $P22$ and $N49$, and the greatly flux--deficient domain between these two poles, relative to the potential field configuration at the time of the flare.
  
  \subsection{Summary}
  We may summarize the previous several sections in the following way.  In order to perform an energy calculation, we need a set of poles $\{P\}^{t_f}$ at a given time final time $t_f$; a set of nulls at the same time $\{N\}^{t_f}$; a set of separators at the same time $\{\sigma\}^{t_f}$; a time--history of the reconnective changes in domain fluxes from the final time, backwards to a previous time $t_i$ where we may assume the field was potential, $\{\Delta^i \Matrix{R}\}$; and a set of domains $\{D_\sigma\}^{t_f}$ linked by each separator.

  Having all these pieces, we then calculate the total domain flux difference in each domain at time $t_f$ relative to the initial presumed potential field.  For each separator, we find all domains linked by that separator.  Next, each linked separator's time--history of reconnective flux changes is summed.  This total flux--change for each linked domain is then multiplied by the Gauss Linking Number for that domain/separator pair.  Finally, the total domain changes are summed for each linked domain to fix the total flux difference of the separator at the final time $t_f$, relative to the potential field at the initial time $t_i$.  Mathematically, this is given by \eqref{eq:totfcr}, reproduced here:
  \begin{equation*}
    \psi_\sigma^{(cr)f} = -\sum_{D}\sum_{j=t_i}^{t_f-1}\Delta^j\Matrix{R}_D
  \end{equation*}
  This flux is generated by the current along the separator, \eqref{eq:crpsi}, which may be found by the inversion \eqref{eq:cr}, and plugged into the energy equation \eqref{eq:wmcc} to find the energy in excess of the potential field energy at time $t_f$ in the Minimum Current Corona model with the FCE constraint applied at time $t_i$.  This energy is a lower bound for any coronal energy model.

  \section{\label{sec:ergapp}Application of MCC Energy Calculation to NOAA AR 11112}
  By time of the flare (\figref{fig:newflux}) the new flux in NOAA AR 11112 has bubbled up and expanded within the interior of the old--flux ring.  The effect is especially pronounced between regions P22 and N49, the boundary of which has a very strong horizontal gradient in the vertical field.  Because this gradient is between old and new flux, prior to reconnection during the flare there is little flux in the domain connecting these two regions, and certainly far less than there would be in the potential configuration at that time.
  
  We expect one action of the flare will be to increase the flux within the P22--N49 domain.  Because its flux is conserved, any flux P22 adds to its N49 domain must come at the expense of its other domains.  The poles with which P22 shares these other domains---primarily N58, N55, and N87 with which P22 emerged---must in turn shuffle their flux into different domains.  In this way, a flare originating in one place propagates its effects throughout the entire magnetic structure until it establishes a new, lower energy equilibrium.

  \begin{figure}[ht]
    \begin{center}
      \includegraphics[width=\textwidth]{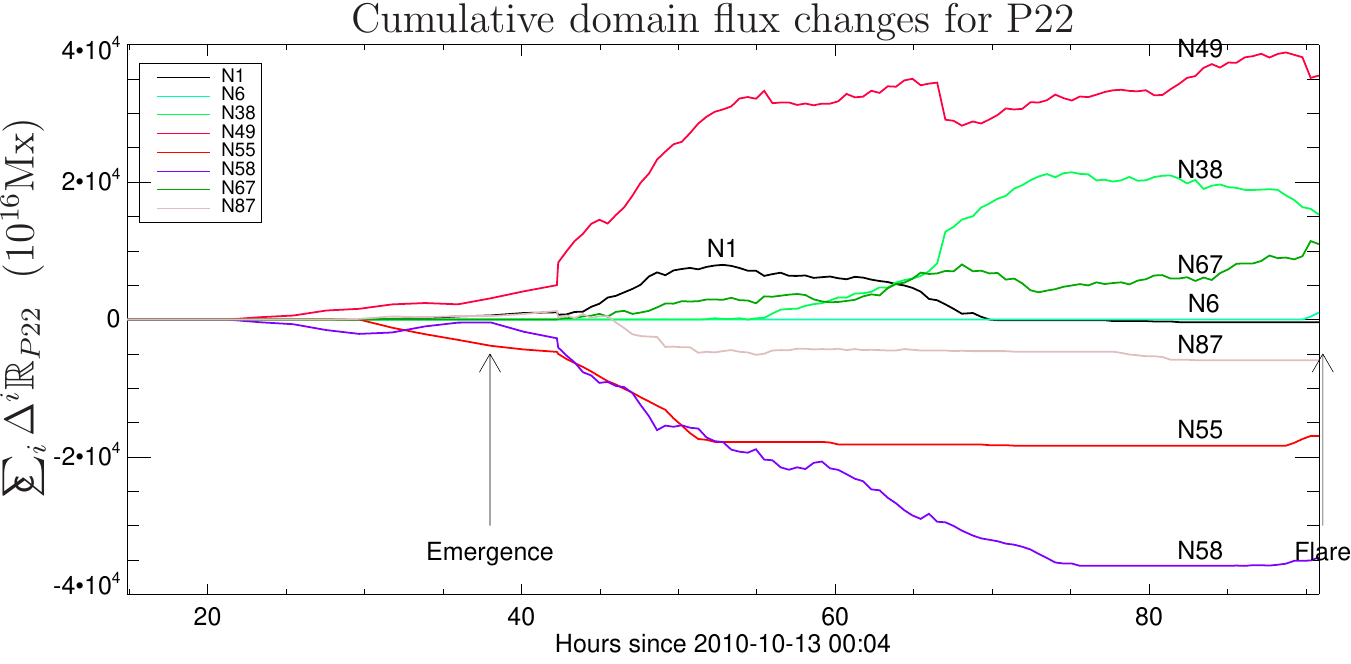}
      \caption[P22 domain flux changes]{\label{fig:p22rmat} Cumulative domain flux changes for various P22 domains, relative to the potential field at t=0.  Arrows indicate the onset of flux emergence and the UT 19:07 M2.9 flare.  There is some ``pre--emergence'' flux due to our 13 timestep smoothing function.}
    \end{center}
  \end{figure}

  We can get a picture of this process by looking at the cumulative domain flux changes for various P22 domains, shown in \figref{fig:p22rmat}.  Here, the flux at each timestep is given by the cumulative sum of all flux changes up to that time, $\sum_{j=0}^i(\Delta^j\Matrix{R}_{P22,*})$.  This figure shows by how much flux a domain must be changed to match the potential field configuration of the initial field, plus the contribution due to emergence.  

  P22 emerges in conjunction with N55, N58, and N87, thus all flux is assigned to domains P22--N55, P22-N58, and P22-N87.  The potential field, however, assigns less flux to these domains and more (i.e. some) to domains such as P22--N1 and P22--N49.  These latter domains are flux--deficient relative to the potential field, and appear above the zero line in \figref{fig:p22rmat}.  The former are flux--excessive, as a result of emergence, and appear below.

  Note that the potential field distributes the flux among domains differently at different times, and our analysis captures these changes.  In particular, P22 emerges to the east of N1, then migrates westward.  As P22 moves past N1, the P22--N1 domain first becomes deficient, then more sufficient as the P22--N38 and P22--N67 domains become increasingly deficient.  The total flux in P22 only slowly increases throughout much this time (it is mostly emerged by 50hrs, as shown in \figref{fig:fluxplot}), so these later changes in the potential configuration are largely due to shear motion in the photosphere, rather than the initial emergence.

  AIA 171\AA\ images during the flare reveal how reconnection works to restore a flux balance closer to potential.  At successive times, we see brightenings of small loops progressing west to east within the newly emerged regions.  After this we see a series of loop brightenings connecting the new negative regions within the ring to the old diffuse positive region to the west.  We believe that this series of loop brightenings is a direct result of domain restructuring due to the emergence of the new flux and its highly nonpotential original field configuration.

  \begin{figure}[ht]
    \begin{center}
      \includegraphics[width=\textwidth]{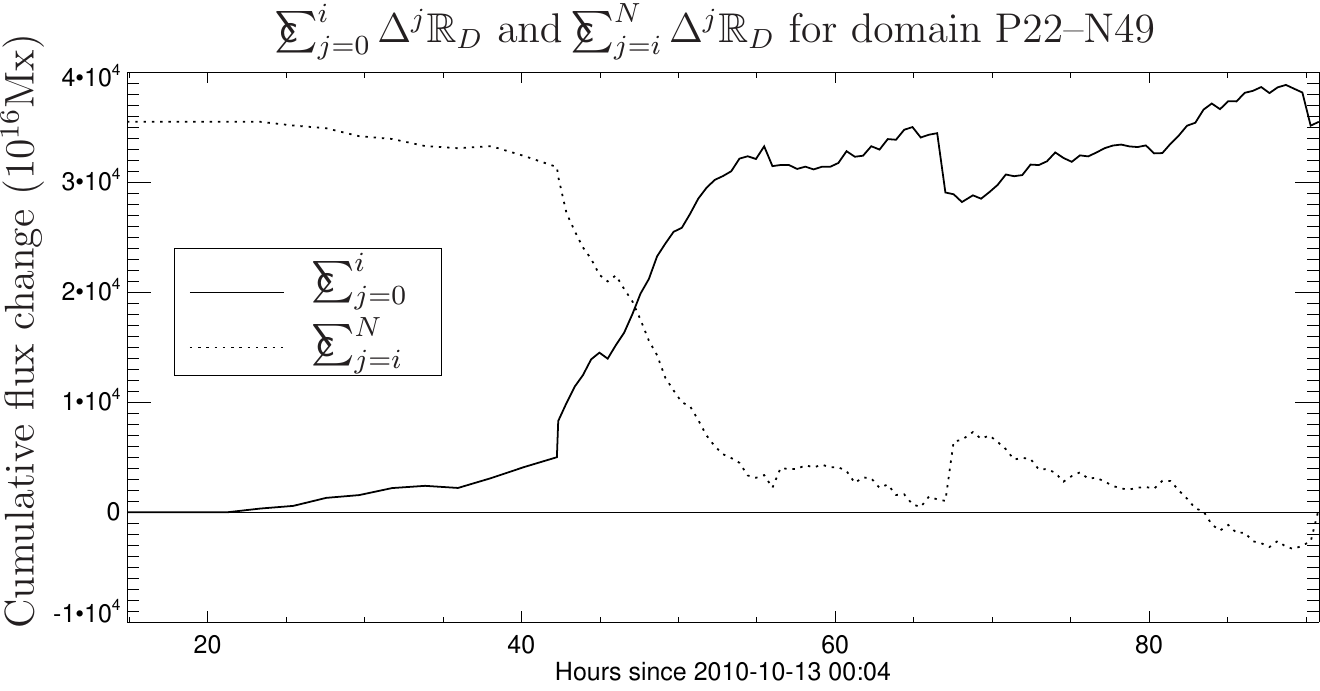}
      \caption[Cumulative domain flux shown two ways]{\label{fig:twosums}Two cumulative time--sums over $\Delta^i\Matrix{R}_{P22-N49}$.  The solid line sums from the initial time, 0, to time $i$ and is the flux required to change the domain at time $i$ to an assumed potential field at the initial time.  The dotted line sums from time $i$ to the final time, $N$, and is the flux required to change the final potential field configuration to a flux configuration fixed at time $i$.  The difference between the initial and final values is the same for both sums, and they are mirrors of each other about the average of this difference.}
    \end{center}
  \end{figure}
  
  The fluxes shown in \figref{fig:p22rmat} provide a good description for understanding the evolution of each domain, and directly follow the calculation outlined in \S~\ref{sec:seps}.  In this chapter, however, we are interested in calculating the free energy due to a difference between domain fluxes fixed at some initial time and potential fluxes at the time of the flare.  For a completely new active region, we would assume all domain fluxes are zero except those which emerge together; in this case, our emergence occurs in a region of substantial old flux.  As such, our final time is fixed by the flare, and we must choose an appropriate time of initial constraint.  Combining equations \eqref{eq:freal} and \eqref{eq:fquest}, we find that $\Delta^i\Matrix{P}_D = \Delta^i\Matrix{F}_D + \Delta^i\Matrix{R}_D$, so that $\Delta^i\Matrix{R}_D$ is the flux that must be added to domain $D$ to account for change in addition to submergence and emergence, which is accounted for within $\Delta^i\Matrix{F}_D$.  \figref{fig:p22rmat} shows the cumulative flux $\sum_{j=0}^i\Delta^j\Matrix{R}_D$.  Wishing to vary the time of constraint, we calculate the reverse of this: $\sum_{j=i}^N\Delta^j\Matrix{R}_D$.  This is the amount of flux that must be added to the domain constrained at time $i$, $\Matrix{F}_D^i$, to match the potential field domain at time $N>i$.  As we will show below, this summation is basically constant prior to emergence: the old flux region is initially stable, and flux emergence drives a departure from this stable configuration.

  \figref{fig:twosums} gives an example of both the forward and reverse summations for domain P22--N49.  We set the fluxes for the energy calculation in this way because the assumption of the MCC model is that, barring reconnection, the domain fluxes are fixed, and prior to the flare there is no reconnection.  In order to compare with a potential field at the time of a flare, we determine our topology immediately before the flare.  This fixes all of the poles, nulls, and separators, and the domains linked by each separator.  The energy calculation then answers the question, ``\emph{Supposing the domain fluxes are fixed at time $i$, by how much must they be changed to match the potential field at the final time?}''  This difference for each domain linked by a separator then determines the amount of current along that separator, and hence the amount of energy stored by that separator.

   \begin{figure}[ht]
    \begin{center}
      \includegraphics[width=\textwidth]{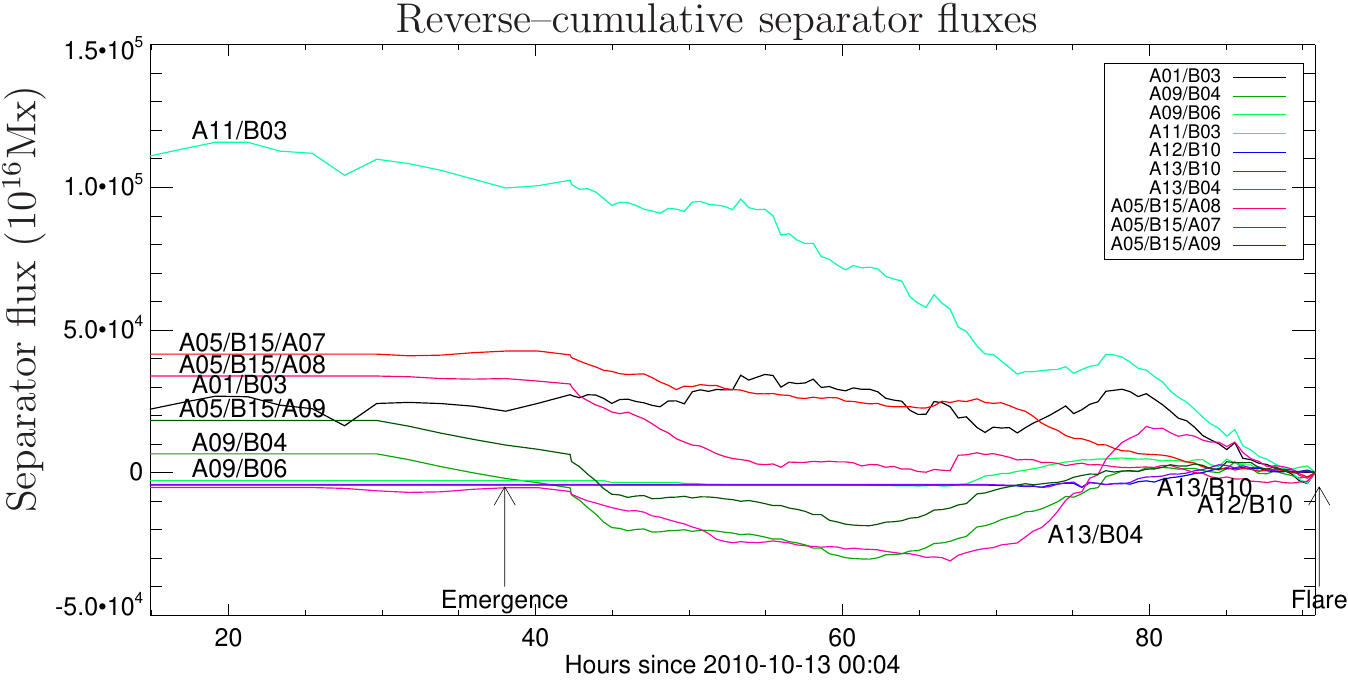}
      \caption[Separator fluxes in AR11112]{\label{fig:sepflux} A reverse--cumulative sum of the total domain fluxes linked by each separator relative to the potential field configuration just prior to the flare, identified by nulls to which they attach.  Separators with three listed nulls are those which pass through the coronal null, B15.}
    \end{center}
  \end{figure}
  
   As mentioned previously, a single separator will generally link multiple domains.  For instance, one of the separators passing through the coronal null B15 in \figref{fig:topo}, connecting nulls A05--B15--A07, links three domains, N58--P22, N67--P22, and N38--P29, with Gauss Linking Numbers -1, -1, and +1, respectively.  The residual fluxes for two of these domains, N58--P22 and N67--P22, are shown in green and purple in \figref{fig:p22rmat}; the N38--P29 domain residual fluxes is similar, peaking at $\sim2.5\times 10^{20}\unit{Mx}$ around 90 hours.  This separator $\sigma$'s flux, including all three domains $\{D\}$ with linking numbers $L_D$, is then given at each time $i$ by
   \begin{equation}
     \psi_\sigma^{(cr)i} = \sum_{D}\sum_{j=i}^N\Delta^j\Matrix{R}_D\times L_{D}
   \end{equation}
   The sum for this separator is plotted in red in \figref{fig:sepflux}.  The curve shows that, as we look further back in time, the flux linked by this domain increasingly departs from the final potential field configuration, until around $t=43$ hours when all changes level off.  This is when region P22 first emerges.  Note that every separator shows relatively little change in the linked flux before this time.

\begin{figure}[ht]
    \begin{center}
      \includegraphics[width=\textwidth]{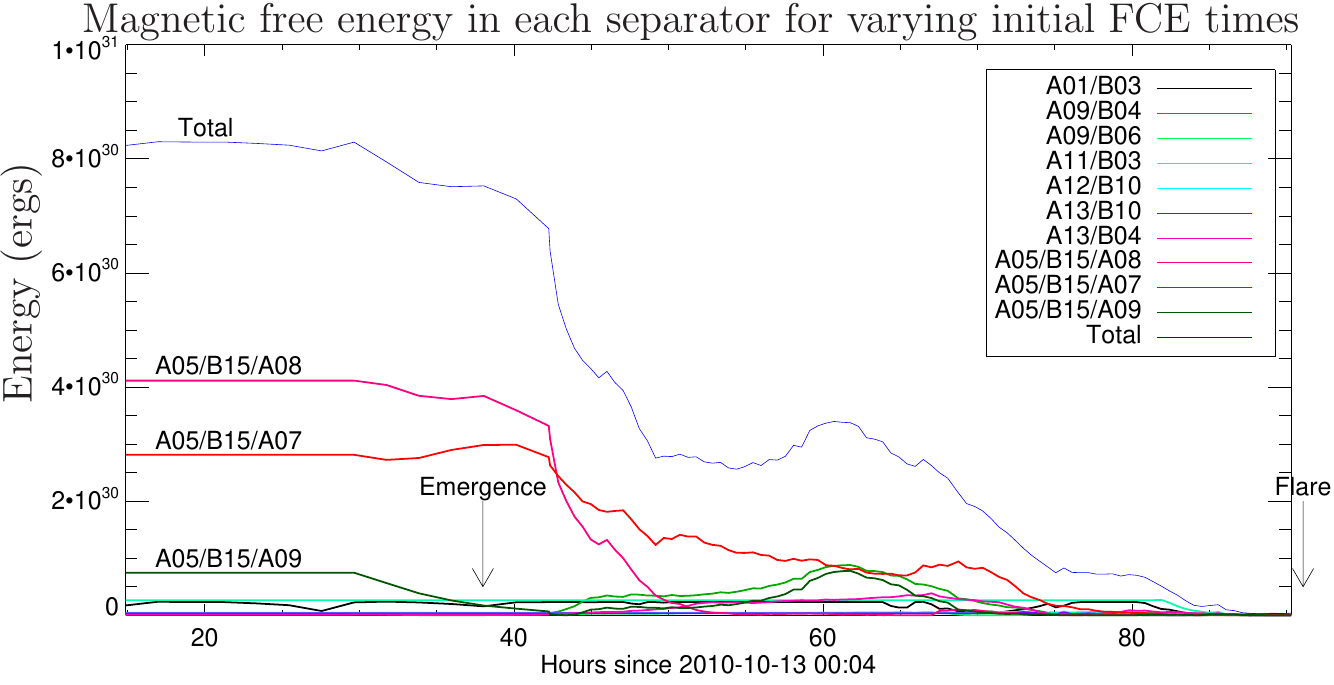}
      \caption[Separator energies for AR11112]{\label{fig:energy} Free magnetic energy relative to a potential field at the final time, and applying the FCE assumption at successively earlier timesteps for each separator, and for the total (blue).  Our energy estimate is then the value of the initial timestep, $\approx 8.25\times 10^{30}$ergs.  Arrows indicates the onset of emergence and the time of the M2.9 flare.  Note that there is some ``pre--emergence'' due to our boxcar smoothing function.}
    \end{center}
  \end{figure}

   Based on \figref{fig:sepflux}, we might expect the A11/B03 separator to store the most energy, as it has the most linked flux.  However, the amount of current flowing in this separator is 1--2 orders of magnitude less than in the three separators connected to the coronal null: the longer a separator, the less current necessary to generate the required self--flux.  \figref{fig:energy} plots the magnetic free energy when applying the FCE assumption at varying timesteps.  We see that the energy of the three separators connected to the coronal null, shown in magenta, red, and dark green, completely dominate the magnetic free energy buildup.  The blue line shows the total for all separators.  We also see that, constraining the domain fluxes at any time before the onset of flux emergence, $t\approx 40$ hours, effects little change in the total energy buildup.  Essentially all of the flux difference relative to the final potential field configuration is due to the emergence itself.  

   The MCC yields a lower bound on the magnetic free energy of a system.  After accounting for all flux linked by all separators, and hence the energy stored in each separator, we conclude that NOAA AR 11112 stores a minimum of $8.25\times 10^{30}$ ergs of free magnetic energy over the 2--3 days leading up to the October 16\tothe{th}, 2010 GOES M2.9 flare.

  \section{\label{sec:conclusion2}Conclusion}
  In this work we have extended the application of the Minimum Current Corona model to the estimation of energy stored along separators in the coronal field in active regions where flux submergence and emergence plays a significant role in the photospheric field's evolution.  This estimation operates in two distinct parts.  In the first, we track the evolution of the photospheric line--of--sight magnetic field using SDO/HMI magnetograms.  We found that the higher cadence and greater resolution of HMI, compared to previous analysis using full disk MDI data at a 96--minute cadence, greatly improved this portion of the analysis.  We found that a half hour cadence was sufficient to capture the evolution of this active region.  Analysis of this region began before the HMI 720 second averaged data were available.  Future investigations will use the high quality averaged dataset.

  The tracking information is stored in a mask array, which assigns unique labels to contiguous groups of data pixels, and matches these labels across all timesteps.  This framework accounts for all processes we see in the data: submergence and emergence, merging, splitting, braiding, and spinning.  In the present work we focus on submergence and emergence of magnetic flux.  All other processes are implicitly addressed in our work except for spinning.  \citet{Kazachenko:2010} have accounted for spinning explicitly in their work and found that its importance is case dependent.  We do not believe spinning plays a significant role in NOAA AR 11112.

  In the second part of our analysis we use the tracking information in the mask array to characterize the system wholly in terms of its topology.  Each region's total flux, together with its flux--weighted centroid, defines a magnetic pole.  A potential field extrapolation using the poles as sources, together with a mirror corona where $\vect{B}(x,y,z) = -\vect{B}(x,y,-z)$, yields all null points of the system.  The potential field extrapolation determines the locations of separators, which are where current sheets form due to free magnetic energy in the coronal field in the MCC model.  The amount of current flowing in each separator, and the amount of free energy in the field, is due directly to the nonpotentiality of each flux domain.

  A lower bound on the free magnetic energy is calculated from the separators and domain fluxes using the method of \citet{Longcope:2004}, who found that the MCC estimate may be less than other MHD free energy estimates by a factor of ten or more.  For NOAA AR 11112 we found a lower bound of $8.25 \times 10^{30}$ ergs.  This method requires that all domains linked by each separator are known.  This has been done by inspection in the past \citep{Kazachenko:2010}, but in the present work we have proposed an efficient automatic method using the Gauss Linking Number.  Provided the specification of some closure for both field lines and separators, and because domains are simply connected spaces, calculating the linking number for each separator and a single field line from each domain will find all linked domains for each separator, and the sense in which they are linked $(\pm 1)$.  It must be noted, however, that domains are not equivalent to connections: a single pair of positive and negatives poles may be multiply connected, with different domains in a single connection distinguished by one or more separators.  We found no such instances of this in the case of NOAA AR 11112, but this scenario may be simply dealt with by tracing a large number of random field lines between each pair of multiply connected poles.  One may then calculate the Gauss Linking Number of each separator with the suite of field lines, and determine the proportion of field lines linked by each separator.  The linked flux may then be divided among the separators, accordingly.

  Our free energy estimation does have some limitations.  Most important, we can only calculate energy storage, not energy release, either by a flare or some more quiescent process.  This is because we fix the domain fluxes at some point in the past and compare them to the potential field domain fluxes at some given time; however the flare is a coronal phenomenon, while the potential field configuration is derived purely from photospheric data.  Using coronal data to quantify how domain fluxes change during a flare would allow for an energy estimation before and after a flare, though we are aware of no method for doing so.

  Second, our energy calculation only takes into account the self--inductance of each current ribbon.  For a set of currents loops, the total energy is given by both self and mutual inductance terms.  When the loops are physically separated, the self inductance terms should dominate.  However, if some subset of loops share a common leg (as is the case when multiple separators are connected to a coronal null point), then mutual inductance will significantly contribute to the total energy and must be accounted for.  This work remains to be done.

  \subacknowledgements
  Graham Barnes graciously provided code for producing the potential field connectivity matrices using a Monte Carlo algorithm with Bayesian estimates, as described in \cite{Barnes:2005}.  Development of the code was supported by the Air Force Office of Scientific Research under contract FA9550-06-C-0019.  We also thank our Summer 2010 REU student Johanna Bridge for her work in assessing the performance of the automatic tracking algorithms.  This work was supported by NASA LWS.

\chapter{Calculating Separate Magnetic Free Energy Estimates for Active Regions Producing Multiple Flares: NOAA AR11158}\label{ch:ar11158}
\begin{manuscriptauths}
  Manuscript in Chapter 3
  \newline
  \newline
  Author: Lucas A. Tarr
  \newline
  \newline
  Contributions: Conceived and implemented study design.  Wrote first draft of manuscript.
  \newline
  \newline
  Co--Author: Dana W. Longcope
  \newline
  \newline
  Contributions: Helped with study design.  Provided feedback of analysis and comments on drafts of the manuscript.
  \newline
  \newline
  Co--Author: Margaret Millhouse
  \newline
  \newline
  Contributions: Prepared the dataset of magnetograms.
\end{manuscriptauths}
\pagebreak
  
\begin{manuscriptinfo}
  \noindent Lucas A. Tarr, Dana W. Longcope, and Margaret Millhouse\\
  The Astrophysical Journal\\
  Status of Manuscript:\\
  \uline{\phantom{5eM}}Prepared for submission to a peer--reviewed journal\\
  \uline{\phantom{5eM}}Officially submitted to a peer--reviewed journal\\
  \uline{\phantom{5eM}}Accepted by a peer--reviewed journal\\
  \uline{\phantom{5eM}}\makebox[0pt]{\hspace{-2em}x}Published in a peer--reviewed journal\\
  \newline
  \newline
  Published June, 2013, ApJ 770, 4
\end{manuscriptinfo}

\begin{abstract}
  It is well known that photospheric flux emergence is an important process for stressing coronal fields and storing magnetic free energy, which may then be released during a flare.  The \emph{Helioseismic and Magnetic Imager} (HMI) onboard the \emph{Solar Dynamics Observatory} (SDO) captured the entire emergence of NOAA AR 11158.  This region emerged as two distinct bipoles, possibly connected underneath the photosphere, yet characterized by different photospheric field evolutions and fluxes.  The combined active region complex produced 15 GOES C--class, 2 M--class, and the X2.2 Valentine's Day Flare during the four days after initial emergence on February 12th, 2011.  The M and X class flares are of particular interest because they are nonhomologous, involving different subregions of the active region.  We use a Magnetic Charge Topology together with the Minimum Current Corona model of the coronal field to model field evolution of the complex.  Combining this with observations of flare ribbons in the 1600\AA\ channel of the \emph{Atmospheric Imaging Assembly} (AIA) onboard SDO, we propose a minimization algorithm for estimating the amount of reconnected flux and resulting drop in magnetic free energy during a flare.  For the M6.6, M2.2, and X2.2 flares, we find a flux exchange of $4.2\times 10^{20}\unit{Mx},\ 2.0 \times 10^{20}\unit{Mx} , \hbox{and } 21.0 \times 10^{20}\unit{Mx}$, respectively, resulting in free energy drops of $3.89\times 10^{30}\unit{ergs}, 2.62\times 10^{30}\unit{ergs}, \hbox{and } 1.68\times 10^{32}\unit{ergs}$.

\end{abstract}

\section{\label{sec:intro3}Introduction} 

Solar flares are the most extravagant examples of rapid energy release in the solar system, with the largest releasing around $10^{32}\unit{ergs}$ on a timescale of hours \citep{Benz:2008}.  This energy, imparted to the plasma confined along coronal magnetic loops of active regions, is distributed between kinetic, thermal, and radiative process in some way that may vary from flare to flare.  While the ultimate source of this energy is likely stresses introduced by convective motion of the plasma at and below the photosphere, we believe the direct source is the conversion of free magnetic energy: magnetic energy in excess of the active region's potential magnetic field energy.

As has been clear for many decades, active regions consist of bundles of flux tubes, concentrated prior to their emergence through the photosphere \citep{Zwaan:1978}.  The free energy builds up as the flux tubes forming an active region are stressed at the photospheric boundary, where plasma forces dominate field evolution (plasma $\beta\equiv 8\pi p/B^2 > 1$, with $p$ the gas pressure).  Moving outward from the solar surface into the corona, the plasma pressure rapidly diminishes and magnetic forces dominate, until a third regime is reached where plasma forces once again dominate.  As noted by \citet{Gary:2001}, even within an active region, the high $\beta$ portion of the upper corona may occur as low as $200\unit{Mm}$ above the solar surface.  We are primarily concerned with lower lying loops and magnetic domains and so will assume a low $\beta$ regime.  Barring any reconfiguration of the coronal field, the active region's magnetic domains are pushed into a highly nonpotential state by the photospheric motions of their footpoints.  Relaxation towards a potential field configuration through magnetic reconnection then allows for the conversion of magnetic free energy into kinetic and thermal energy through, e.g., field line shortening, shock formation, electron acceleration, or (possibly) ion acceleration \citep{Longcope:2009a, Guidoni:2010, Fletcher:2008, Hudson:2012}.

The number of quantitative estimates of this energy buildup using observations has recently increased, but results remain varied.  Nonlinear force--free models \citep[NLFF:][]{Sun:2012, Gilchrist:2012} have received much attention during the past decade, strongly driven by both increases in computing power and the arrival of vector magnetograms from space--based telescopes onboard Hinode (SOT/SP) and SDO (HMI).  While these models are a promising avenue of research, they come with their own set of problems, as discussed in \citet{DeRosa:2009}.  The lower boundary conditions are, in general, incompatible with the force--free assumption \citep{Metcalf:1995}.  Several methods exist to overcome this difficulty \citep{Wheatland:2009}, leading to different energy estimates for a single vector magnetogram, even when using a single extrapolation code \citep{DeRosa:2009}.  

A further problem is that the models amount to a series of independent fields at consecutive timesteps.  At each time, a new NLFF field is generated from the boundary data, uninformed by the solution from the previous timestep.  Contrasting with this are flux transport and magneto--frictional models, which do include a memory \citep{Yang:1986, Mackay:2011}.  These methods primarily focus on the global coronal response to active region emergence, destabilization, and eruption as opposed to the detailed analysis of processes within an active region, which is the topic of this investigation \citep{Yeates:2008}.  One reason for this is that the coronal portions of these models evolve the large--scale mean field using an induction equation with an effective magnetic diffusivity \citep{VanBallegooijen:2000}, so that the formation of fine--scale current sheets is beneath their resolution.  Most dynamical simulations without magnetic diffusion show a tendency toward fine layers \citep{VanBallegooijen:1985}.

We describe the coronal field using the Magnetic Charge Topology (MCT) model \citep{Baum:1980,Longcope:2005} at each time.  The system is described by a set of unipolar regions.  The distribution of magnetic flux between each pair of oppositely signed regions defines the system's connectivity.  As the active region evolves its connectivity will generally change.  To relate each time with the next, we employ the Minimum Current Corona model \citep[MCC:][]{Longcope:1996,Longcope:2001}.  By itself, MCT describes only potential fields, which contain no current.  The MCC introduces currents, and the resulting energetics, into the MCT model by asserting that the coronal field move through a series of Flux Constrained Equilibria (FCE).  In that case, the connectivity of the real field will be different from the potential field's connectivity.

One shortcoming of the MCC method as currently used is its inability to account for violation of these flux constraints, which are the topological manifestations of reconnection and the resulting energy release.  Previous studies \citep{Tarr:2012, Kazachenko:2012, Kazachenko:2010, Kazachenko:2009} have therefore only reported the total free energy difference between the MCC and a potential field configuration.  Our goal here is to relax those flux constraints at any timestep, while also allowing the system to continue evolving thereafter.  In this way, we may model multiple reconnection events for a single active region.

We present here a method for identifying the magnetic domains activated in successive flares based on observations of flare ribbons in the AIA 1600\AA\ channel.  This allows us to separately calculate the free energy available to each successive, nonhomologous flare.  If we further assume that all magnetic flux topologically capable of transferring during a reconnection event does transfer, then we may also estimate the actual energy release during a flare.

In the following sections, we will describe the data used (\S \ref{sec:data_ch3}), our methods for modeling the photospheric and coronal fields (\S\ref{sec:model_ch3}), how one may estimate the MCC free energy based on those models (\S\ref{sec:energy}), and the use of observations of flare ribbons to determine those domains activated in successive flares (\S\ref{sec:ribbons}).  We will conclude with a discussion of the results of our analysis (\S\ref{sec:disc}).

\section{\label{sec:data_ch3}Data}
To construct the MCC model of magnetic field evolution we use a series of 250 line--of--sight (LOS) magnetograms taken by the Helioseismic and Magnetic Imager \citep[HMI:][]{Schou:2012,Scherrer:2012,Wachter:2012} onboard the Solar Dynamics Observatory (SDO).  The data are at a 24 minute cadence between Feb.~11 2011 08:10:12 UT and Feb.~15 11:46:12 UT, and are taken from the JSOC \textsf{hmi.M\_720s} (level $1.5$) data series.  The region considered, NOAA AR11158, produced the first GOES X--Class flare of solar cycle 24, and has therefore already been analyzed in a variety of ways by numerous authors \citep[see][and references therein]{Petrie:2012}.

In addition we have used images of flare ribbons observed with Atmospheric Imaging Assembly (SDO/AIA) in the 1600\AA\ channel \citep{Lemen:2011}.  We obtained three sets of 1600\AA\ images via the SSW cutout service maintained by Lockheed Martin\footnote{\url{http://www.lmsal.com/get_aia_data/}} for $\approx 30\unit{min}$ during each flare with peak magnitude greater the M1.0: an M6.6 flare peaking at Feb 13, 17:28; M2.2 peaking at Feb 14, 17:20; and X2.2 flare peaking at Feb 15, 01:44.  All AIA data were prepared to level $1.5$ using the standard \fnc{aia\_prep} routine in SolarSoftware IDL \citep{Freeland:1998}.  

We coalign each set of AIA 1600\AA\ images to the magnetogram closest to the peak time of each flare.  To do so, we use the solar rotation rate to shift the first AIA image in a sequence to the time of the magnetogram.  It so happens that the 75G contour of HMI LOS magnetograms (after assuming a radial field and correcting for pixel foreshortening, discussed below) outlines the bright network patches in the 1600\AA\ band.  We shift the AIA image, by eye, until the contour and bright network patches align.  This could be automated by a cross--correlation between the magnetogram contour and a corresponding contour in AIA, but we have not yet implemented this procedure.  The AIA timesets are internally aligned, so we apply the same by--eye offset to each image in the sequence, after shifting each by solar rotation to the time of the chosen magnetogram.

The HMI magnetograms contain known (but as yet unmodeled) diurnal variations, due to the velocity of the spacecraft's orbit \citep{Liu:2012}.  The amplitude of the variations is around $2.5\%$ of the unsigned flux within an active region.  We are concerned with flux emergence trends over the course of days, over which time the effects of these variations should largely cancel.  We simply accept this as an additional source of error in our model.

\section{\label{sec:model_ch3}Modeling the Magnetic Evolution}

We apply the methods described in \S~\ref{ch:ar11112} and references therein to generate the magnetic modeling for this series of events.  The analysis splits into two sections, detailed below.  First, we characterize the photospheric field by partitioning the observed magnetograms into a set of unipolar regions.  Pairs of oppositely--signed regions $\{j,k\}$ may be linked through emergence when each region's flux increases between two timesteps.\footnote{Regions may also submerge or diffuse, as P1 does after $t\approx 25\unit{hrs}$.  Algorithmically, there is no distinction between these processes.  If any of a pole's flux change between two timesteps cannot be paired with another region, it is formally paired with a source of opposite sign located at infinity.  We discuss this in more detail at the end of \S\ref{sec:mod-phot}.}  At each time $i$, the amount of flux change of each pairing is recorded in a photospheric--field--change matrix $\Delta^i\Matrix{S}_{j,k}$.  This set of matrices is therefore a time history of the flux with which each region emerged with every other region.

In the second part of our analysis, we develop a topological model of coronal domains immediately prior to each major flare.  Our flux emergence matrix discussed above is the real flux in each coronal domain, which we may compare to the flux in each domain in a potential field extrapolation.  The difference between the two is the nonpotentiality of each domain.  The equilibrium with minimum magnetic energy that still includes this difference in domain fluxes, called the flux constrained equilibrium (FCE), contains current sheets on each of its separators \citep{Longcope:2001}.  Our topological model determines the location of all current sheets within the active region complex, the strength of each related to the nonpotentiality of its associated domains.  Finally, this provides us with an estimate of the energy in the FCE, which is itself a lower bound on the magnetic free energy stored in the actual magnetic field.

\subsection{\label{sec:mod-phot}Modeling the Photospheric Field}
We characterize the photospheric field by determining the magnetic flux through each pixel.  First, we convert each line of sight (LOS) pixel to vertical by assuming a radial field: we divide the flux by $cos(\theta)$, where $\theta$ is the polar angle from disk center; second, we account for reduced flux due to pixel foreshortening dividing the flux in each pixel by $\cos(\theta)$ again.  We smoothe our data by extrapolating the vertical field to a height of $3\unit{Mm}$ as detailed in \citet{Longcope:2009}, and reduce noise by ignoring any pixels below a $75\unit{G}$ threshold.  Pixels above this threshold are partitioned using the downhill tessellation algorithm of \citet{Barnes:2005}, creating a \emph{mask}.  Each pixel is assigned an integer, and contiguous groups of like--signed pixels of the same integer compose a \emph{region}.  Pixels below our threshold belong to no region and are assigned a mask value of 0.

This tessellation scheme can generate thousands of small regions at each timestep, so adjacent regions of like polarity are merged when the saddle point value of the magnetic field between them is less than $700\unit{G}$.  Finally, we exclude any region whose total flux is less than $2\times10^{20}\unit{Mx}$.  This process is carried out at each timestep, independently.  We call the resulting set of masks a \emph{mask array}.  Regions in one timestep are associated with those in the surrounding timestep first by a bidirectional association between timesteps, and then applying the \fnc{rmv\_flick} and \fnc{rmv\_vanish} algorithms, described in detail in the previous chapter.  

The focus of these methods is to distinguish between regions of flux that emerge from below the photosphere at different times and in different places, and to keep track of these individual regions as they undergo shear motion on the solar surface after emergence.  To this end, after the automatic algorithms listed above have run to convergence, we manually shift the boundaries between regions to ensure we consistently track flux emergence and migration over the entire timeseries.  We allow individual regions to emerge, submerge, change shape, translate, split, and merge.  To further reduce the number of regions, we exclude regions which have non--zero flux for fewer than 5 timesteps ($\approx 100\unit{min}$).  

\begin{figure}[ht]
  \begin{center}
    \includegraphics[width=0.8\textwidth]{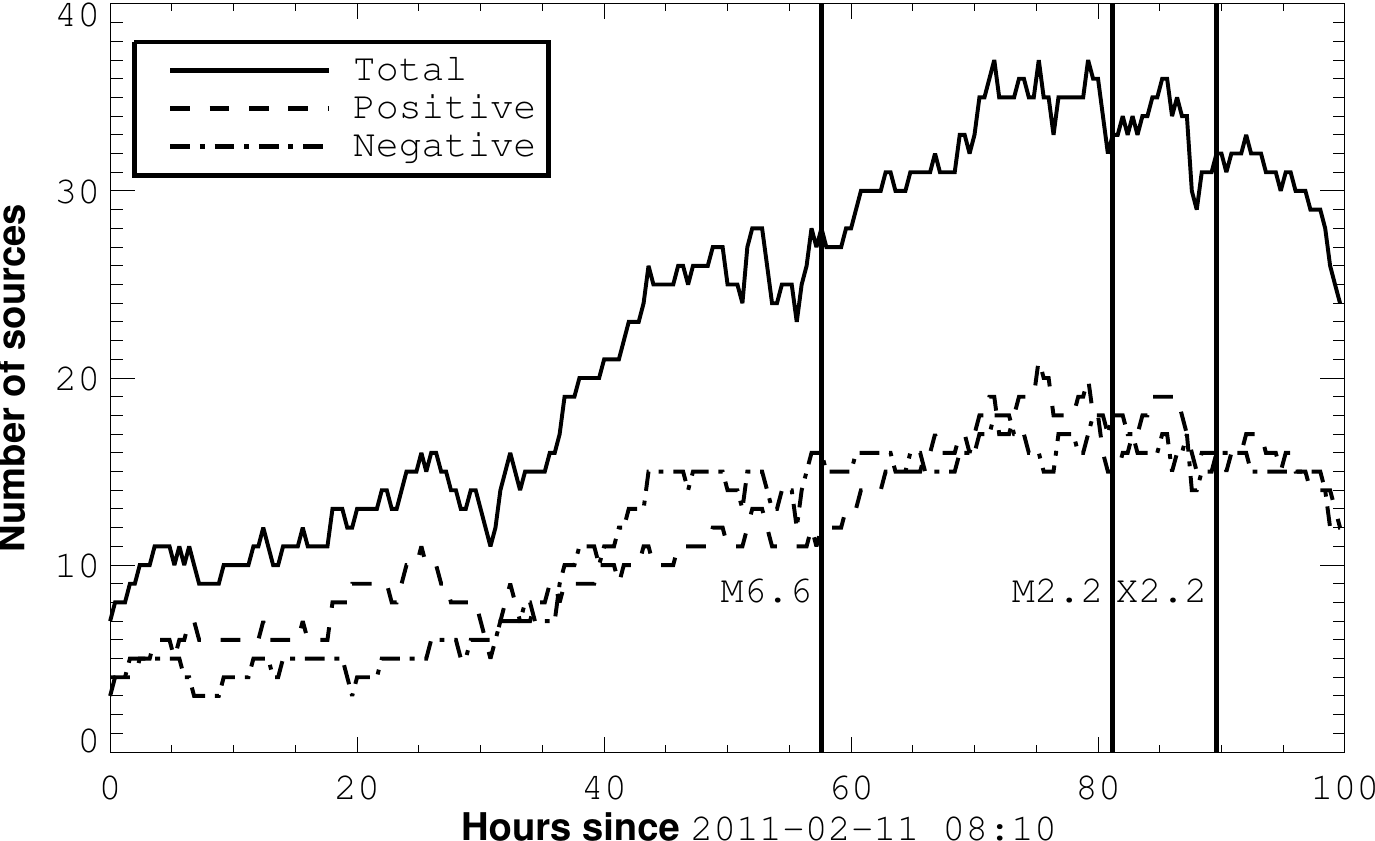}
    \caption[AR11158 MCT Sources]{\label{fig:sources} Number of distinct positive, negative, and total mask regions at each time.}
  \end{center}
\end{figure}
\figref{fig:sources} shows the number of mask regions of each polarity over the entire timeseries.  In our final analysis, we track 118 regions for $\approx 100$ hrs.  The number of distinct regions at any time varies between 10 and 35, generally growing at a steady rate as the active region complex emerges between 2011-02-11T08:10:12 and 2011-02-14T16:10:12, then dropping slightly as the fully emerged system continues to concentrate\footnote{\figref{fig:sources} shows that for AR11158, neither polarity is consistently or significantly more fragmented than the other.  This contrasts with observations of most other active regions, where the leading polarity is substantially more concentrated than the trailing.  This can show up in MCT models as the leading polarity's flux distributed amongst fewer poles than the trailing polarity.  This does not seem to be the case for AR11158, which we find curious, although it does not affect our analysis.}.

\begin{figure*}[ht]
  \begin{center}
  \includegraphics[width=0.8\textwidth]{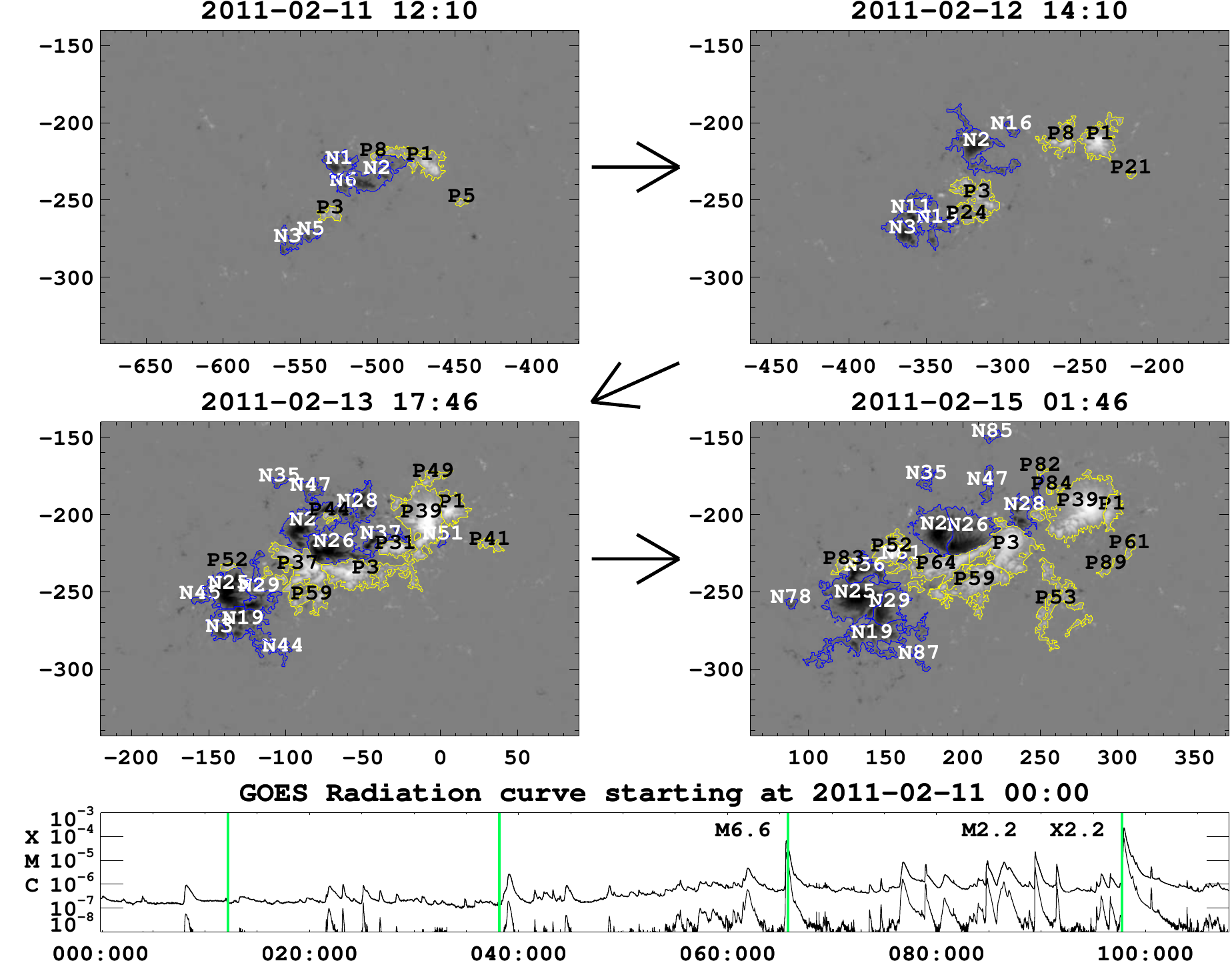}
  \caption[AR11158 example magnetograms]{\label{fig:partition}  \emph{Top:} HMI LOS magnetograms of NOAA AR 11158 overlaid with their respective masks.  The grayscale saturates at $\pm max(\vert B\vert )=\pm 1500.0$G, and the axes are in arcseconds from disk center.  Arrows indicate the time sequence.  \emph{Bottom:} Full--disk integrated GOES electron radiation curve.  The four vertical lines correspond to the times of the four displayed magnetograms.}
  \end{center}
\end{figure*}

Four examples of the LOS magnetogram overlaid with masks are shown in \figref{fig:partition}.  The top four images are each a single frame from an animation of the full 250 timesteps viewable in supporting media.  Arrows indicate the chronological sequence.  Below the magnetograms is a full--disk integrated GOES X--Ray curve $(1.0-8.0\text{\AA} )$ over the course of the time series, with vertical lines showing the times of the four magnetograms.

\begin{figure}[ht]
  \begin{center}
    \includegraphics[width=0.8\textwidth]{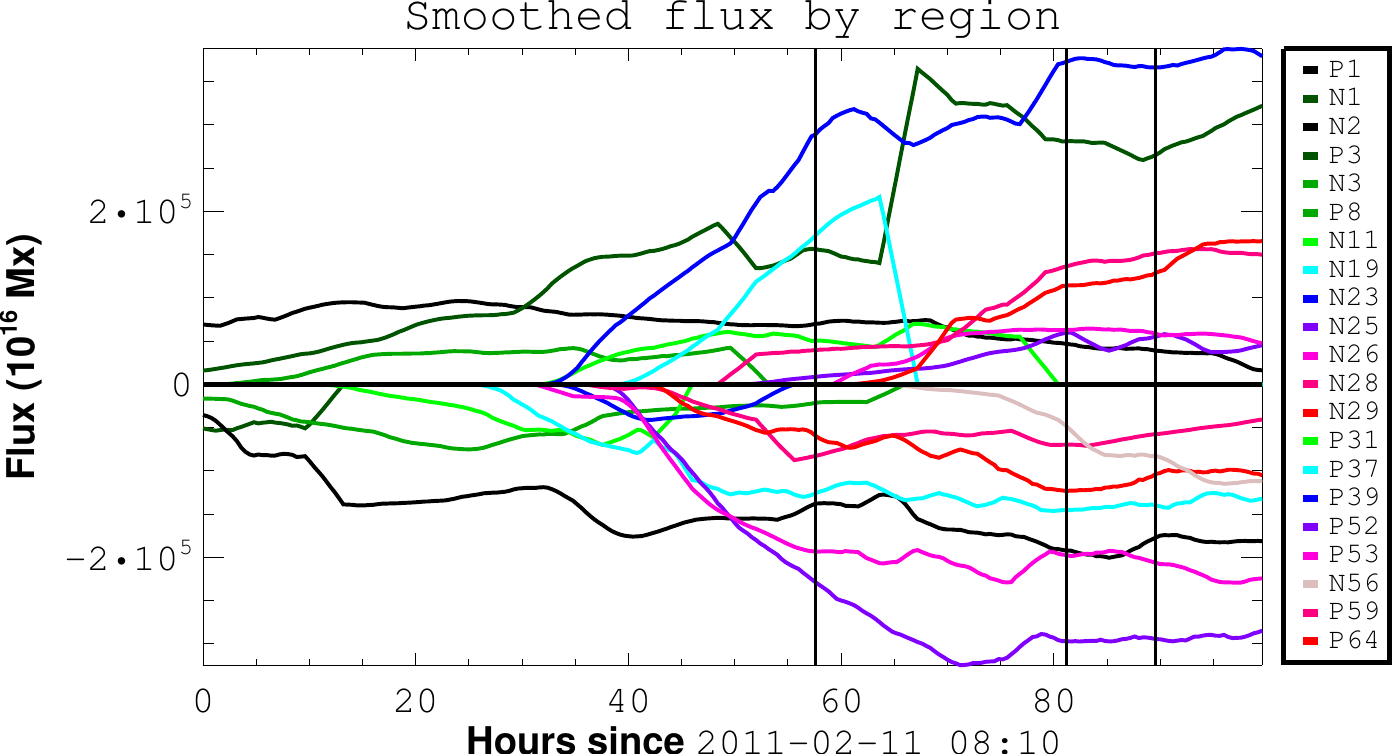}
    \caption[Smoothed region fluxes in AR11158]{\label{fig:sflux} Smoothed flux by convolution with a 3.6\unit{hr} boxcar function for each region having at least $4\times 10^{20}\unit{Mx}$ at some time.}
  \end{center}
\end{figure}

In preparation for the transition to an MCT--MCC analysis of the system, we represent each distinct region as a magnetic point source, or \emph{pole}, in the local tangent plane at each timestep.  For consistency from timestep to timestep, our point of tangency at each time is taken as the center of charge at the initial timestep, migrated through solar rotation to the present timestep.  Pole $j$ at time $i$ is defined by its associated region's total flux $\psi^i_j$ and flux--weighted centroid $\bar{\vect{x}}_j^i$:
\begin{gather}
  \label{eq:pole}\psi^i_j = \int_{\mathcal{R}^i_j} B_z(x,y)\, dx\, dy \\
  \label{eq:centroid}\bar{\vect{x}}_j^i = (\psi_j^i)^{-1}\int_{\mathcal{R}_j^i} \vect{x} B_z(x,y)\, dx\, dy \quad.
\end{gather}
The vertical magnetic field $B_z(x,y)$ accounts for both line--of--sight effects, by assuming a radial field, and pixel foreshortening, as described above.  To reduce the effects of noise in our masking algorithms, we additionally smoothe the flux in each region over the four day series by convolution with a 9 timestep (3.6 hour) boxcar function, using edge truncation.  For instance, pole $j$'s flux at time $i=2$ is averaged to $\bar{\psi}_j^2=\frac{1}{9}(3\times\psi_j^0+\sum_{i=1}^6\psi_j^i)$.  The resulting smoothed fluxes are shown in \figref{fig:sflux} for regions that have $\vert\psi\vert > 4\times 10^{20}\unit{Mx}$ at any timestep in the series.  Discontinuous jumps, for instance around 65 hrs, indicate merging, in this case P37 into P3.  This occurs when the distinction between separately emerged regions becomes ambiguous.

It is evident from viewing the full animation associated with magnetograms in \figref{fig:partition} that the active region complex emerges in several distinct episodes.  The complex has two distinct primary locations of flux emergence: one to the north, which leads another to the south by about $35\unit{Mm}$.  The first emergence episode is ongoing at the beginning of our timeseries, at which point several smaller emerged regions are still consolidating (eg., N3/N5 in the south; N1/N2/N6, and P1/P6 in the north).  The second episode begins around our 60th timestep (30 hrs from t=0), on Feb.~12th at 16:00UT, and continues very steadily until timestep 80 (t=60 hrs, Feb.~13th, 20:00).  At this time, the northern emergence ceases, while it appears that the southern emergence continues, possibly in repeated $(\approx 6 hr)$ bursts.  These bursts may be an artifact of the daily variations in HMI's reported flux, noted above.

\begin{figure}[ht]
  
  \begin{center}
    \includegraphics[width=0.8\textwidth]{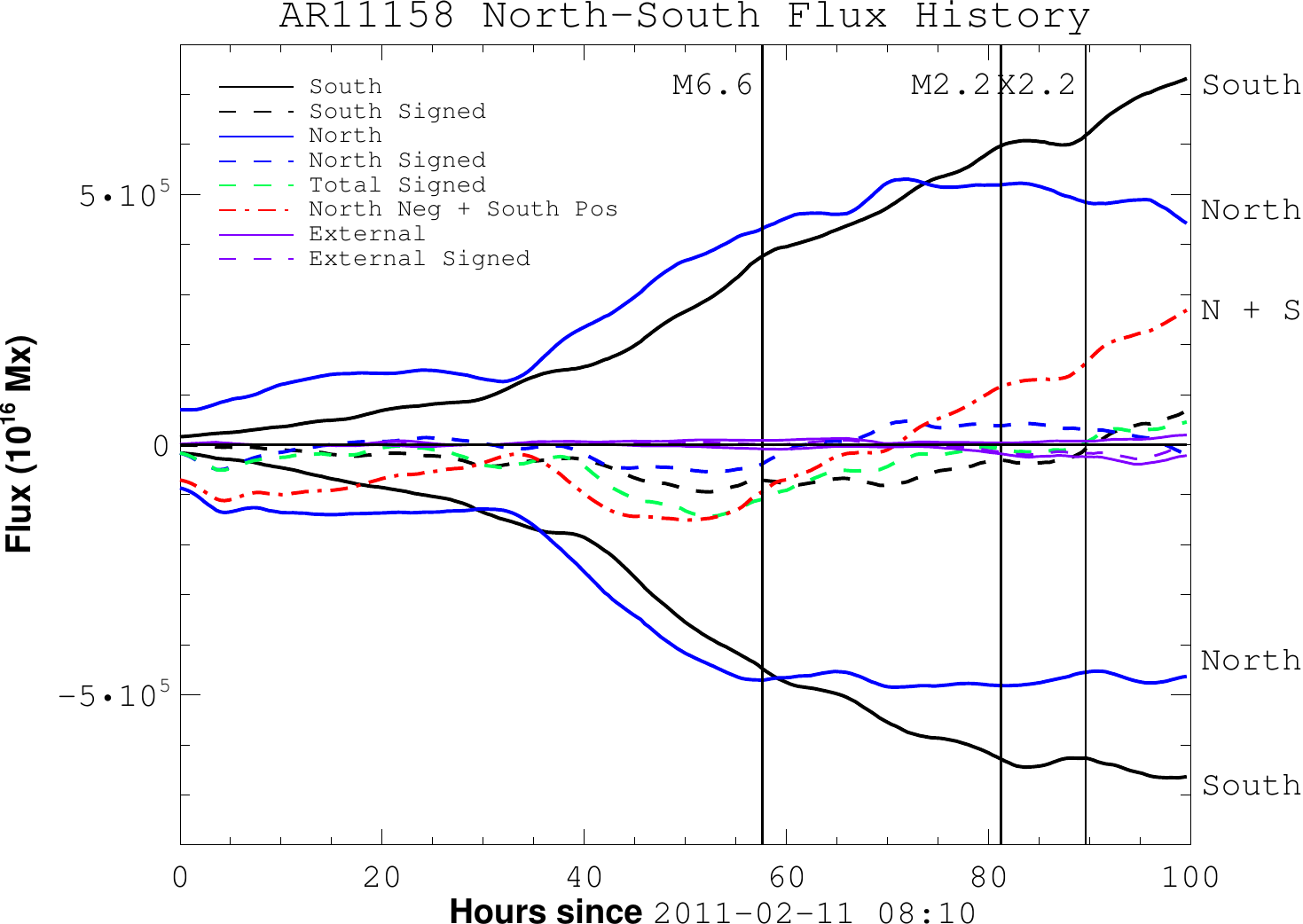}
    \caption[North--South Emergence Patterns in AR11158]{\label{fig:f-imbal} Signed flux in the northern (solid blue) and southern (solid black) emergence zones. Dashed lines show total signed flux in north (black), south (blue), and all (green) regions, and the combination of northern negative with southern positive regions is shown as dash-dot (red).  This readily shows the two distinct emergence patterns of the northern and southern regions.}
  \end{center}
\end{figure}

\figref{fig:f-imbal} shows the evolution of signed flux in those regions belonging to the northern emergence (blue), southern emergence (black), and a set of surrounding quiet sun (network) regions that drifted above our thresholds at various times (purple).  Of the 118 regions tracked over the timeseries, 54 belonged to the north (36 positive, 18 negative), 37 the south (13 positive, 24 negative), and 27 external to the active region complex (13 positive, 14 negative).  Dashed lines in the figure show the flux imbalance for each of the three sets, and also the flux imbalance of the total system (green).  Note that qualitatively, the northern regions undergo a very different emergence evolution compared to the southern regions.  This is easily seen with the dashed red line, which shows the total signed flux of the complex's central region: northern negative flux summed with southern positive flux.

The system's connectivity is defined by the amount of flux connecting each pole to every other pole \citep{Longcope:2009}.  This constitutes a \emph{graph}, where each pole is a vertex and each domain an edge, with the weight of each edge given by the domain flux.  The total flux of a single pole is the summed weight of all edges connected to it, and the total flux of the system of poles is the summed weight of all edges in the graph.  If there is an overall imbalance of flux, the remainder must be connected to a source located (formally) at infinity.  In general, a pair of vertices may be connected by more than one edge, and are then called multiply connected.  \citet{Parnell:2007} has described such situations in detail.  We have found several such instances of multiply connected vertices in AR11158, though we show below that, in this case, their effect on the system's energetics is negligible.

In determining the distribution of flux emergence within the active region, we use the method described in the previous chapter to define a connectivity graph for the flux difference between consecutive timesteps.  This change must come in pairs, as positive and negative poles emerge and submerge together.  The total flux change between times $i$ and $i+1$ for a single pole $j$ is given by \eqref{eq:update}, reproduced here
\begin{align}
  \label{eq:update-pol}\psi^{i+1}_j - \psi^i_j = \sum_{b}\Delta^i\Matrix{M}_{j,b} + \sum_k\Delta^i\Matrix{S}_{j,k} \quad ,
\end{align}
where $\Delta^i\Matrix{M}_{j,b}$ describes any shift in the boundary between like--signed region $b$ adjacent to $j$ and $\Delta^i\Matrix{S}_{j,k}$ describes any change in the photospheric field itself.  The former is a graph with edges connecting like signed regions of opposite flux change (flux that one region loses, another gains), while the latter is a graph connecting opposite signed regions with same sensed flux change.  The algorithms for determining these are fully described in \S~\ref{ch:ar11112}.  To accurately deal with the two regions of emergence, we first calculate the matrix $\Delta^i\Matrix{S}$ for northern and southern regions separately, then combine the two resulting connectivity graphs.  Finally, we allow for connections between north and south using any remaining flux change.

We may quantify our success at capturing the flux--change processes by reconstructing the total flux of a region using its initial flux and elements of the surface change matrix.  At time $i$, we estimate a region $j$'s flux as $\psi_{j}^i = \psi_j^0 + \sum_{l=0}^{i-1} \sum_k\Delta^l\Matrix{S}_{j,k}$.  Summing these reconstructed fluxes over a set of like--signed regions connected by internal boundaries, say all the positive flux in the southern emergence zone, should represent the total emergence of the collected regions.  We find that our method always underestimates this emergence.  Over the entire time series, we find a maximum discrepancy of between $8\%$ and $25\%$, depending on the group we reconstruct (northern positive, northern negative, southern positive, southern negative).  We believe this conservative attribution of flux change to emergence or submergence processes stems from a greedy boundary--change algorithm, asymmetries in the concentration of newly emerged positive and negative flux, and the diurnal variations due to spacecraft motion.  These factors combine to force our algorithms to attribute of $8-25\%$ of flux change with sources formally at infinity.  

Finally, we note that there is quite a bit of variation in our underestimate of emerging flux.  Our algorithm has the greatest underestimate when reconstructing the northern positive flux emergence regions.  At one timestep it is only able to pair up $75\%$ of the actual flux change, but is able to pair at least $85\%$ of the change for more than 125 out the 250 timesteps.  The flux change formally paired with sources at infinity generally rises over the timeseries, and peaks at $13.4\%$ of the total instantaneous unsigned flux 12 hours before the M6.6 flare, then varies between $11-13.25\%$ for the rest of the series, ending 10 hours after the X2.2 flare.  This variation is consistent with the $2.7\%$ daily variation in unsigned flux found by \citet{Liu:2012}. The flux change assigned to sources at infinity at the times of the M6.6, M2.2, and X2.2 flares are $11.4\%$, $12.2\%$, and $13.3\%$ of the instantaneous unsigned flux, respectively.

\subsection{\label{sec:mod-cor}Modeling the Coronal Field}
Having quantified the connectivity graph for flux change, we may use the topological methods of \S~\ref{sec:erg} to calculate the free energy stored in coronal fields.  At every timestep we determine the potential field connectivity matrix, $\Matrix{P}^i$, using the Monte Carlo method of \citep{Barnes:2005}.  At the timesteps immediately preceding each M and X class flare, we calculate the system's potential field topology in terms of poles, nulls, and separators \citep{Longcope:2002}.  In our analysis, deviations from a potential field configuration take the form of differences in the amount of flux (either more or less) in the real field's domains versus the potential field's domains.  In the MCC model, the departure of a domain from a potential field configuration gives rise to currents in associated separators.  Every domain that is topologically linked by a separator contributes to that separator's current.  To determine which domains each separator links we use the Gauss Linking Number method of \S\ref{sec:gauss}.  Completing the free energy estimate for each separator, we use the direct connection between currents flowing along separators and free magnetic energy given by \citet{Longcope:2004}.

As shown in \S~\ref{sec:erg}, the self--flux of a separator (current ribbon) $\sigma$, denoted by $\psi_\sigma^{(cr)}$ and generated by currents flowing along it, is equal to the difference between the linked--domain fluxes in the real and potential fields:
\begin{align}
  \label{eq:fce_ch3}\psi_\sigma^{(cr)i} &= \psi_\sigma^i - \psi_\sigma^{(v)i} = \sum_D\Matrix{F}^i_D - \sum_D\Matrix{P}^i_D\quad ,\\
  \label{eq:totfcr_ch3} &\equiv - \sum_D\sum_{j=0}^{i-1}\Delta^j\Matrix{R}_D\quad .
\end{align}
$\psi_\sigma^i \hbox{ and } \psi_\sigma^{(v)i}$ are the separator fluxes in the real and potential fields, respectively.  These may be written as summations over linked domains $D$, elements of the connectivity matrices.  Here, $\Matrix{F}^i_D$ is the real domain flux of a domain $D$ at time $i$, given by the initial potential field flux and the summation over time of the surface flux change matrix, defined above: 

\begin{equation}
  \Matrix{F}^i_D = \Matrix{P}_D^0 + \sum_{j=0}^{i-1}\Delta^j\Matrix{S}_D \qquad .
\end{equation}
The difference between the real and potential field fluxes at each time gives $\Delta^j\Matrix{R}_D$, the total amount of flux which may be redistributed between domains in a reconnection event.  The sum of $\Delta^j\Matrix{R}_D$ over all times up to $i$, and over all linked domains $D$, results in Equation \eqref{eq:totfcr_ch3}.

We may extend this model to include reconnection by considering the effect of reconnection at some time $k$ on the connectivity matrices described above.  The total flux through the photosphere does not change during a flare, so the potential connectivities $\Matrix{P}^k$, which are uniquely determined by the photospheric boundary at any time, do not change.  The only effect is to transfer flux between domains in the real field.  We accomplish this by adding some flux transfer matrix $\Matrix{X}^k$ to $\Matrix{F}$ at time $k$, so that
\begin{align}
  \label{eq:xmat} \Matrix{F}^k_{\text{postflare}} = (\Matrix{F}^k_{\text{preflare}}+\Matrix{X}^k).
\end{align}
According to Equations \ref{eq:fce_ch3} and \ref{eq:totfcr_ch3}, there is an opposite assignment of flux in the redistribution matrix $\Matrix{R}^k_{\text{postflare}}= \Matrix{R}^k_{\text{preflare}}-\Matrix{X}^k$.  We may model the effect of multiple reconnections by adding/subtracting flux transfer matrices $\Matrix{X}^l,\ \Matrix{X}^m,\ \Matrix{X}^n,\ldots$ as necessary.  Therefore, the separator self--flux at any time $i$, including all past reconnection events, is given by 
\begin{align}
  \psi&_\sigma^{(cr)i} = - \sum_D\sum_{j=0}^{i-1}\Bigl(\Delta^j\Matrix{R}_D + H(j-k)\Matrix{X}_D^k \notag\\
  \label{eq:fcr}& + H(j-l)\Matrix{X}_D^l +H(j-m)\Matrix{X}_D^m \ldots \Bigr) \ ,
\end{align}
where $H(j)= \{0, j<0; 1, j>0\}$ is the Heaviside step function.  In the following section, we will propose a minimization scheme for estimating the reconnection matrix $\Matrix{X}$ at the time of a flare.

Having thus determined each separator's self--flux at any time, we follow \citet{Longcope:2004} to relate that self--flux \eqref{eq:fce_ch3} to the separator current by 
\begin{gather}
  \label{eq:crpsi_ch3}
  \psi^{(cr)i}_\sigma = \Frac{I L}{4 \pi}\ln\Biggl(\Frac{e I^*}{\vert I\vert}\Biggr)\quad .
  \intertext{The fuctional inversion}
  \label{eq:cr_ch3}
  I(\psi^{(cr)i}_\sigma) = I^*\Lambda^{-1}(4\pi \psi^{(cr)i}_\sigma/LI^*)
\end{gather}
with $\Lambda(x) = x\ln(e/\vert x\vert) $ allows us to represent the current in terms of the fluxes.  In this, $I$ is the separator current, $L$ its length, and $e$ the base of the natural logarithm.  $I^*$ is a characteristic current, related to the separators geometry and magnetic shear along its length; for a complete definition and derivation, please see \citet{Longcope:2004}.  Finally, from \citet{Longcope:2004} equation (4) we can calculate the energy in the MCC model in excess of the potential field magnetic energy,
\begin{equation}
  \label{eq:wmcc_ch3}
  \Delta W_{MCC} = \Frac{1}{4\pi}\int_{\Psi_{potl}}^{\Psi}I d\Psi = \Frac{L I^2}{32 \pi^2}\ln\Bigl(\Frac{\sqrt{e} I^*}{\vert I\vert}\Bigr)
\end{equation}
which, via equation \eqref{eq:cr_ch3}, is a function of the calculated separator fluxes $\psi^{(cr)i}_\sigma$.

We determine the coronal topology for the M6.6 flare at 17:22 on Feb.~13, 6 minutes before flare onset, and 16 minutes before GOES peak intensity.  At this time, our model consists of 27 sources (16 negative, 11 positive) and 26 nulls.  Following \citet{Longcope:2002}, these numbers satisfy both the 2D and 3D Euler characteristics, so we believe we have found all nulls.  Every null is prone, and there are no coronal nulls.  One null is asymptotic in the sense of \citet{Longcope:2009}, lying along the direction of the region's dipole moment computed about the center of unsigned flux, $\mu$, at a distance $r_0 = 2\mu/q_\infty$, where $q_\infty$ is the net charge.  This null's separatrix surface forms a boundary between the region's closed flux and surrounding open flux.  7 additional source--null pairs are part of unbroken fans: P57/B22, P52/B21, N51/A20, N45/A18, N43/A16, N42/A25, and N38/A26.  Using these values and the equation between (26) and (27) of \citet{Longcope:2002}, we expect to find 17 separators in the corona (along with 17 mirror separators), which we do find.  We therefore believe we have completely specified the system's topology on the eve of the flare.  

We perform a similar analysis just prior to the M2.2 and X2.2 flares.  While we do not find every topological element in these later flares, we believe we find all that play a significant role in each case.

For the 17 separators at the time of the M6.6 flare, we find 90 linked domains.  Two of the separators have the same endpoints, nulls A07/B01, and therefore enclose multiply connected source pairs \citep{Parnell:2007}.  These are known as redundant separators.  In this case, the two separators lay nearly along the same path, implying a slight wrinkle in the intersecting fan surfaces.  This creates one additional flux domain enclosed by the two separators.  Because the cross--sectional area in this case is small, the enclosed flux is small relative to the total flux enclosed by each separator, and the corresponding energy due to the redundant separator is negligible.  The Monte Carlo estimate of fluxes enclosed by the different separators used 500 field lines.  There was no difference in the number linked by the separators, so we conclude that the fluxes they link are identical to a fraction of a percent.  We use only one of the two in our calculation, and arrive at the same result independent of this choice.

\section{\label{sec:energy}Energy Estimates}
As stated above, one shortcoming of current MCT/MCC analyses are their inability to account for violation of the flux constraints.  As such, they have no way to account for reconnection.  We here present a method for relaxing those constraints at any timestep, while allowing the system to continue evolving after reconnection.

During reconnection, flux is exchanged across the field's separators.  Each separator lies at the boundary of four flux domains, and the separators involved in the flare identify the set of domains which exchange flux.  Some of these domains are flux superfluent, containing more flux than in a potential field, and some deficient.  Not every separator needs to be involved in every flare, and not all flux is necessarily transferred in every flare, even within the subset of involved domains.  

Reconnection does not simply involve the transfer of flux from surplus to deficit domains.  Two domains on opposite sides of a separator\footnote{A separator connects two null points of opposite sign, each of whose two spines connect to sources of the same polarity.  The four domains form all possible connections between the two positive and negative spine sources.  We designate two domains ``opposite'' if they share no spine sources.} (X--point in 2D) reconnect fieldlines, transferring flux to the remaining two domains.  There is no physical reason why opposing domains must both have more (or less) flux than a potential field configuration.  Instead, the state of the current domain depends on the history of its poles: where and with whom they emerged, whom they reconnected with in the past, and what their current geometric orientation is.  The only requirement for reconnection is that the two flux--donating domains contain some flux (nonzero elements of $\Matrix{F}$ in Equation \eqref{eq:xmat}).  In such cases, this poses the interesting question of whether the potential field configuration is always attainable through reconnection, or if there exists some local minima in configuration space.  We briefly consider this below, but leave a more detailed analysis of the question for another investigation.

We use a simple iterative minimization algorithm to model the redistribution of flux across a set of separators involved in a particular event.  As detailed below, at each iteration, the algorithm exchanges flux across each separator, and then picks the exchange that results in the greatest drop in the total system's free energy.  The iterations continue until any attempted exchange of flux across any separator increases the total system's free energy.  The free energy is calculated by the summation of Equation \eqref{eq:wmcc_ch3} over every separator.  

The amount of exchanged flux, $d\psi$, is a fraction of a percent of any domain's flux, so that thousands of iterations may be required for convergence.  This allows the algorithm to fully explore the route of steepest decent.  For instance, it might exchange flux across Separator 1 for 50 iterations, then find that Separator 2 provides greater drops in free energy for 2 iterations, after which exchange across Separator 1 is again the path of steepest descent, and so on.  This is because each domain may be directly involved in the reconnection for multiple separators, sometimes donating, sometimes receiving.  In fact, a domain's role as a donor or receiver for a particular separator may change over the course of the minimization, as flux exchange across other separators changes the path of steepest free energy decent.  The algorithm was designed to capture the result of this kind of subtle interplay.  We have performed the minimization using $d\psi$ of multiple magnitude, and found that the resulting flux configuration is stable for $d\psi < 10^{18}\unit{Mx}$, while the smallest domain flux is around $2.5\times 10^{19}\unit{Mx}$.  To be conservative, we set $d\psi$ at $10^{17}\unit{Mx}$, $0.4\%$ of the smallest region at the time, and $0.01\%$ of the largest.

When every possible flux exchange increases the total system's free energy, the distribution of flux amongst the coronal domains has reached a local minimum in terms of free energy.  This is not necessarily the potential field state, and indeed, for the 3 events we consider in this chapter, the system never reaches the potential field configuration.

For clarity, we detail the algorithm as pseudocode:
\begin{enumerate}
  \item Repeat the following:
    \begin{enumerate}
      \item Calculate the system's current free energy $W_i$
      \item For each separator $\sigma$:
        \begin{enumerate}
          \item Transfer flux $d\psi$ across separator $\sigma$
          \item Calculate the system's new free energy $W_\sigma$, and record that value.
        \end{enumerate}
      \item Find the transfer which resulted in the greatest drop in free energy: max($W_i-W_\sigma$).
      \item Add that flux exchange in the reconnection matrix $\Matrix{X}$ \\
            $[$For example, if the greatest free energy drop was due to a transfer across separator $\sigma_G$ with the flux--donating domains  $\{j,k\}$ and $\{l,m\}$, and the two receiving domains $\{n,o\}$ and $\{p,q\}$, then $(\Matrix{X}_{jk/lm} -= d\psi)$ and $(\Matrix{X}_{no/pq}+= d\psi). ]$
    \end{enumerate}
  \item \ldots Until the proposed transfer of flux in step i. increases the total system's free energy for any separator---e.g., max$(W_i-W_\sigma)< 0\ \forall\ \sigma$.   
\end{enumerate}

\section{\label{sec:ribbons}Observations of Flare Ribbons}

Having determined how to model reconnection, we now turn our attention to determining when to apply a minimization.  At present we have no model for the mechanism in the actual corona which initiates reconnection at a current sheet.  All we can infer, from observations of actual flares, is that at some instant the reconnection does begin at certain separators.  We therefore rely on observations of this kind to determine the separators undergoing reconnection, and when this reconnection occurs.  

For the present study, we perform a minimization for every flare associated with AR11158 with GOES class of M1.0 or greater.  This pares the number of separate minimizations down to a manageable amount for a proof of concept, while still allowing us to understand some of the large scale processes at work in the active region's evolution.

We employ chromospheric data to select a subset of coronal domains involved in each flare.  We associate the chromospheric flare ribbons observed in the 1600\AA\ channel of AIA data with specific spine field lines of the topological skeleton.  While the spine lines do not always geometrically match the ribbons, there is a topological correspondence \citep{DesJardins:2009, Kazachenko:2012}.  The spines are the photospheric footpoints of reconnecting loops, and therefore indicate which flux domains are involved in each flare.  Magnetic reconnection across a separator couples the flux redistribution in the corona to the photospheric spines of the separator's null points.  As reconnection occurs, the involved separators' chromospheric footpoints move along the spine lines \citep{DesJardins:2009}.  The highlighting of spine lines by flare ribbons thus indicates those separators involved in each flare.  To make use of this information, we relax flux constraints (allow for reconnection between four domains) using only those separators associated with the highlighted ribbons.

\begin{figure}[ht]
  \begin{center}
    \includegraphics[width=0.8\textwidth]{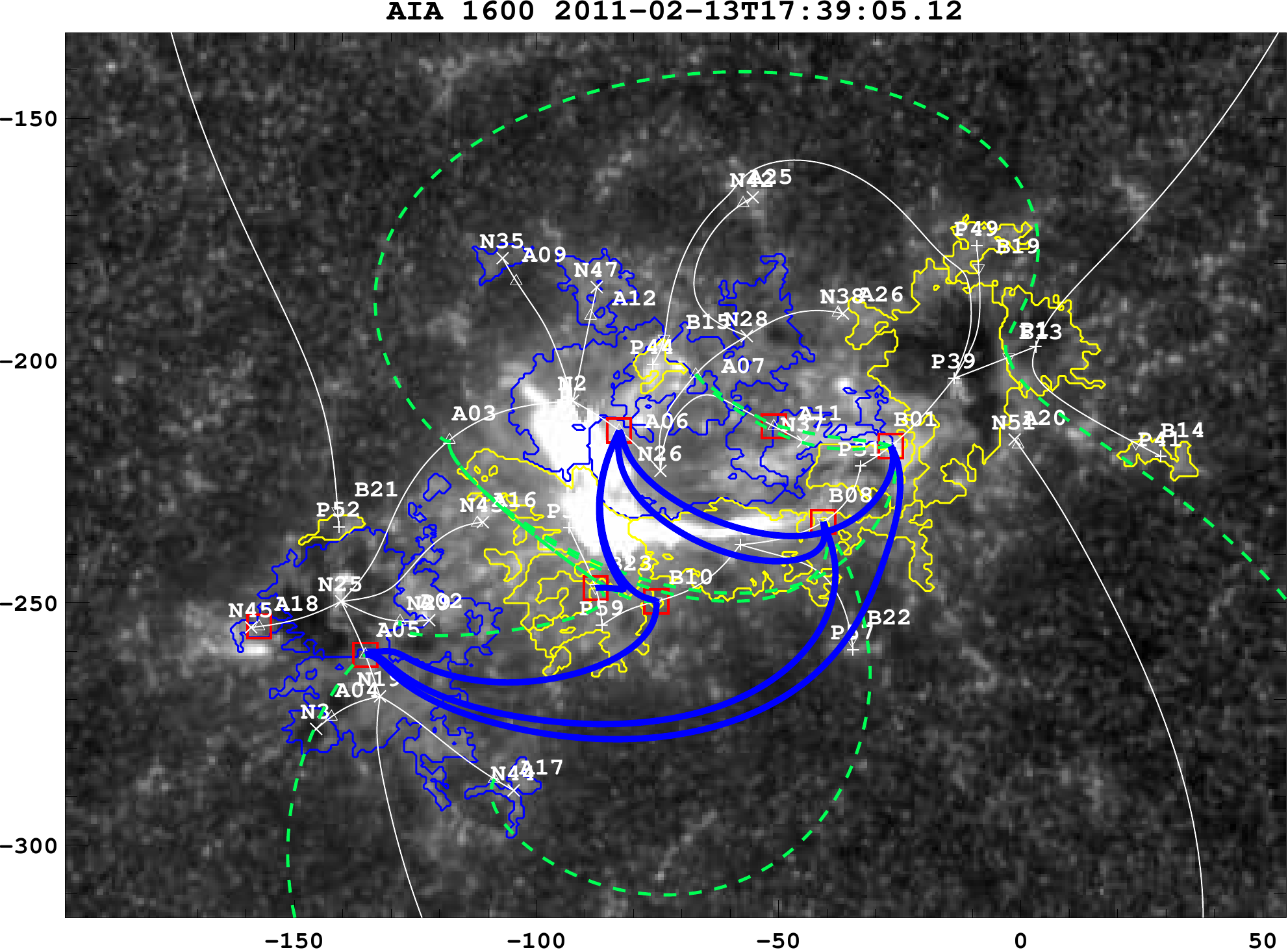}
    \caption[M6.6 Flare Ribbons with topology]{\label{fig:143ribbon} Log--scaled AIA 1600\AA\ image during the GOES class M6.6 flare, with coordinates given in arcseconds from disk center.  The grayscale saturates at 6000 counts, roughly half the pre--flare maximum pixel value.  The potential field skeleton is overlaid: pluses and crosses are positive and negative poles, respectively; triangles are positive ($\vartriangle$) and negative ($\triangledown$) nulls; thin solid white lines depict spines.  The energy calculation only attempts reconnection across those separators having two boxed nullpoints as footpoints.  These seven separators are displayed as thick solid blue lines.  The remaining ten separators are displayed as green dashed lines.}
  \end{center}
\end{figure}

Figure \ref{fig:143ribbon} shows the AIA 1600\AA\ data for a selected timestep during the M6.6 flare.  The AIA image is displayed in a logarithmic grayscale, and shows a relatively simple two ribbon flare.  Overplotted are contours of the magnetogram at +75G (yellow) and -75G (blue), as well as the topological skeleton (see caption for details, and supplimentary media for an animation covering the time of the flare).  Clearly visible are the two primary flare ribbons, located on either side of the central polarity inversion line (PIL between southern--emerged positive flux and northern negative flux).  Evident in the animation are several other, smaller flare ribbons, located in the southern negative and northern positive regions.

There are spine fieldlines associated with each ribbon.  The northern primary ribbon corresponds to the spine lines of null A06, between poles N2 and N26, near $(-90,-210)^{\prime\prime}$.  The southern ribbon is only morphologically similar to the potential field MCT model.  We separate the more diffuse P59 region from the more concentrated P3 and P37, which forces the creation of two null points (B23, B10) with associated spine lines, near $([-90/-75],-250)^{\prime\prime}$, respectively.  We believe the real field likely has a null directly between P3 and P37, creating a more direct spine line between the two.  

The spines involved in this flare have the red--boxed nullpoints in Figure \ref{fig:143ribbon} as their spine sources.  This indicates that flux is transferred only across those separators connecting two of the boxed nulls.  The projection onto the photosphere of seven such separators in this flare are shown as thick blue lines.  The remaining separators are shown as dashed green lines.  The free energy of these other separators may still change during the flare despite having no reconnection across them, provided the Gauss Linking Number between the separator field line and any domain which does participate in reconnection is nonzero, as indicated by Equation \eqref{eq:fcr}.

\begin{figure}[ht]
  
  \begin{center}
    \includegraphics[width=0.8\textwidth]{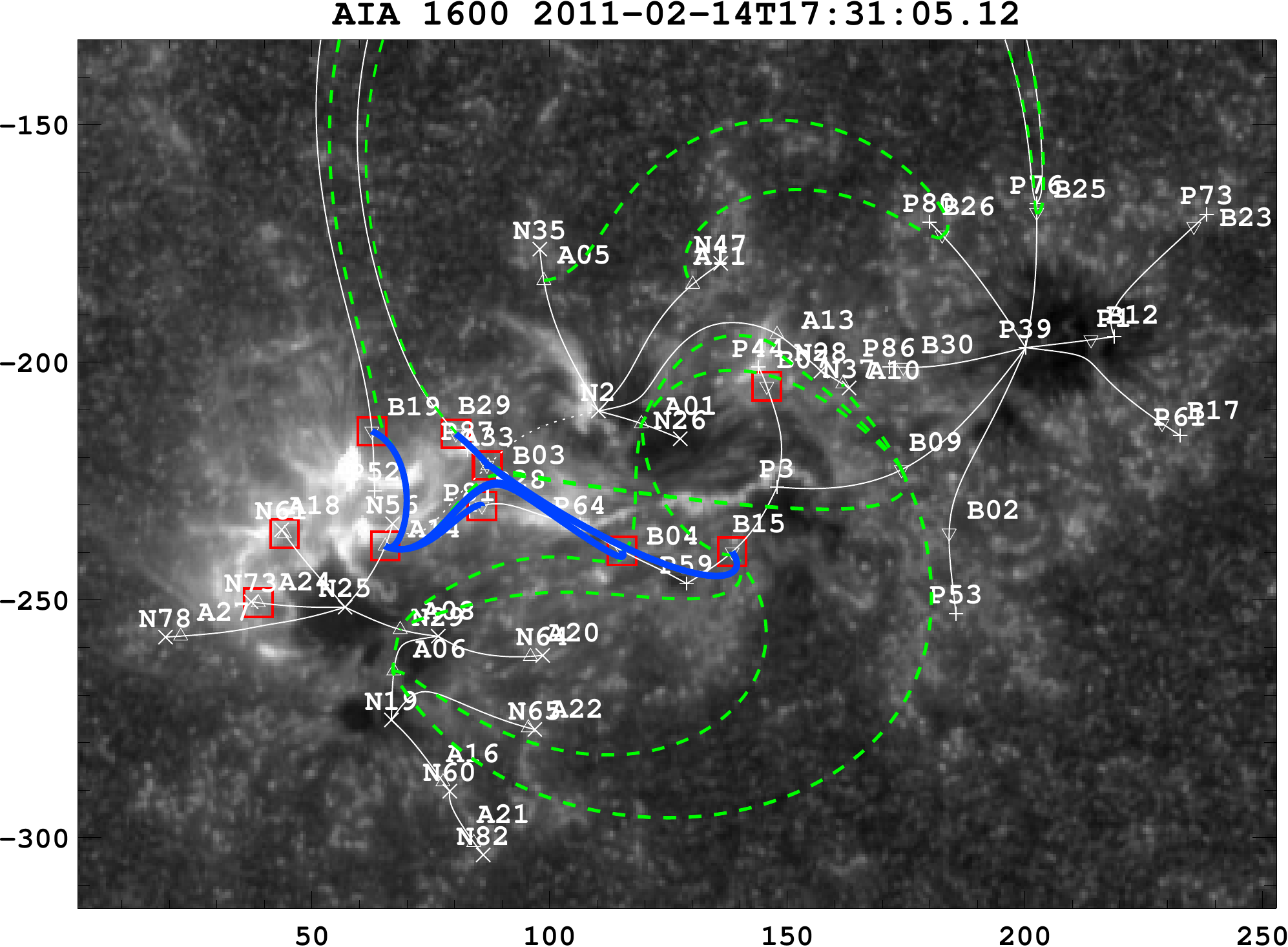}
    \caption[M2.2 Flare Ribbons with topology]{\label{fig:202ribbon} Same as \figref{fig:143ribbon}, for the M2.2 flare.  Solid thick blue lines show separators connected to nulls with spines laying approximately along paths of flare ribbons observed in AIA 1600\AA\ channel.  Locations of other separators are shown as dashed green lines.}
  \end{center}
\end{figure}

\begin{figure}[ht]
  
  \begin{center}
    \includegraphics[width=0.8\textwidth]{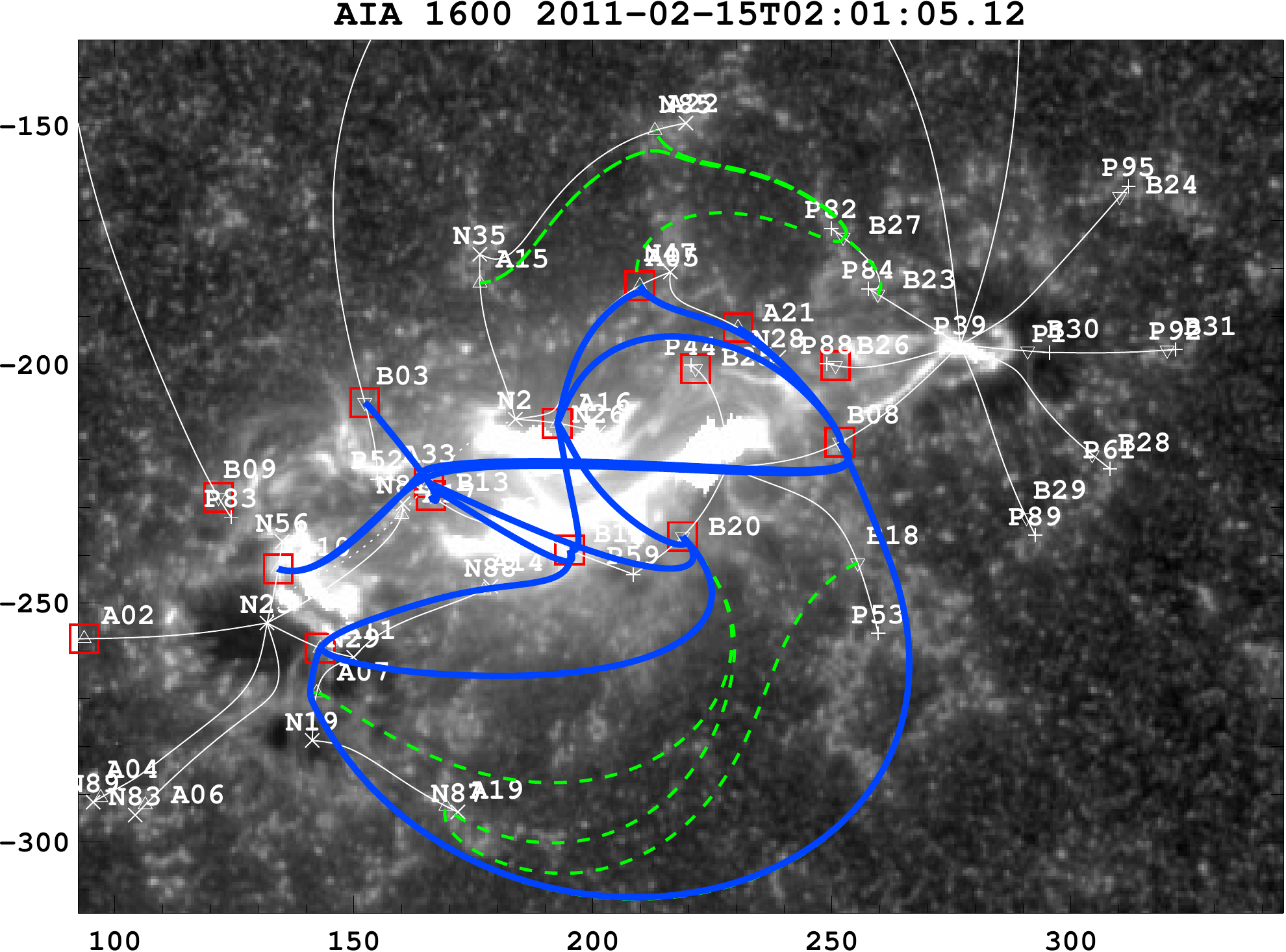}
    \caption[X2.2 Flare Ribbons with topology]{\label{fig:224ribbon} Same as \figref{fig:202ribbon}, for the X2.2 flare.}
  \end{center}
\end{figure}

Figures \ref{fig:202ribbon} and \ref{fig:224ribbon} are similar to Figure \ref{fig:143ribbon}, during the M2.2 and X2.2 flares, respectively.  We have left off the contours of the magnetogram in these figures for clarity.  The X class flare in particular is more complex than previous flares, involving more and disparate parts of the active region complex.  This increased activity is likely influenced by the creation of a coronal null point just prior to the M2.2 flare, whose fan surface effectively separates the northern and southern emergence zones.  This null, A33, is found at $(90, -225)^{\prime\prime}$ in \figref{fig:202ribbon}, with a spine field line shown as a dotted line connecting N2 to N56.  In \figref{fig:224ribbon}, it is found at $(165, -225)^{\prime\prime}$.

The energy buildup prior to the M6.6 flare is particularly dependent on the emergence of N26 in the north.  This generates the null point in the north to which the four northern involved separators attach.  In the 25 hours between N26's emergence at 2011-02-12 16:00 UT and the M6.6 flare, we calculate an increase in free energy due to currents along these four separators of $2.87 \times 10^{31}\unit{erg}$, about one third of the total MCC free energy in the system at this time.  These separators link domains N26/P37, N26/P59, N26/P31, N26/P39, N26/P44, N28/P31, N37/P3, and N37/P39.

The 1600\AA\ flare ribbons indicate that 8 separators are involved in the M2.2 flare (\figref{fig:202ribbon}).  Four connect to null A14 between regions N25 and N56, and 4 connect through the coronal null A33; their projections in the photospheric plane are shown as solid blue lines, with the remaining separators shown as dashed green lines.  As shown in Table \ref{tab:minimize}, in this case, our minimization algorithm exchanges flux across all of these separators.

For the X2.2 flare shown in \figref{fig:224ribbon}, the correspondence between 1600\AA{} flare ribbons and the topological skeleton indicate that 16 separators are involved.  5 connect to the coronal null point at A33.  The uninvolved separators are again shown as dashed green lines.  Of these 16 separators, the minimization algorithm exchanges flux across 10, including 4 of those connected to the coronal null.  While both the M2.2 and X2.2 minimizations include separators that better match those expected from observations of the flare ribbons, they still show the same puzzling behavior as in the M6.6 flare, which we will discuss in detail in the following section.

\begin{figure*}[ht]
  
  \begin{center}
    \includegraphics[width=0.8\textwidth]{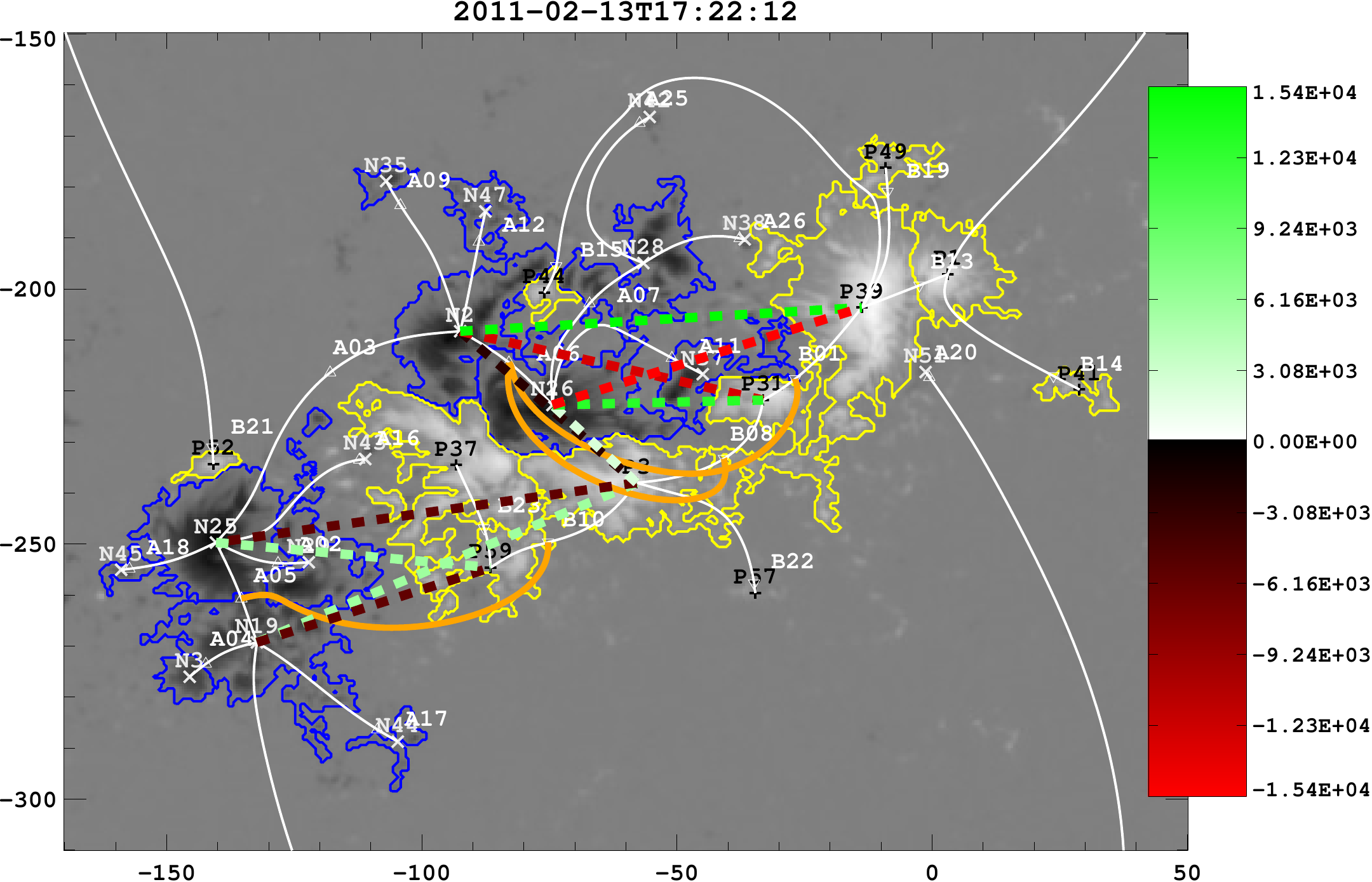}
    \caption[M6.6 Minimization]{\label{fig:m66min} Flux redistributed as a result of energy minimization.  The background image shows the magnetogram, mask, poles, nulls, and spine field lines.  Thick orange lines show the three separators utilized during the minimization.  Dashed lines indicate domains involved in the minimization, with colors representing the amount of flux gained (white to green) or donated (black to red).  The colorbar scale is in units of $10^{16}\unit{Mx}$.}
  \end{center}
\end{figure*}
\figref{fig:m66min} illustrates the result of the free energy minimization for the M6.6 flare.  Thick orange lines show the projections of separators across which flux was transferred by the minimization process.  Note that this involves three separators of the seven identified via flare ribbons in \figref{fig:143ribbon}, which are themselves a subset of the 17 total separators at this time.  The dashed lines overlaid on the magnetogram, mask, and topological footprint background show those domains that exchanged flux during minimization---these are not fieldlines, just identifications of the involved domains.  The amount of flux loss or gain is indicated by the colorbar, with white to green indicating increasing amounts of flux gain, and black to red increasing flux donation.  

It is immediately apparent that the minimization does not exactly match our expectation from the flare ribbons, despite our specification of ``involved separators.''  In particular, the four domains involving N2, N26, P37, and P59 are essentially nonparticipants in the modeled flare, whereas they are clearly the dominant players in the actual flare.  In our model of emergence, these domains are not simply flux deficient relative to the potential field, but have \emph{zero} initial flux.  We will discuss this in more detail in \S\ref{sec:disc}.

In total for the M6.6 flare minimization, we find that $4.2\times 10^{20}\unit{Mx}$ of flux was exchanged between $10$ domains across $3$ separators.  This exchange took $1922$ iterative minimization steps, resulting in a total energy drop within the entire system of $E_{\unit{drop}}=3.9\times10^{30}\unit{erg}$, $2.5\%$ of the pre--minimization MCC free energy ($E_{\unit{MCC}}=1.5\times 10^{32}\unit{erg}$) and $1.1\%$ of the potential field energy ($E_{\unit{potl}}=3.9\times 10^{32}$).  These results are summarized in Table \ref{tab:minimize}, together with those for the M2.2 and X2.2 flares, and we discuss them in more detail in the next section.

\begin{table}[ht]
  \begin{center}
    \footnotesize
    \caption{\label{tab:minimize}Summary of energy minimization for each flare}
    \begin{tabular}{lcccc}
      \tableline\tableline
     Flare & \textcolor{blue}{Flux ($10^{20}$Mx)} & Domains & Separators & Steps \\
     \hline
     M6.6 & \textcolor{blue}{4.2}  & 10 & 3  & 1922  \\
     M2.2 & \textcolor{blue}{2.0}  & 15 & 8  & 327   \\
     X2.2 & \textcolor{blue}{21.0} & 17 & 10 & 29504 \\
     \hline
     \hline
     \\
          & Initial $E_\text{MCC}$ & \textcolor{blue}{$\Delta E_\text{MCC}$} &  $E_\text{potl}$ \\
     \hline
     M6.6 & $1.53\times 10^{32}$ & \textcolor{blue}{$3.89\times 10^{30}$} & $3.83\times 10^{32}$ \\
     M2.2 & $1.65\times 10^{32}$ & \textcolor{blue}{$2.62\times 10^{30}$} & $5.77\times 10^{32}$ \\ 
     X2.2 & $2.94\times 10^{32}$ & \textcolor{blue}{$1.68\times 10^{32}$} & $5.55\times 10^{32}$  \\
    \end{tabular} 
    \tablecomments{Columns Definitions: 1) GOES class; 2) Flux exchanged by the minimization algorithm; 3) Number of domains involved in the minimization; 4) Number of separators across which flux is exchanged; 5) Number of algorithm steps; 6) Initial free energy of the MCC; 7) Energy drop due to minimization; and 8) Potential energy using the magnetogram as a lower boundary.}
  \end{center}
\end{table}
 
\section{\label{sec:disc}Discussion}

This investigation builds on the previous chapter in adding the observational history of an active region's flux evolution, in particular its flux emergence, to the MCC model.  Here we have relied on line--of--sight magnetograms provided by SDO/HMI and generated our flux histories by assuming a radial field.  With the arrival of the HMI Active Region Patches (HARPs) dataseries to JSOC, future investigations can use the actual vertical flux determined by HMI's vector magnetograms.  

We were fortunate in the present case to have HMI observe the entire history of AR11158 from its emergence around $50^\circ$ Solar east on Feb.~10th, 2011, to its rotation off-disc some 9 days later.  In the more common case where we do not observe the entire emergence of an active region, we could use a NLFFF extrapolation to generate an initial connectivity state, which would then be updated in time via the methods of \S\ref{sec:mod-phot}.

We have gone several steps further than previous energy estimates using the MCT/MCC framework, such as \citet{Tarr:2012} or \citet{Kazachenko:2012}.  Most importantly, we now allow for violation of the no--reconnection flux constraints of the MCC.  This enables us to not only consider energy storage due to currents along a subset of separators in a flare, but also allows for an estimate of flux transfer and energy conversion during a flare.  Further work may yield an interesting comparison between the flux transferred in our minimization to estimates of reconnected flux based on analysis of flare ribbons, as was done in \citet{Longcope:2007}.

For each flare, we choose a subset of separators that are allowed to transfer flux during our minimization based on observations of chromospheric flare ribbons.  Each separator bounds 4 domains, though some of these domains are bounded by more than one separator.  Thus, the number of domains which may exchange flux via reconnection is typically less than 4 times the number of involved separators.  

It is interesting that the minimization scheme we have proposed does not utilize every allowed domain.  As mentioned above, in order to undergo reconnection, two domains on opposite sides of a separator must both contain flux (have nonzero elements in the connectivity matrix $\Matrix{F}$).  Focusing again on the M6.6 flare, the two separators that connect to null A06, between poles N2 and N26, and have pole P59 as a spine source of their B--type nulls (nulls B10 and B23) contain very strong currents and border flux domains with highly nonpotential connectivities.  And yet, they do not participate in the minimization.  In this case, there is no pathway of free energy loss that results in donatable flux on opposite sides of these separators.  The lack of any flux in the P59/(N2, N26, N28) domains effectively cuts off any involvement of P37 in our model of this flare.

Between poles N2, N26, P37, and P59, flux transfer may occur in either of two directions.  Domains N2/P59 and N26/P37 may donate flux, with N2/P37 and N26/P59 receiving, or N2/P37 and N26/P59 may donate flux with N2/P59 and N26/P37 receiving.  In either case, one of P59's domains must donate flux.  Since both the P59/N2 and P59/N26 domains have \emph{no} flux in our model (not to be confused with having less than the potential field configuration), this reconnection cannot occur.

We do, however, observe the primary flare ribbons involving just these domains.  So, where has our model of this active region gone awry?  First, we note that there are many small flare events smaller than M1.0 prior to the first flare we consider.  Any of these events may transfer flux into the domain necessary for the M6.6 event.  

Second, it is likely that a great deal of reconnection takes place that is unassociated with any GOES class flare.  This may be a ``steady'' reconnection, as is observable in EUV and X--ray images of emerging magnetic regions.  For instance, in the previous chapter on AR11112 from October 2010, we saw that newly emerged flux steadily reconnected with surrounding preexisting flux, as evidenced by the encroachment of the bright core seen in Hinode/XRT data into the surrounding flux.  This is discussed in detail in the next chapter.  Although AR11112 did produce several flares later in its evolution, this steady reconnection occurred independently of any observable flares.  We believe that a similar process is ongoing in AR11158's evolution.  In particular, we believe this type of steady reconnection occurs along the central PIL, where southern emerged positive flux has collided with and sheared relative to northern emerged negative flux.  Such a steady, low--level reconnection would populate these central domains with flux.  The free energy of this portion would increase as these domains continued to shear, culminating in the series of explosive reconnections observed at later times.  One may see evidence of this in EUV images from AIA, showing low laying loops that cross the central PIL, together with very high loops apparently connecting the most westward positive flux to the most eastward negative flux concentrations.

Another interesting point is that, in our minimizations, no separator expelled all of its current and thereby reached the potential field state, as one might expect to be the case in an MCC model\footnote{In other types of models \citep[e.g.,]{Regnier:2007}, the potential field state is \emph{not} generally accessible via reconnection.  For instance, in models where helicity is conserved, reconnection drives the system towards the linear force free field with the same helicity.}. Instead, in each of the three reconnection events modeled here, the system reaches a state where reconnection across one separator reduces one domain's flux to zero.  Reconnection across a second separator can repopulate the zero--flux domain, allowing further reconnection across the first separator.   However, it often happens that the two separators each require flux donation from that same domain, and so no further reconnection is possible across either one.  \citet{Longcope:2010} also found that the total free energy derived from the MCC model was greater than their calculated energy losses observed during the Feb. 24, 2004 X class flare.  They attributed this to incomplete relaxation through reconnection, but had no way to assess why this might be the case, as we have developed here.

\begin{figure}[ht]
  
  \begin{center}
    \includegraphics[width=0.8\textwidth]{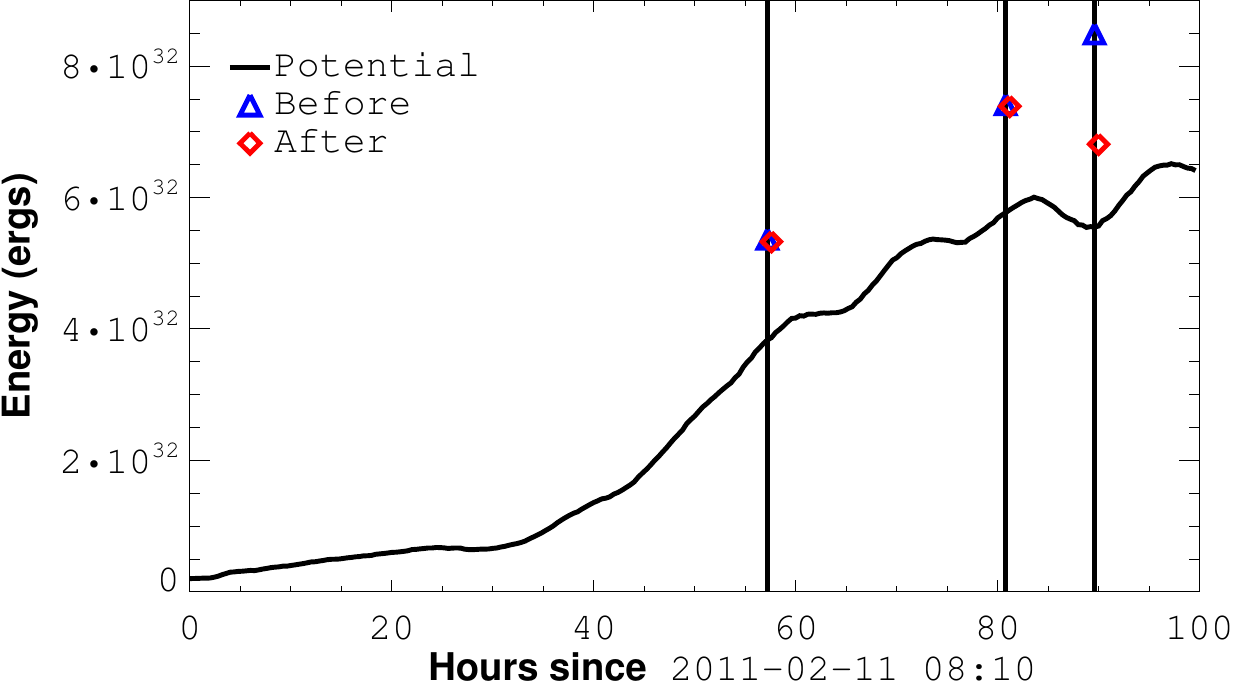}
    \caption[Potential, pre--, and post--minimization energies for AR11158]{\label{fig:Eplot} The potential energy (solid line) and MCC free energies both before (triangles) and after (diamonds) each minimization, the times of which are indicated by vertical lines.  There is no way to calculate a potential energy in the MCT/MCC model, so the free energy has been plotted above the potential energy derived from a Fourier transform method.}
  \end{center}
\end{figure}

We may compare the energies associated with each event in a variety of ways.  The solid line in \figref{fig:Eplot} shows the potential energy calculated from each magnetogram using the Fourier Transform method of \citet{Sakurai:1989}.  This is similar to, but not directly comparable with, the potential energy shown in Figure 4(c) of \citet[][(hereafter S12)]{Sun:2012}.   When not stated explicitly in the text, we have determined approximate values for potential and free energies from their Figure 4.  

Because S12 use the vertical flux determined from the HMI vector magnetograms whereas we use the LOS field, deprojected assuming a radial field at every pixel, we have different lower boundaries for determining the potential energy, so that our calculated values differ.  Our potential energy appears consistently lower than that of S12, and also seems to have more substantial variations.  These variations become more pronounced after $t=65\unit{hrs}$, when the northern region ceases substantial emergence and the active region crosses disk center.

The free energies of the MCC model $E_{\unit{MCC}}$ are shown in \figref{fig:Eplot} as blue triangles at the time of each flare and red diamonds after minimization.  We have added the MCC values to the calculated potential field energies at each time in order to emphasize that these are energies in excess of the potential field.  However, this addition should be taken with caution, because the free and potential energies are calculated using two incompatible boundary conditions: the former using the point sources of the MCT model with an imposed FCE constraint, the later using continuously distributed magnetograms.  The potential energy of a point source is infinite, and the MCC is only able to determine the difference between energies of the FCE and potential fields.  Even given these issues, it is still a useful comparison to make.

The initial free energy of the MCC increases from flare to flare, while the calculated energy drop due to flux exchange during minimization scales with the size of each flare.  This is most easily seen in the $\Delta E_{\unit{MCC}}$ column of Table \ref{tab:minimize}.  The general trend of the amount of flux exchange, and resulting energy drop, in each flare follows our expectations given the GOES class.  An order of magnitude more flux is exchanged for the X2.2 flare compared to the M flares, yet this leads to a two order of magnitude greater free energy drop.  This highlights the important point that free energy does not scale linearly with flux--difference from potential, even for a given separator.

S12 made extensive use of HMI's vector magnetograms to present a detailed discussion of magnetic energy in AR11158, so it is useful to compare the results of that NLFFF model to ours.  Those authors extrapolate a NLFF field at a 12 minute cadence using the HMI vector magnetogram as a lower boundary.  Each extrapolation is independent of the others, maintaining no memory of previous magnetic flux or connectivity.  It can therefore be difficult to ascribe any rise or drop in the free energy from one time to the next to any particular event.  They state a spectropolarimetric noise of $\approx 4\times 10^{30}\unit{ergs}$, but acknowledge that errors due to extrapolation are unknown and possibly greater.  We note that a persistent change in the free energy from before to after an event is likely to represent a real change in the actual coronal field, despite the calculation being based only on photospheric data.

For the X2.2 flare, S12 find an initial free energy of $\approx 2.5\times 10^{32}\unit{ergs}$, and a persistent free energy drop from before to after the flare of $0.34\pm 0.04\times 10^{32}\unit{ergs}$.  This may be compared to our model, which sets the initial free energy at $2.94\times 10^{32}\unit{ergs}$ and a pre--to--post minimization drop of $1.68\times 10^{32}\unit{ergs}$.  That both of these results are greater than those of S12 is a little surprising because MCC provides a lower bound on the free energy of linear force free fields evolving quasistatically \citep{Longcope:2001}.  While it is not necessary that a LFFF have less free energy than a NLFFF derived from the same boundary, one often assumes it to be the case.  

At the same time, the MCT/MCC model does not involve an extrapolation, except to determine the topological structure of the region.  The non--potentiality of the region is determined simply by using observations of emergence and submergence to fix the regions' connectivity.  Our minimization scheme only determines the total amount of energy loss provided that all possible free energy minimizing reconnections take place.  Not only may the algorithm terminate in a local free energy minimum, as discussed above, there is no physical reason why all possible reconnections need to take place in a single event.

It is much more difficult to make such a comparison between these two models for the M6.6 and M2.2 flares.  From Figure 4(d) of S12, we estimate initial free energies of $1.2\times 10^{32}\unit{ergs}$ and $2.0\times 10^{32}\unit{ergs}$ for these flares, respectively, compared to our results of $1.53\times 10^{32}\unit{ergs}$ and $1.65\times 10^{32}\unit{ergs}$.  Our minimization gives pre--to--post reconnection drops of $3.92\times 10^{30}\unit{ergs}$ and $2.62\times 10^{30}\unit{ergs}$ in each case, which is the approximate level of the spectropolarimetic noise in the NLFFF extrapolations.  Indeed, again looking at Figure 4(d) of S12, neither the M6.6 or M2.2 flare appears cotemporal with a decrease in free energy, and certainly not with a persistent decrease, as is the case with the X2.2 flare.

As a final energetic comparison, we calculate the energy loss due to radiation in each flare using the GOES light curves via the method of \citet{Longcope:2010, Kazachenko:2012}.  For the M6.6, M2.2, and X2.2 flares we find radiative losses of $1.2\times 10^{30} \unit{ergs}$, $0.5\times 10^{30}\unit{ergs}$, and $4.2\times 10^{30}\unit{ergs}$, respectively.  The above mentioned papers show that other energetic losses, such as thermal conduction and enthalpy flux, tend to dominate the radiative losses during flares.  The total energetic loss is difficult to precisely quantify, but may exceed the GOES estimated radiative loss by a factor of $\approx 2-15$.  Given that, the results of our minimization for the M6.6 and M2.2 compare favorably with the GOES estimates.  However, our estimation for energy loss during the X--class flare may be greater than the observed energetic losses by an order of magnitude.  This is not surprising because the great extent of the X--class flare ribbons encompassed more separators and more domains, ultimately allowing more pathways for energy minimizing reconnection.  This allowed the minimization algorithm to exchange much more flux before halting in a local minimum.

\subacknowledgments
Graham Barnes provided code for producing the potential field connectivity matrices using a Monte Carlo algorithm with Bayesian estimates, as described in \cite{Barnes:2005}.  Development of the code was supported by the Air Force Office of Scientific Research under contract FA9550-06-C-0019.  We would like to acknowledge our use of NASA's Astrophysics Data System.  This work is supported by NASA under contract SP02H3901R from Lockheed--Martin to MSU.  
 
\chapter{Quiescent reconnection rate between emerging active regions and preexisting field, with associated heating: NOAA AR11112}\label{ch:rxrate}
\begin{manuscriptauths}
  Manuscript in Chapter 3
  \newline
  \newline
  Author: Lucas A. Tarr
  \newline
  \newline
  Contributions: Conceived and implemented study design.  Wrote first draft of manuscript.
  \newline
  \newline
  Co--Author: Dana W. Longcope
  \newline
  \newline
  Contributions: Helped with study design.  Provided feedback on analysis and comments on drafts of the manuscript.
  \newline
  \newline
  Co--Author: David McKenzie
  \newline
  \newline
  Contributions: Helped with study design.  Provided feedback on analysis and comments on drafts of the manuscript.
  \newline
  \newline
  Co--Author: Keiji Yoshimura
  \newline
  \newline
  Contributions: Data preparation of the XRT datasets, and co--alignment between XRT and AIA datasets.  
\end{manuscriptauths}
\pagebreak
  
\begin{manuscriptinfo}
  \noindent Lucas A. Tarr, Dana W. Longcope, David McKenzie, and Keiji Yoshimura\\
  Solar Physics\\
  Status of Manuscript:\\
  \uline{\phantom{5eM}}\makebox[0pt]{\hspace{-2em}x}Prepared for submission to a peer--reviewed journal\\
  \uline{\phantom{5eM}}Officially submitted to a peer--reviewed journal\\
  \uline{\phantom{5eM}}Accepted by a peer--reviewed journal\\
  \uline{\phantom{5eM}}Published in a peer--reviewed journal\\
  \newline
  \newline
  To be submitted September, 2013
\end{manuscriptinfo}

\begin{abstract}
  When flux emerges from beneath the photosphere it displaces the preexisting field in the corona, and a current sheet generally forms at the boundary between the old and new magnetic domains.  Reconnection in the current sheet between the old and new domains relaxes this highly stressed configuration to a lower energy state.  The oft-studied rapid reconnection results in flares, but just as much flux may be transferred in steady reconnection, orders of magnitude more slowly than that resulting in flares.  In this study we quantify this quiescent reconnection rate for the case of emerging Active Region 11112.  The bright, low lying kernel of coronal loops above the emerging field, observed with the \emph{Atmospheric Imaging Assembly} onboard the \emph{Solar Dynamics Observatory} and the \emph{X--ray Telescope} onboard \emph{Hinode}, originally show magnetic connectivity only between regions of newly emerged flux when overlaid on HMI magnetograms. Over the course of several days, this bright kernel advances into the preexisting flux. The advancement of this easily visible boundary into the old flux regions allows the measurment of the rate of reconnection between old and new magnetic domains.  We compare the reconnection rate to the inferred heating of the coronal plasma. To our knowledge, this is the first measurement of steady, quiescent heating related to reconnection.  We determine that the newly emerged flux reconnects at a fairly steady average rate of $0.38 \times 10^{16}\unit{Mx/s}$ over two days, while the radiated power varies between $2\sim8\times 10^{25}\unit{erg/s}$ over the same time.  We find that as much as $40\%$ of the total emerged flux at any given time may have reconnected in this way.  The total amount of transferred flux ($\sim1\times10^{21}\unit{Mx}$) and radiated power ($\sim3\times10^{32}\unit{ergs}$) are comparable to that of an X--class flare, but are stretched out over 45 hours instead of 30 minutes.
\end{abstract}

\section{\label{sec:intro4}Introduction} 

When magnetic flux emerges through the photospheric boundary, the coronal field must respond in some way.  The exact type of response will depend on the configuration of the preexisting field as well as the rate and total amount of flux emergence.  The response itself may be broken down into several stages \citep{Heyvaerts:1977}: a preflare heating phase, an impulsive phase of particle acceleration and rapid increase in H$\alpha$ emission, and finally a main phase where such emission decreases.  Such a model naturally accounts for many diverse observations, from quiet Sun X--ray Bright Points to the largest observed flares with associated coronal mass ejections (CMEs).  

Much of the work of the last several decades has focused on the impulsive and decay phases.  This is partly because of their extravagant nature and direct impact on space weather at Earth, but also, and importantly, because of their relatively short timescale, typically less than one hour, and corresponding ease of observation.  In contrast, the preflare heating phase of flux emergence, a period marked by continuous magnetic reconnection between the new and old flux, may last for days.  An accurate description of the process requires simultaneous observations of the magnetic field, to capture the emergence itself, and of EUV and X--ray emission, to capture the coronal response.  Only in the last few years have such simultaneous and continuous observations at the needed spatial resolution been possible.

Magnetic reconnection is one of the most likely direct sources for coronal heating \citep{Archontis:2008, Reale:2010}.  In terms of the coronal energy balance, reconnection converts free magnetic energy---energy in excess of the potential field defined by photospheric sources of flux---into kinetic and thermal energy of the plasma.  It is easy to demonstrate that free magnetic energy rapidly increases during flux emergence if no reconnection occurs: the coronal field develops a tangential discontinuity, or current sheet, at the interface between the preexisting and newly emerged flux systems.  Field lines on one side of the discontinuity have footpoints wholly within the preexisting flux system and field lines on the other side wholly within the newly emerged system.  At the location of reconnection field lines from the two sides exchange footpoints, so that two new field lines are created, each connecting new to old flux.  These new field lines retract, adding kinetic energy, and compress, adding thermal energy to the plasma \citep{Guidoni:2010}.

\citet{Longcope:2005b} studied the amount of reconnection between two seperately emerged active regions.

The properties of the current sheet are determined by the amount of emerging flux and its configuration relative to the preexisting coronal field \citep{Heyvaerts:1977,Archontis:2008}.  The current sheet itself determines the rate of magnetic reconnection and therefore the rate at which magnetic free energy is converted into kinetic and thermal energy.  In this way, the observed heating and reconnection rates of an emerging flux tube will help us understand the coronal response to emerging flux.

In Section \ref{sec:data} we describe the observations.  Section \ref{sec:tracking} explains our use of the magnetic field data to characterize the emerging flux region.  Section \ref{sec:euv} details how we use the EUV observations to determine the actual coronal flux domains.  In Section \ref{sec:rxrate} we combine the EUV and magnetic field observations to find the amount of reconnected flux at every time in our data series.  In Section \ref{sec:xrt} we describe the use of XRT filter ratios to calculate the radiated power within our region of interest.  Finally, in Section \ref{sec:discussion} we discuss the results of our observations and how this informs our model of the flux emergence process.

\section{\label{sec:data}Data}
We begin with measurements of the photospheric vector magnetic field from the SDO/HMI instrument, obtained via the \textsf{hmi.sharp\_720s} series at JSOC, SHARP number 0211.  The $180^\circ$ ambiguity in the azimuth has been disambiguated using a variant of the Metcalf minimization scheme as described in \citet{Sun:2012}.

Second, we use SDO/AIA EUV images from the 211\AA{} channel.  These are prepped to level 1.5 using the standard \fnc{aia\_prep.pro} utility in SolarSoftware (SSW) \citep{Freeland:1998}.

Finally, we use two filters from Hinode/XRT: the the titanium--on--polyimide and thin--aluminum--on--mesh filters, colloquially referred to as 'Ti--poly' and 'Al--mesh.'  Because we use a ratio of the two filtergrams at each time to form temperature and emission measure maps, both filters must be present.  Our analysis is therefore limited to the AL--MESH filter's normal cadence of 1 image per hour over the course of these observations.  There are occasional multi--hour lapses as XRT either performs synoptic observations or points towards other features on the solar disk.

\subsection{\label{sec:ca}Co--alignment}
Spatially, coalignment between the SDO/AIA and XRT images was accomplished by calibrating the differences in ``roll angle'', ``plate scale'', and ``pointing'' between the two instruments.  SDO data was first prepped to level 1.5 using the \fnc{aia\_prep.pro} package available in SSWIDL.  The differences in the roll angles and plate scales were measured accurately through a cross calibration technique using full disk solar images (details are in ``Yoshimura and McKenzie (in preparation)'').  The pointing differences were corrected by applying cross correlation technique using AIA 335\AA{} and XRT thin filter images.  Since all the instruments onboard SDO were well calibrated each other by the Venus transit observation in June, 2012, we can coalign the XRT data with any data from AIA and HMI.  The error of the coalignment in this study is estimated to be smaller than 1 arcsec.

Our analysis uses approximately a one hour cadence.  If there is a time difference between two observations we translate all relevant images to the time of the nearest magnetic field data using the differential solar rotation rate described in \citet{Snodgrass:1990}.

\section{\label{sec:tracking}Tracking Photospheric Flux Concentrations}
\begin{figure}[ht]
  \begin{center}
    \includegraphics[width=\textwidth]{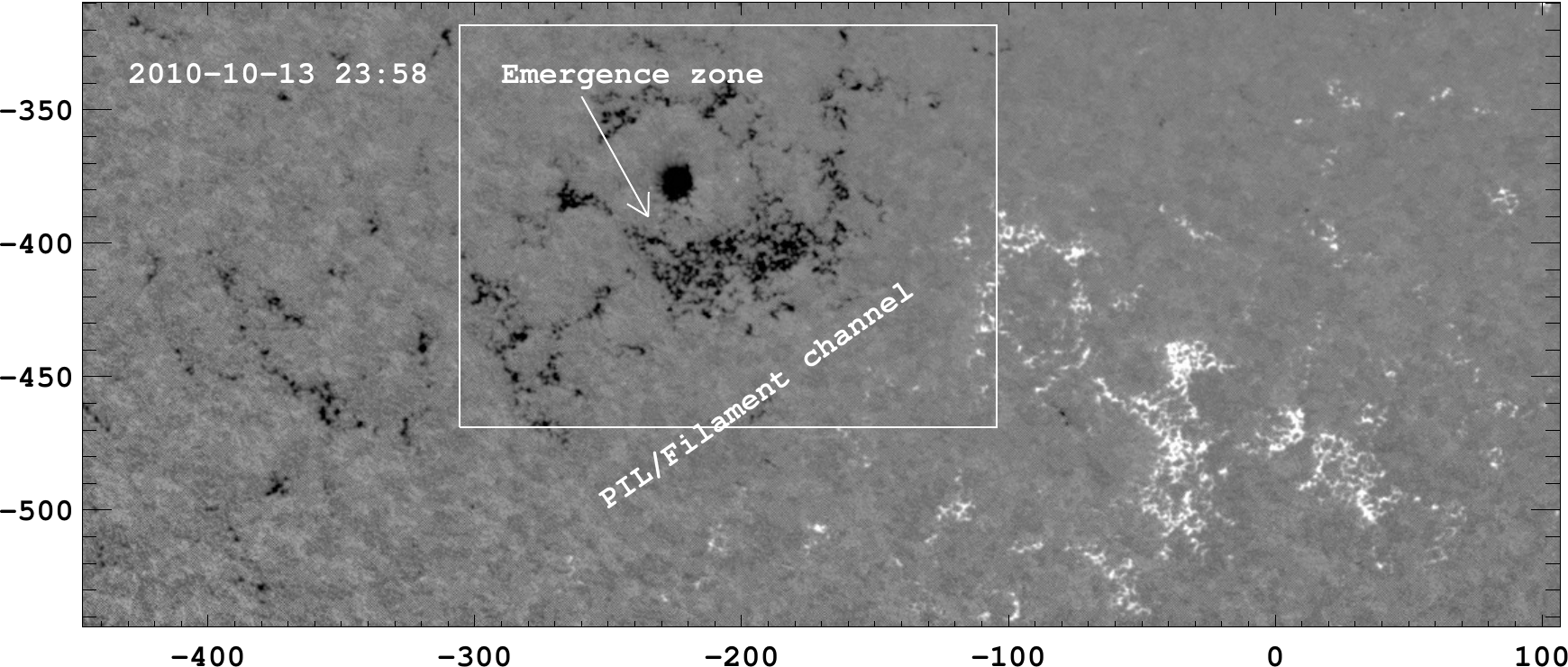}
    \caption[AR11112 radial magnetic field context]{\label{fig:magcontext} Context magnetogram showing radial magnetic flux through each pixel, displayed in POS coordinates, with X-- and Y--axes in arcseconds from disk center.  The greyscale saturates at $\pm 6.66\times 10^{17}\unit{Mx}$ (an average field strength of $500\unit{G}$ for a pixel located at disk center).  Light(dark) pixels are positive(negative) polarity.  The rectangle shows the field of view used throughout the analysis.  The large scale PIL bisects the area, and the emergence we discuss occurs within the diffuse negative polarity regions, centered slightly south and east of the ``Bull's eye'' feature.}
  \end{center}
\end{figure}
\figref{fig:magcontext} shows a large field of view magnetogram of the preexisting decayed active region.  A large polarity inversion line (PIL) separates the negative flux to the east from positive flux to the west.  EUV images from this time show a simple arcade of field lines arching over the PIL.  Visible in various AIA channels is a large filament laying along the PIL, apparently underneath the arcade.  The boxed region shows the field of view (FOV) we will use for the remainder of our analysis.  All emergence occurs within this unipolar, negative polarity region.

We repeat the flux--tracking analysis of Chapter 2.  That analysis used the 45 second cadence LOS magnetograms from HMI available at the time.  We now use the actual radial field derived from the vector magnetograms.  Our former study was a detailed analysis of the emerging field's topology and connectivity and the resulting energetics.  In contrast, in the current investigation we use the  magnetic field to distinguish between all new and old flux of each polarity, which we can do with ease, but do not need in such detail.  We therefore use a simplified mask array, and correspondingly different region labels.  

\begin{figure}[ht]
  \begin{center}
    \includegraphics[width=0.8\textwidth]{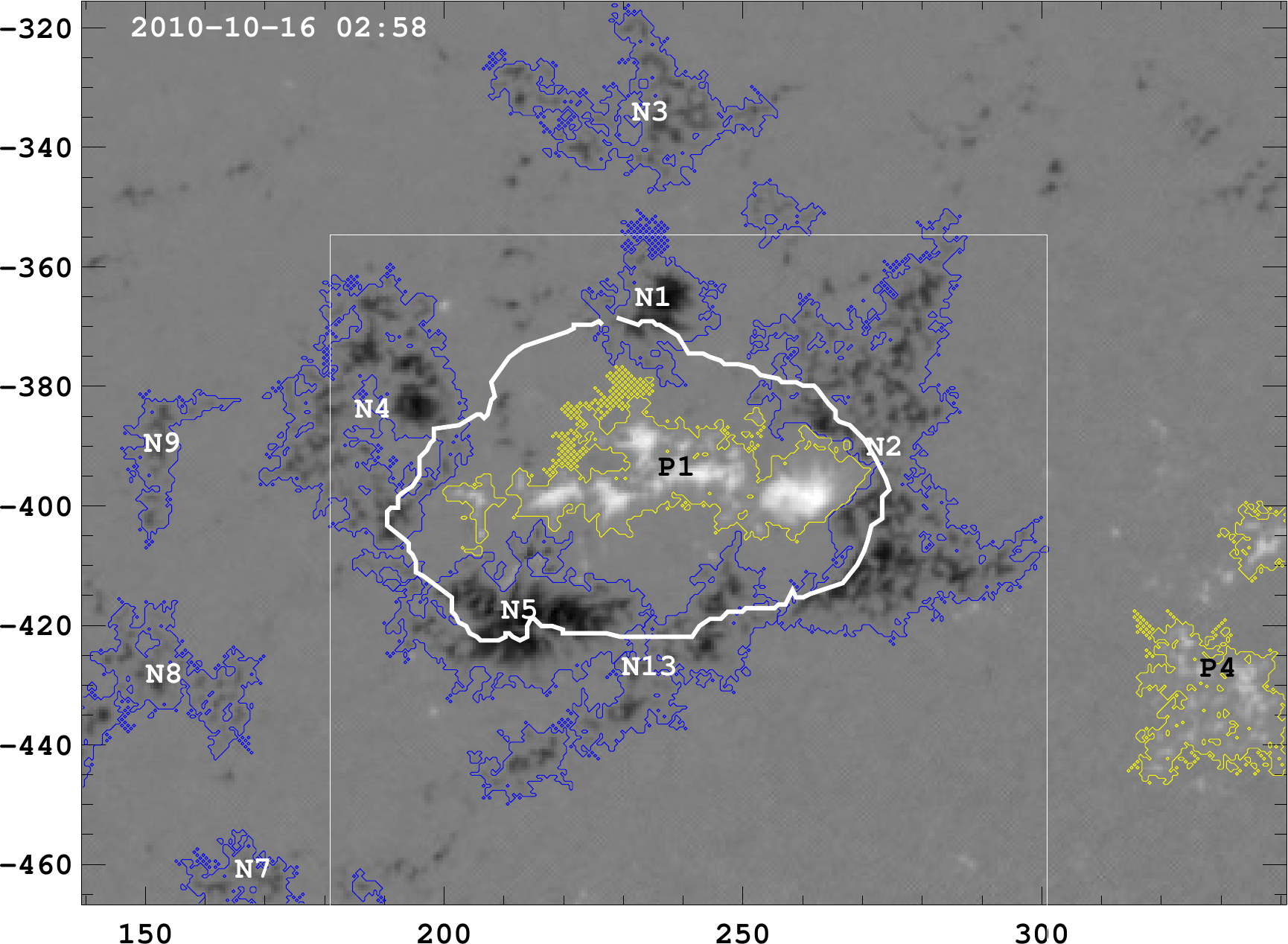}
    \caption[AR11112 radial magnetic field detail]{\label{fig:magevo} Same as \figref{fig:magcontext}, for the smaller FOV.  This greyscale saturates at $\pm 2.66\times 10^{18}\unit{Mx}$ (an average field strength of $2000\unit{G}$ for a pixel located at disk center).  Thin lines show the regions tracked in our mask array.  The thick line is the boundary derived from AIA 211\AA{} images, shown in \figref{fig:euvbnd}.  Region labels are marked at the flux--weighted--centroid of each region.  Note that the large region of preexisting negative flux to the west of the emergence region, N2, now forms a crescent shape and its label therefore lies on the boundary between P1 and N2.  The box shows the field of view of \figref{fig:channelcomp}.}
  \end{center}
\end{figure}

With a slightly different analysis goal in the present chapter, we use slightly different algorithm parameters compared to \S\ref{sec:photosphere}.  For this study, we use a lower threshold of $50\unit{G}$ average field strength over each pixel, and extrapolate the lower boundary to a height $h=3.0\unit{Mm}$ to create a smoother mask stucture at each time.  Each mask region must contain at least $2.66\times 10^{19}\unit{Mx}$ ($2\times 10^4 \unit{G*px}$), and exist for at least 5 hours, for inclusion in our analysis.  For comparison, near the end of our time series (Oct 16, 23:00UT) the total unsigned flux (above 50 gauss) for the large FOV magnetogram (cf. \figref{fig:magcontext})is $3.04\times10^{22}\unit{Mx}$, while the total unsigned flux of newly emerged field is $4.09\times 10^{21}\unit{Mx}$, about $13\%$ of the total.

The primary regions we discuss in this study are described as follows (see \figref{fig:magevo}).  In the region of emergence, we label all emerging positive flux as P1, rather than distinguishing between the different concentrations in successive episodes of emergence.  N1 is the circular ``Bull's Eye'' concentration of strong negative flux just north of emergence.  N4 and N5 are regions to the east and south of emergence, respectively, that contain difficult--to--distinguish mixes of new and old flux, although by the end of emergence both contain primarily new flux.  N13 is the oblong region just west of N5.  It contains primarily preexisting flux, but is located near the center of emergence, so we have grouped it separately.  

Finally, we will focus the most attention on the region labeled N2, just west of the emergence zone.  Prior to emergence, this region is roughly circular, consisting of numerous small flux concentrations.  The emerging field drives into the middle of this patch of plage, pushing parts of N2 to the north and south and deforming it into the crescent shape seen in \figref{fig:magevo} (cf. the boxed portion of \figref{fig:magcontext}).  The total flux of N2 remains relatively constant at $2\times 10^{21}\unit{Mx}$, varying above and below this value by about $0.14\times 10^{21}\unit{Mx}$, or $7\%$ of the average value.  This variation is mostly due to patches of network flux joining to and detaching from the western boundary of the N2 mask region, and is 3 orders of magnitude less than the amount of emerged flux.  We are therefore very confident that the mask region N2 contains only flux that predates emergence, and that any coronal loops we observe connecting N2 to the emerging postive flux P1 are due solely to reconnection between the two flux domains.

The topology of the region is described in detail in Chapter 2.  There, the we use a Magnetic Charge Topology (MCT) to describe the field at each time, together with a Minimum Current Corona (MCC) constraint to model time evolution.  The essential element is that the ring of negative polarity field (regions N1, N2, N4, N5, and N13)\footnote{Note that the region labels are different in this simplified study compared to Chapter 2.} creates a ring of spine field lines connecting each source to  nullpoints located between the sources.  This ring surrounds positive region P1.  P1 is one of the spine sources for a coronal null point, whose other spine source is located in the diffuse polar region to solar west, in the vicinity of P4, P7, and P8.  The fan (or separatrix surface) of the coronal nullpoint intersects the photosphere along the spine lines of the ring of negative polarity field, forming a dome over P1.  Field lines inside the dome connect to the newly emerged P1, while fieldlines outside the dome connect to preexisting positive field.  

The animation of \figref{fig:magevo} shows the evolution of the radial magnetic field and our mask array.  While we transform the vector magnetic field data to find the radial, latitudinal, and longitudinal components at every pixel, throughout this chapter we display images in Plane-Of-Sky (POS) coordinates.  That is to say, in the magnetograms, the lightness or darkness of a pixel represents the value of total flux in $\unit{Mx}$ within the pixel, but we have not distorted the pixel's shape from CCD coordinates.  This just makes the comparison of features observed in the HMI, AIA, and XRT data much more straightforward.  

\section{\label{sec:euv}Magnetic Domains Observed in EUV}
\begin{figure}[ht]
  \begin{center}
    \includegraphics[width=\textwidth]{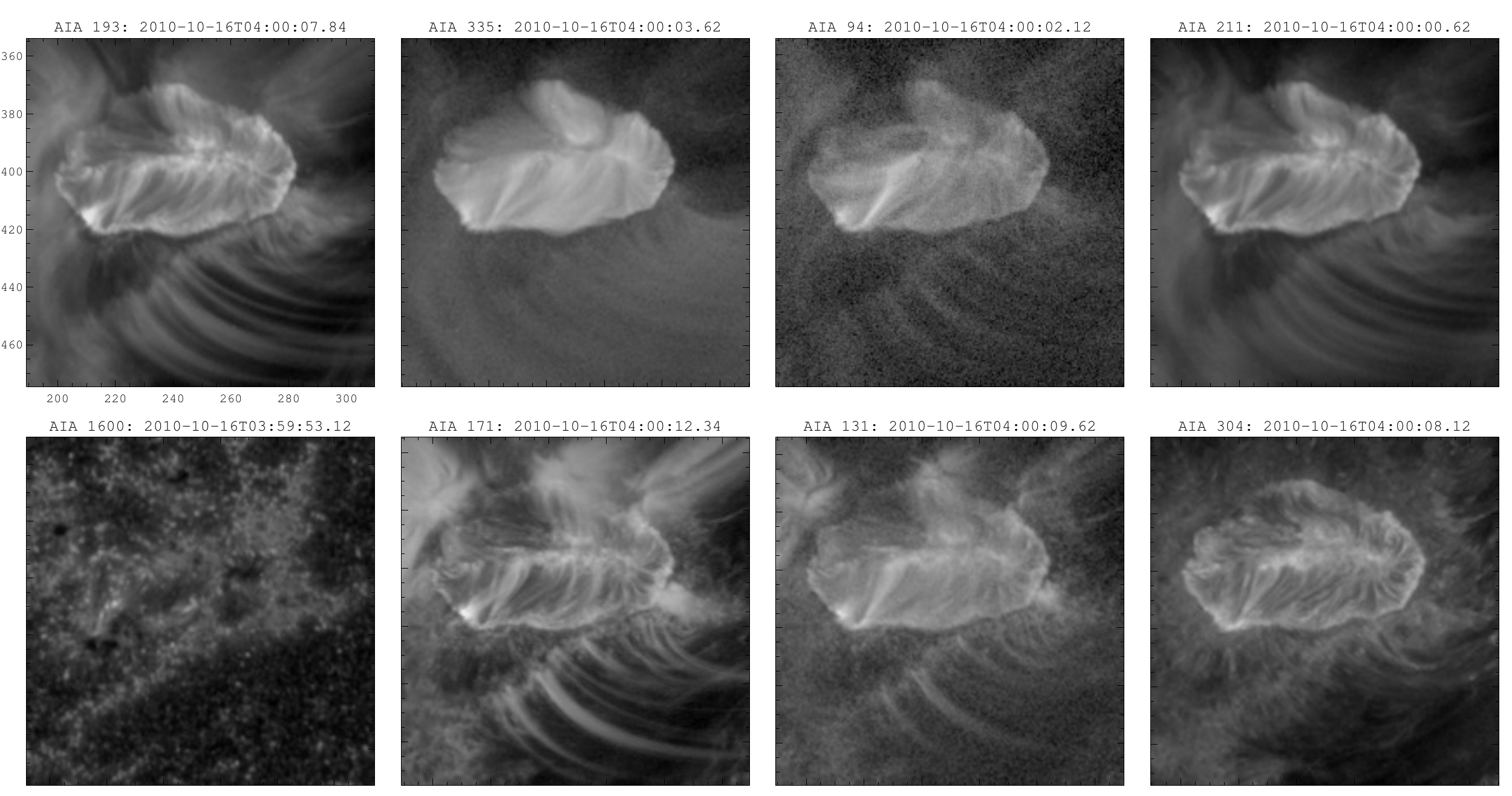}
    \caption[AR11112 detail of EUV channels]{\label{fig:channelcomp} Log--scaled UV and EUV data from AIA, for the timestep shown in \figref{fig:magevo} and FOV shown as the rectangle in the figure.  The dark band dividing the two flux domains is visible in every EUV band, and from \figref{fig:magevo} we see that it surrounds the positive emerged flux, and lies completely within the negative flux regions.}
  \end{center}
\end{figure}

As magnetic flux emerges from beneath the photosphere, we observe new magnetic loops in the EUV and X--ray images taken with AIA and XRT.  These appear as a bright cluster of short loops connecting the newly emerged positive and negative flux, as seen in \figref{fig:channelcomp} and the animation thereof.  These new loops displace the previously extant field in the coronal volume immediately above the photospheric emergence.  The displacement takes two forms.  The first is through horizontal shearing of the photospheric plasma to which the coronal field is anchored; the second is through generation of electric currents on the boundary between the old and new magnetic domains, expelling the previous field and supporting the new field within the coronal volume.

The new flux emerges completely within an area of preexisting negative unipolar flux, with the PIL located to the Solar west.  As discussed in \S\ref{sec:tracking}, this means the positive emerged flux is surrounded by a sea of negative flux, generating a magnetic dome topology.  One of the null's spine sources is located underneath this dome, while the other sits in the unipolar region of positive flux across the diffuse PIL.  Within the dome fieldlines connect to the newly emerged positive flux, while outside the dome fieldlines connect to the old positive flux, thus defining two topologically distinct magnetic domains.

The boundary between the two domains is easily identified in the EUV observations as a ``dark band'' that partially rings the newly emerged flux.  As \figref{fig:channelcomp} shows, the band is mostly easily seen along the southern portion of the active region core, and appears in all EUV wavelengths.  EUV loops to one side of the band are short, bright, and connect to the newly emerged positive flux, while those on the other are long, diffuse, and appear to connect in the large area of unipolar positive flux across the diffuse PIL.  We call the set short bright loops the \emph{kernel}.

The properties of the band that separate the two domains are discussed in detail in \citet{Scott:2013}.  Notable among these are continuous, persistent (lasting several days) blueshifts of $\sim 25\unit{km/s}$ observed in Hinode/EIS.  The band itself is approximately cospatial with the intersection of the coronal null's separatrix surface with the photosphere.  In the MCT, this ring is a circuit of spine lines passing through both old and new negative flux concentrations.

\begin{figure}[ht]
  \begin{center}
    \includegraphics[width=\textwidth]{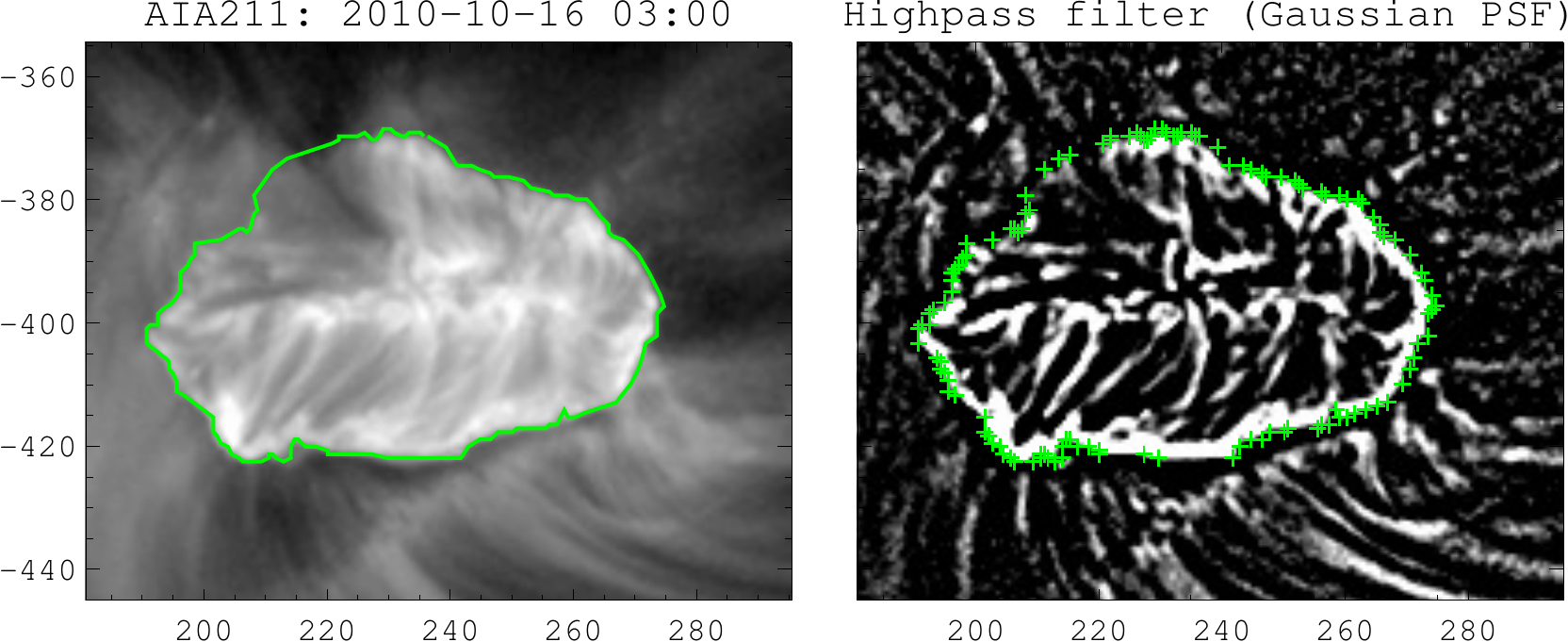}
    \caption[AR11112 EUV boundary]{\label{fig:euvbnd} Boundary of kernel flux and external flux, determined from EUV images.  Right: 211\AA{} data after convolution with a Gaussian highpass filter.  Green plusses show the by--eye chosen pixels that define the boundary.  Left: log--scaled 211\AA{} image with flux domain boundary superposed as a solid green line.}
  \end{center}
\end{figure}

To find precise spatial boundaries of the kernel at each timestep, we use data from the AIA 211\AA{} channel at every hour beginning at 2010-10-14 00:00 UT and ending on 2010-10-17 11:00 UT.  We trace the boundary by eye in a highpass filtered version of each EUV image, as shown in \figref{fig:euvbnd}.  The highpass filtered image is created by convolving the image with a 10 pixel by 10 pixel Gaussian kernel with a standard deviation of 5 pixels and subtracting that from the original image.  After this process, the dark band itself is easily visible along the east, south, and west sides of the active region kernel.  The northern boundary is usually obscured by both the loops within the kernel and loops connected to the patch of highly concentrated preexisting negative flux just north of the emergence zone, labeled N1.  In the animation \figref{fig:channelcomp}, we see that the diffuse set of loops connected to N1 pass across the northern portion of the AR kernel like the beam from a lighthouse.

\figref{fig:euvbnd} shows the AIA 211\AA{} channel from \figref{fig:channelcomp}.  Displayed is the 2010-10-16 03:00 UT image in a logarithmic greyscale on the left and after covolution with the highpass filter on the right.  The selected points are shown as green plus marks in the right--hand image.  We select points on the inside of the dark band, at the edge of loops connecting to P1.  The boundary defined by these points is shown as a solid green line in the left--hand image.  A similar boundary is created for each timestep in our data series.

We can first identify the boundary at 2010-10-14 22:00 UT.  This is approximately 17 hours after the first signs of emergence are visible in the magnetic field data, and 13 hours after the bright loops first become visible in the AIA:211\AA{}, XRT:Ti--poly, and XRT:Al--mesh data.

\section{\label{sec:rxrate}Inferred Reconnection Rate}
We overlay the boundary between the two coronal domains (loops connected to emerged positive flux, and loops connecting elsewhere) determined in \S\ref{sec:euv} over the closest--in--time radial magnetic field at each timestep.  See, for example, the thick line in \figref{fig:magevo}.  The intersection of this boundary with the contour of pre--emergence negative flux to the west of emergence (N2) provides us with a lower--bound estimate of the amount of reconnection between the old and newly emerged flux.  This is quite a conservative estimate because we exclude any reconnection to N13 from our analysis, owing to that region's proximity to the emergence zone; and further, after Oct 15th at 23:00 UT we begin to see the EUV boundary pass through the contour for N1, north of the emergence zone, indicating additional reconnection with that preexisting concentration.  We will return to these considerations in more detail below.

\begin{figure}[ht]
  \begin{center}
    \includegraphics[width=\textwidth]{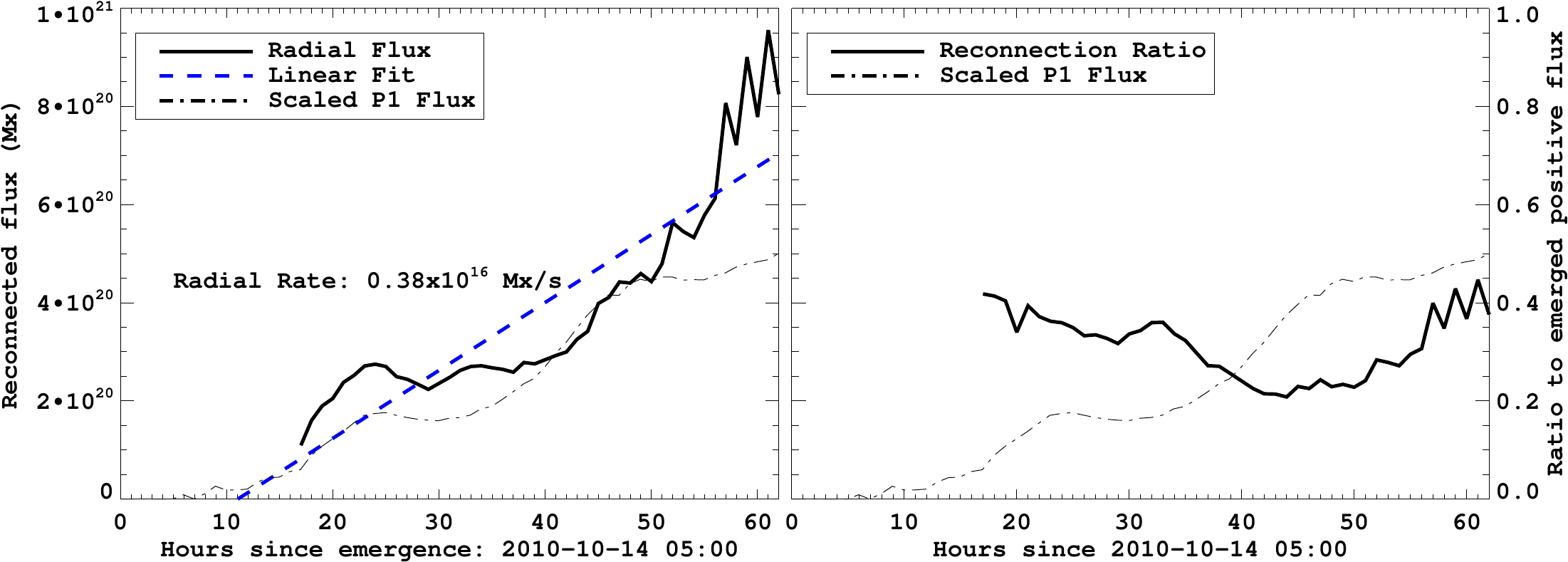}
    \caption[AR11112 inferred reconnection rate]{\label{fig:phi_rx} Left: Estimated reconnected flux determined by the overlap between EUV boundary and the boundary of N2 from radial magnetic flux (solid black line) and a linear fit (dashed blue).  Right: Ratio of the reconnected flux from the radial field to the total emerged positive flux (solid).  In both plots, the total emerged positive flux is scaled to half the range of the y--axis (dash--dot).}
  \end{center}
\end{figure}
\figref{fig:phi_rx} plots the amount of flux in the intersection of the EUV boundary and N2 boundary at each timestep.  Because the EUV boundary delineates fieldlines connecting to the newly emerged positive and those connecting to preexisting positive flux, and because N2 is a region containing only preexisting negative flux, the amount of flux in the intersection is a measure of the amount of reconnected flux at a given time.  From the linear fit to the amount of reconnected flux (dashed blue line, left panel of \figref{fig:phi_rx}), we determine a relatively steady reconnection rate of $0.38\times 10^{16}\unit{Mx/s}$ over about 2 days.  This corresponds to a characteristic EMF of $\mathcal{E}=\frac{d\Phi}{dt}=38\times 10^6$ volts within the reconnection region.

We can estimate the fraction of emerged flux that has reconnected at a given time by taking the ratio of the reconnected flux in N2 to the total amount of newly emerged positive flux P1.  This ratio is shown as the solid line in the right side plot of \figref{fig:phi_rx}, and varies between $25\%$ and $46\%$ of the emerged flux.  The amount of total emergence is depicted as a dashed--dot line, which is the flux of P1 scaled to half the y--axis range.  The absolute amount of emergence is not important for this comparison so much as the relative rate of emergence at different times.  The ratio may vary in phase with the rate of emergence, and therefore with the rate at which the free energy density increases in the coronal field, though this is difficult to precisely establish.  

Consider the decrease in the ratio from $t=30$ to $t=50$ hours.  During this time, the fraction of reconnected flux drops by less than half, while the amount of emerged flux more than doubles.  We see that the reconnection rate may be tracking the emergence rate to some degree, but that coupling is not simple.  Indeed, from the linear fit of the reconnection rate in the left plot of \figref{fig:phi_rx}, it may take at least 11 hours after flux emergence before any existing field reconnects with new field.

Secondly, some of the variations we see in the amount of emerged flux are consistent with the diurnal variations in the measured magnetic field of HMI data \citep{Liu:2012}.  See in particular the ``humps'' in the dashed--dot line of normalized P1 emergence in \figref{fig:phi_rx}, around $t=25$ and 50 hours.  Because we are primarily concerned with the average reconnection rate over time, and because these variations equally affect the measured magnetic field throughout the field of view, we will not concern ourselves with this added source of uncertainty.

Finally, around $t=42$ hours after emergence, loops begin unequivocally to form between P1 and N1 (see the animation of \figref{fig:channelcomp}).  These are excluded from our analysis for the reasons discussed above.  Reconnection may already have occurred by $t=42$ between the two regions, although it is difficult to tell.  Parts of N1 form the footpoints of loops that appear to arch over the top of the emerging flux to connect with positive flux across the large scale PIL, obscuring the interaction between N1 and P1 until $t=42$.  For this reason we provide the conservative estimate limited to reconnection with N2.  

Regardless of these ambiguities, the total amount of reconnection between P1 and N1 is small, about $11\%$ of that between P1 and N2.  By the end of the time range we consider, $62$ hours after first emergence, the intersection between the EUV boundary and N1 contains $1.0\times 10^{20}\unit{Mx}$, compared to $8.9\times10^{20}\unit{Mx}$ for the intersection with N2.  The intersection of the EUV boundary with N13 contains $1.7\times10^{20}\unit{Mx}$.

As a final note, the amount of emerged flux shown as the dashed line in the right hand plot of \figref{fig:phi_rx} begins around $t=5$ hours after first emergence.  This start time is simply due to the thresholds we set in the creation of our mask structures.  It is unlikely to affect our linear fit to the reconnection rate, which shows first reconnection at $t=11$, because we are unable to measure any reconnected flux until $t=17$, 9 hours after P1 rises above the mask thresholds, and 17 hours after first noted emergence in the vector field data.

\section{\label{sec:xrt}Temperature, Emission Measure, and Radiated Power}
We generate maps of temperature $T$ and volumetric emission measure $EM$ at each pixel in the XRT field of view using the filter ratio method described in \citet{Narukage:2011}.  This analysis assumes an isothermal plasma.  Ratios of different filters provide more or less tightly constrained temperatures, with some ratios being ill--defined (multivalued) in certain regions, or highly dependent on the assumed isotope abundances.  The ratio of Ti--poly to Al--mesh filters provides one of the cleanest temperature and emission measure signals for expected plasma parameters in nonflaring active regions.  As discussed in that paper, the filter ratio method for a broadband instrument like XRT should correspond to a mean plasma temperature, weighted by the DEM of the emitting coronal plasma.  For this reason, it should result in a reliable measure of the radiated power.

We use a modified version of the \fnc{xrt\_teem.pro} program available in SSW to derive the $T$ and $EM$ maps.  The modified version, described in \citet{Takeda:2012}, allows for different elemental abundances based on whether the emitting plasma is coronal, photospheric, or some mix of the two.  The different choices of abundances affect the calculation of XRT temperature responses and the resulting filter ratios.  Because the present chapter focuses on actively emerging flux, we use the hybrid model, with abundances based on \citet{Fludra:1999}.  The EM in each pixel is given by 
\begin{align}
  \label{eq:em}
  EM= \int n_e^2 dV\qquad,
\end{align}
where $n_e$ is electon density within the pixel and the integral is over the volume of emitting plasma $V$.  Assuming the density is constant throughout the volume, the emission measure is given simply by $n_e^2 V$.  If we further assume a radiative loss function $\Lambda(T)$ we can estimate the power $P$ radiated from each pixel by
\begin{equation}
  P = \Lambda(T)n_e^2 \times V = \Lambda(T) EM(T)\qquad.
\end{equation}
$\Lambda(T) = a(T) T^{f(T)}$ is the radiative loss function of \citet{Klimchuk:2001}, summarized in Table \ref{tab:radloss}.  We then calculate the observed radiated power $P$ by
\begin{equation}
  \label{eq:power}
  \log (P) = \log(EM) + \log (a) + f(T) \log(T)\qquad .
\end{equation}

\begin{table}[ht]
   \centering
     \caption{\label{tab:radloss}  Parameters of the radiative loss function $\Lambda(T)= a(T) \times T^{f(T)}$ given by \citet{Klimchuk:2001}.}
   \begin{tabular}{ccc}
     \tableline\tableline
     Temperature range (K) & $a(T)$ & f(T)\\
     \tableline 
     $\phantom{T_0< }T  < T_{0} = 1.0000\times10^4 $ & $0.0000$      &$ 0            $  \\
     $T_0< T < T_1 = 9.3325\times10^4$ & $1.0909\times10^{-31}$      &$+ 2.          $  \\
     $T_1< T < T_2 = 4.67735\times10^5$   & $8.8669\times10^{-17}$   &$-1.           $  \\
     $T_2< T < T_3 = 1.51356\times10^6$   & $1.8957\times10^{-22}$   &$ 0            $  \\
     $T_3< T < T_4 = 3.54813\times10^6$   & $3.5300\times10^{-13}$   &$-\frac{3}{2}  $  \\
     $T_4< T < T_5 = 7.94328\times10^6$   & $3.4629\times10^{-25}$   &$+\frac{1}{3}  $  \\
     $T_5< T < T_6 = 4.28048\times10^7$   & $5.4883\times10^{-16}$   &$- 1.0         $  \\
     $\phantom{T_0< }T_6< T$\phantom{$_6 = 4.28048\times10^7$} & $1.9600\times10^{-27}$   &$+\frac{1}{2}  $  \\
     \tableline
   \end{tabular} 
 \end{table}

XRT suffers from a time--varying contamination that collects on the CCD, obscuring about $5\%$ of the area in the form of condensation formed spots \citep{Narukage:2011}.  While the locations of these spots are known, their spectral response and time dependent thickness are not.  In order to compensate for this in our analysis, we smoothly interpolate the radiated power along the boundary of each spot into the spot's interior.  To do so we iteratively call the standard IDL \fnc{smooth} function on the power data, which convolves the data with a 2D boxcar function.   At each iteration we apply a 3 pixel by 3 pixel boxcar kernel, and we only update the contaminated pixels so that no uncontaminated data are affected.  The smoothing converges by 500 iterations, at which point the value of each pixel within a contaminated spot is the average of the value at each pixel on the spot's boundary, weighted by the distance to each boundary pixel.  This provides a more reasonable estimate of the radiated power than simply setting the value within the spots to zero.  

\begin{figure}[ht]
  \begin{center}
    \includegraphics[width=\textwidth]{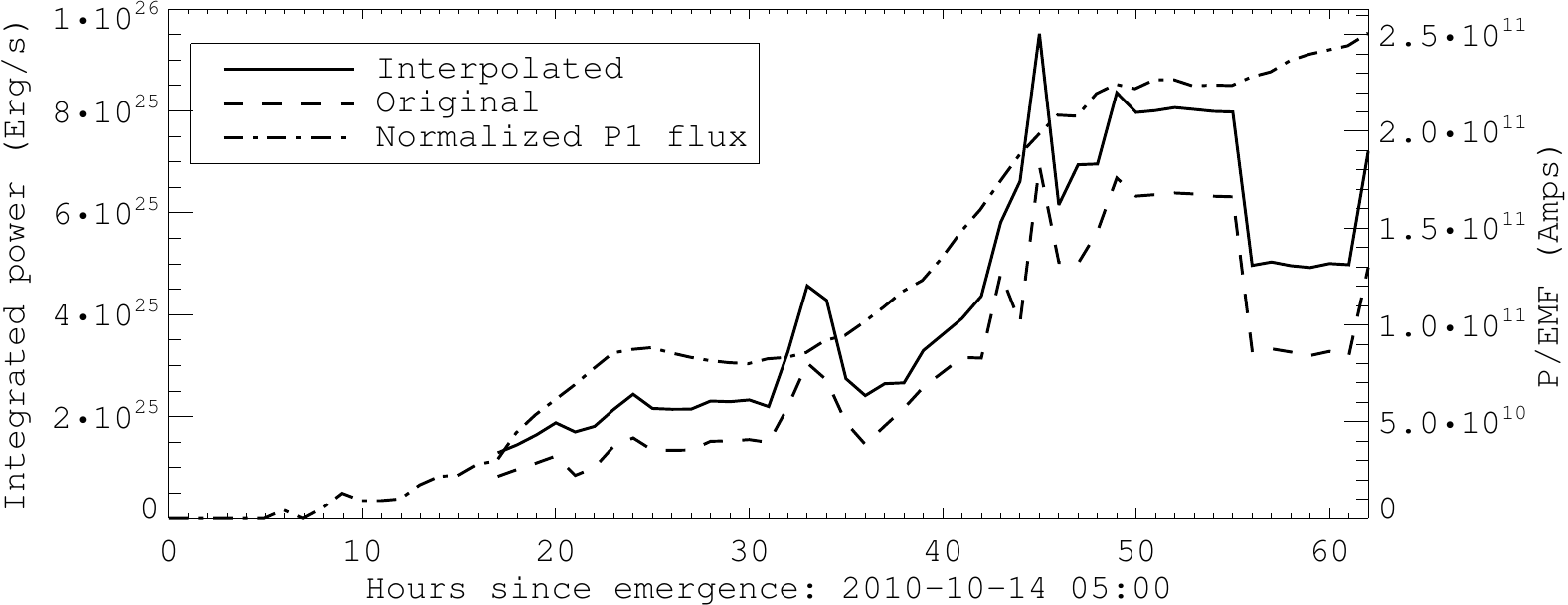}
    \caption[AR11112 inferred radiated power]{\label{fig:power} Left axis: Radiated power derived from XRT filter ratios, from \eqref{eq:power}, in erg/s.  The solid line uses data interpolated into contaminated pixels, and the dashed line uses the original data.  Right axis: characteristic current in the reconnection region assuming a constant EMF of $0.38\times 10^{16}\unit{Mx/s} = 38\times 10^{6}\unit{V}$.  For easy comparison with \figref{fig:phi_rx}, the scaled emergence of P1 is again overlaid as a dash--dot line.  See text for a discussion of the drop near t=55.}
  \end{center}
\end{figure}

\figref{fig:power} shows the total power radiated by the plasma from pixels within the same boundary defined using the EUV data.  The solid line is the total after the interpolation described above, while the dashed line is the total setting each contaminated pixel to zero.  Interpolation increases the estimated radiated power by a roughly constant factor of $50\%$, consistent with the ratio of contaminated to uncontaminated pixels within the EUV defined boundary.  The power rises from $2$ to $8\times10^{25}\unit{erg/s}$ as the active region emerges.  

Note that there was an 11.5 hour gap in the XRT data between $t=50$ and $61.5 \unit{hrs}$ as XRT observed a different area of the Sun.  This creates the two flat portions of the calculated power for timesteps close to those two data points.  Additionally, at $t=61.5$ the kernel was obscured by a particularly large patch of contaminated CCD pixels, which likely led to the significantly decreased radiated power determined at those times.

The dash--dot line in the figure is the normalized emergence of P1, as in \figref{fig:phi_rx}.  The radiated power tracks well with the total emerged flux.  However, because of the large sources of error in our estimation of the radiated power, in particular due to the spots on the XRT CCD and lapses in the timeseries, we cannot currently perform a more quantitative comparison.

The total energy radiated away by the system over the course of our analysis is determined by summing the radiated power over time.  The system radiates $\sim3.2\times 10^{32}\unit{ergs}$ over the 45 hours between establishment of the EUV boundary and the GOES M3.0 flare.

\section{\label{sec:discussion}Discussion}

In this work we present, to our knowledge, the first observations of quiescent reconnection between an emerging flux bundle and surrounding, preexisting field.  The rate of reconnection is approximately steady at $0.38\times 10^{16}\unit{Mx/s}$ over the course of two days, though as noted above this is an underestimate because we focus only on one region of unambiguously preexisting negative flux.  We observe EUV loops connecting to other negative regions as well, but often the EUV boundary is difficult to determine in these locations or there is some mix between emerged and preexisting negative flux. 

While AR11112 does produce an M3.0 flare on Oct 16th, 2010, the implied reconnection we focus on here predates the flare by several days.  It produces little if any observable flaring active of its own, such as increases in the GOES light curve or chromospheric flare ribbons.  Further, we do not observe any sudden jumps in the amount of reconnected flux.  Based on these observations, we conclude that this is indeed an instance of quiescent reconnection.

\begin{figure}
  \begin{center}
    \includegraphics[width=\textwidth]{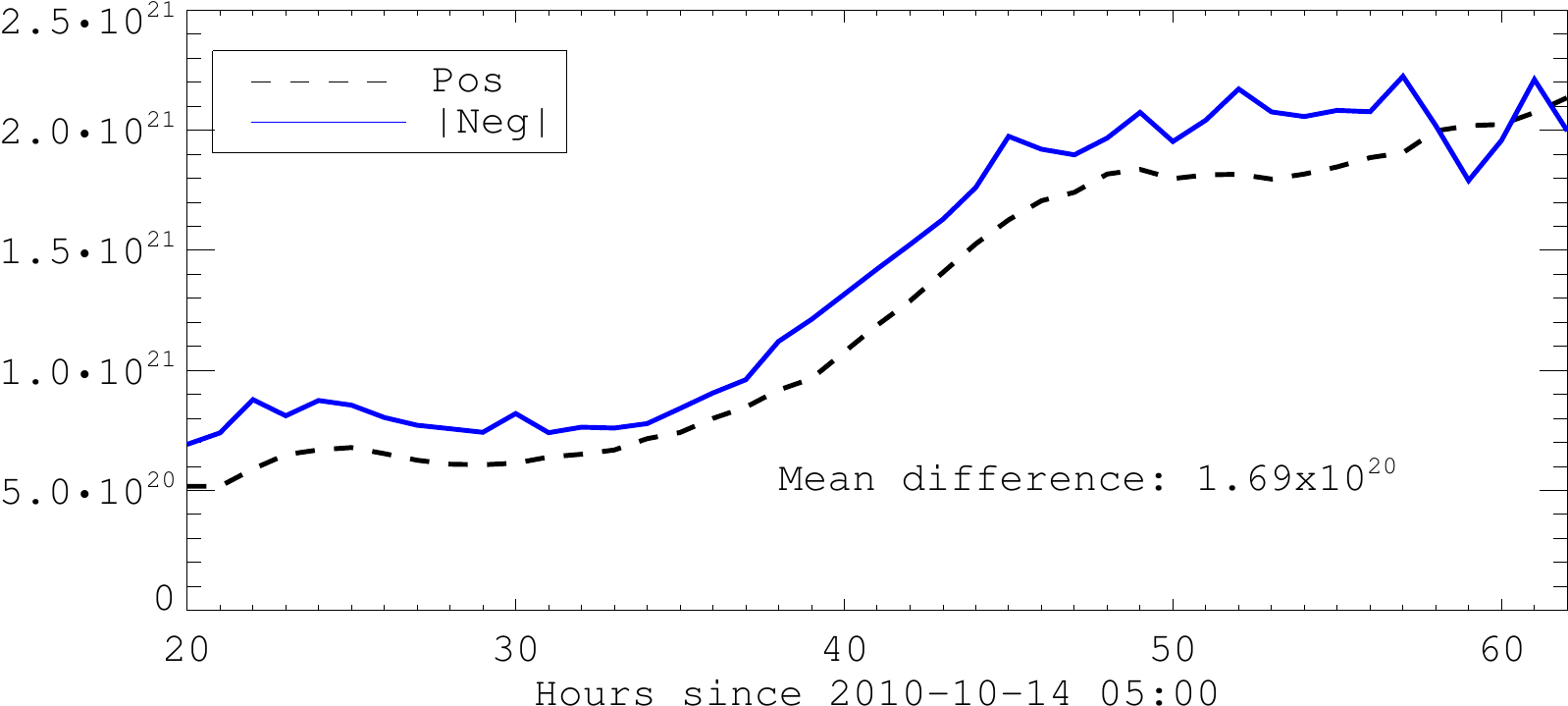}
    \caption[Flux balance within EUV boundary for AR11112]{\label{fig:fluxbalance}  Total positive (black dashed) and absolute value of negative (solid blue) flux within the EUV boundary.  All positive flux within the boundary is due to emergence, while the negative flux is a mix of preexisting and emerged flux.  The mean difference between the two curves is $1.69\times10^{20}\unit{Mx}$ with a variance of $\approx 1.0\times10^{20}\unit{Mx}$, between $5-10\%$ of the total.}
  \end{center}
\end{figure}

The total amount of signed flux within the EUV boundary is well balanced, as shown in \figref{fig:fluxbalance}.  This value is only determined for times when we can observe the boundary.  We find a mean signed flux within the boundary of $1.69\times10^{20}\unit{Mx}$, with a variance of $\approx 1.0\times10^{20}\unit{Mx}$.  The consistancy of the imbalance over time as new flux emerges indicates that we have accurately captured the evolution of this system.  Because our boundary passes through negative flux regions consisting of both emerged and preexisting flux, and because reconnection necessarily transfers flux 1--to--1 between flux domains, this is a further indication that the feature we observe in EUV is in fact the separatrix surface of the coronal null; flux within such a surface must be balanced, even as the surface changes due to continued emergence and reconnection.

Our analysis is conceptually similar to those attempting to determine the amount of flux involved in two ribbon flares.  \citet{Qiu:2010} perform a similar measurement by summing the flux for magnetogram pixels cospatial with UV flare ribbons during the well studied Bastille Day (2000) two ribbon flare.  Their event lasts for less than an hour and their measured rate varies by an order of magnitude, between $10^{18}$ and $10^{19}\unit{Mx/s}$ during this time, resulting in about $10^{22}\unit{Mx}$ total transferred flux.  \citet{Kazachenko:2010} apply a similar method to another heavily studied flare, the Halloween flare of 2003, again finding on order $10^{22}\unit{Mx}$ of flux transferred during the flare.  These two events are among the strongest flares ever recorded for the Sun \citep{Schrijver:2012}---GOES classes X5.7 and X17, respectively---and several orders of magnitude larger in GOES class than any event in AR11112 during the time of our analysis.  Direct comparison of reconnection rates and total flux transfers between these cases is thus problematic.

In Chapter 3, we used the conceptually unrelated free energy minimization scheme to estimate flux transferred during the M6.6, M2.2 and X2.2 flares of February 2011 in AR11158.  We found an amount of flux involved in each flare of $4.2\times 10^{20}\unit{Mx},\ 2.0 \times 10^{20}\unit{Mx} , \hbox{and } 21.0 \times 10^{20}\unit{Mx}$, respectively, 1--2 orders of magnitude less than that for the Bastille Day and Halloween flares.  This shows that the reconnected flux we measure here of $\approx 1\times10^{21}\unit{Mx}$ may be consistent with the amount of flux involved in a smaller GOES X--class flare.  At the same time, the total amount of radiated flux originating within the EUV boundary over the time series is $3\times10^{32}\unit{ergs}$, also consistent with an X--class flare.  In terms of the reconnection and radiative processes ongoing throughout AR11112's emergence, this entire event \emph{is} an X--class flare, but simply takes 45 hours instead of 30 minutes.

Perhaps one of the most puzzling results is the apparent delay between the time of photospheric emergence and any measureable amount of reconnected flux, both during the initial emergence and subsequent surges.  Recall that we see our first signatures of flux emergence in the photosphere around 2010-10-14 05:00 UT as rapidly expanding regions of highly inclined magnetic field.  4 hours later we detect our first signatures in the 211\AA{} images as locally enhanced emission measure.  3 hours after that we see the first well defined EUV loops.  Finally, at UT 22:00, 17 hours after emergence, we observe a fully formed kernel of bright loops bounded by a persistent dark band.  By taking a linear fit to the next 45 hours of our measured reconnected flux, we infer that reconnection began 11 hours after first emergence.  The authors of \citet{Longcope:2005b} use a rather different method (yet also using EUV loops) to estimate a reconnection rate between two active regions, one emerging one preexisting.  They find evidence of reconnection between 6 and 24 hours after emergence.

Another important point is that flux is constantly emerging through the photosphere, even in quiet sun regions during solar minimum.  It is likely that even this small scale emergence operates in a similar fashion to the larger, active region scale phenomena described above.  Much work has recently been devoted to the numerical study of emergence--induced x--ray jets and bright--points, and the related problem of network flux ``recycling'' times: the rate at which emerging small scale flux reconnects to form new footpoints for open magnetic field \citep{Schrijver:1997,Archontis:2008, Hagenaar:2008, Cranmer:2010}.  The latest of these investigations, \citet{Cranmer:2010}, focuses on energetic consequenses of flux emergence and reconnection in the magnetic carpet \citep[first described in]{Schrijver:1997}.  Those authors compare the recycling time for flux emergence (time required for flux to emerge from below the photosphere) and the rate at which closed flux becomes open.  The two timescales compare favorably (see Figure 10(a) of \citet{Cranmer:2010}), and for regions of highly imbalanced initial flux, such as we have for AR11112, that timescale is about 11 hours.  This comparison should not be taken too far, as we have shown the exact relation between their simulations and our observations, but it is a comparison that should be considered further.

\chapter{\label{ch:retro}Retrospective}
The major goal of this disseration has been to model active region evolution in the case of substantial flux emergence.  The importance of flux emergence in stressing the coronal field so that it produces flares and CMEs has been appreciated since the 1970s \citep{Rust:1975}.  While the dynamics of the coronal field are ultimately governed by the MHD equations \eqref{eq:cont}--\eqref{eq:induct}, their difficult nature means that analytic solutions are generally impossible to obtain.  At the same time, for coronal parameters, solving them numerically is both computationally expensive and difficult to interpret.  

For this reason, we have modeled active region evolution in the context of an MCT/MCC model.  To do so in cases where flux emergence drives active region dynamics, the ideal evolution imposed by the FCE constraint of the MCC must be modified.  In this, we succeeded.  We applied the modified model to two interesting cases of active region emergence: AR11112 in Chapter 2, and AR11158 in Chapter 3.  Although physically simple when compared to solving the full MHD equations, our topological model produces the correct order of magnitude free energy required to power solar flares.  It agrees favorably with the energies determined by much more computationally expensive methods, but most importantly agrees with observations of the total power radiated during flares.  Our model also incorporates the boundary data in a less incorrect way compared to NLFFF models or MHD simulations.

The relative simplicity of the model makes it a powerful tool for studying the coronal stresses induced by photospheric evolution and the topology of the field.  This aspect allowed us to introduce our new method for realistically relaxing the stressed state of coronal field by allowing reconnection between domains.  Formerly, relaxations within the MCC model brought the field all the way to a potential field state \citep{Close:2004, Kazachenko:2010, Longcope:2010}.  Total relaxation is unlikely, especially for active regions with repeated large flares in a relatively short time.  \citet{Longcope:2010} noted that the free energy of the MCC, which they assumed was all dispelled during their flare,  was actually greater than the energy loss calculated directly from oberservations of light curves for their flare.  They determined that the potential field was, in fact, unobtainable, but they could not comment further due to the limitations of their model.  Our model for the relaxation in Chapter \ref{ch:ar11158} provides the answer.

The relaxation algorithm therefore leads to the interesting result that the minimum energy state is generally unobtainable by reconnection, at least in the process we described.  At some point, several domains  have transferred away all of their flux, so that no further transfer is possible that decreases the free energy.  For the flares modeled in AR11158, some of these domains clearly contained flux and were involved in each flare, owing to the strong chromospheric flare ribbons running through their footpoints.  However, these domains paired photospheric regions originating in distinctly separate emergence regions, which left the question of how they received their flux in the first place unanswered.

That question leads us finally back to AR11112, to determine directly from simultaneous polarimetric (with inferred magnetograms) and EUV observations the amount of reconnection between separately emerged photospheric magnetic concentrations.  This final study was a particularly useful exercise, in that it quantified for the first time an idea generally assumed to be true throughout the field of solar physics, namely, that emerging flux reconnects with surrounding field.  This work is certainly related to the photospheric and coronal field recycling times \citep{Hagenaar:1997, Close:2005, Hagenaar:2008}, but tackles the problem in the context of active regions.  \citet{Longcope:2005b} undertook a similar study, but were limited by the spatial and temporal resolution of the instruments available at the time.  Indeed, in our study, we could have attempted to quantify new loops in the N5--P4 domain (see \figref{fig:magevo}).  This domain consists of long ($\sim100\unit{Mm}$) diffuse loops, whose relative brightness varies little due to the reconnection given the amount of plasma across which transferred energy is distributed.  Focusing on the P1--N5 domain, with its short $(\lesssim10\unit{Mm})$, well defined loops and easily identified boundary circumvents this problem, and was enabled by the high spatial and temporal resolution, and continuous cotemporal observations, of the suite of SDO instruments.

One obvious use of the extended MCC framework, and particularly the relaxation algorithm, is to study the stability of various coronal field configurations.  For instance, in the case of AR11158, the small tertiary bipole (P52/N56) that emerges just prior to the M2.2 flare appears to destabilize the active region, culminating in the X2.2 flare.  Numerous authors have already considered this question \citep{Sun:2012b, Petrie:2012}.  We are forced to consider how important the details of that small emergence actually are.  For instance, if there was no such emergence but the region had continued to shear between the two large scale bipoles, could the region still produce an X--class flare?  Or, if such a bipole had emerged in the Northwest of the active region complex, would that trigger a large eruption, or was the location of emergence in the East--central portion important?  Our relaxation method allows us to explore these questions.  Such a study, with a goal of determining the relative stability of active regions, should certainly be carried out, and has strong implications for solar flare prediction.

Expanding the work of Chapter \ref{ch:rxrate} to a statistical analysis would also be very useful.  As discussed in \citet{Archontis:2008}, the relative orientation of emerging flux to the background field greatly affects the amount of reconnection.  A statistical survey of that reconnection process would certainly advance our understanding of coronal heating and the transport of flux from below the photosphere to the corona.  It would also provide useful constraints for numerical MHD models.  Such a study would require the time--consuming development of new automatic tracking algorithms, as current algorithms in the Heliophysics Event Knowledgebase (HEK) \citep{Hurlburt:2012} prove insufficient to this specialized task (Derek Lamb, private communication).
\backmatter

\bibliographystyle{apj}
{
  \singlespace

}
\end{document}